\documentclass[11pt]{article}

\usepackage{amsmath,amsbsy,amsfonts,amssymb,dsfont,units}
\usepackage{amsthm}
\usepackage[alphabetic]{amsrefs}
\usepackage{graphicx}
\usepackage{authblk}
\usepackage{todonotes}
\usepackage{stmaryrd}
\usepackage[margin=1in]{geometry}
\usepackage[bottom]{footmisc}

\usepackage{etoolbox}

\usepackage{algorithm,algorithmic,mathtools}

\usepackage{color,cases}
\usepackage{tikz}
\usetikzlibrary{calc,shapes}
\usepackage{subfigure}
\usepackage{hyperref}

\newtheorem{theorem}{Theorem}[section]
\newtheorem{thm}[theorem]{Theorem}
\newtheorem*{thm*}{Theorem}

\newtheorem{cor}[theorem]{Corollary}

\newtheorem{lemma}[theorem]{Lemma}
\newtheorem{lem}[theorem]{Lemma}
\newtheorem*{lem*}{Lemma}

\newtheorem{claim}{Claim}[section]

\theoremstyle{definition}

\newtheorem*{rem*}{Remark}
 \newcommand{\IGNORE}[1]{}

\newcommand{\citep}{\cites}
\newcommand{\citet}{\cites}

\newcommand\E{\mathbb{E}}
\newcommand\EE{\mathbb{E}}

\newcommand\R{\mathbb{R}}
\newcommand\RR{\mathbb{R}}

\newcommand\cF{\mathcal{F}}

\DeclareMathOperator{\var}{Var}

\newcommand{\wb}{\overline}
\newcommand\wh{\widehat}

\newcommand\wt{\widetilde}
\newcommand\eps{\varepsilon}
\newcommand\veps{\varepsilon}

\newcommand{\Or}{\mathcal{O}}

\newcommand{\rd}{\mathrm{d}}
\newcommand{\ud}{\,\mathrm{d}}

\newcommand{\abs}[1]{\lvert#1\rvert}
\newcommand{\norm}[1]{\lVert#1\rVert}

\newcommand{\ifstoc}[2]{\iftoggle{stoc}{#1}{#2}}

\newtoggle{final}
\toggletrue{final}
\newtoggle{arxiv}
\toggletrue{arxiv}
\newtoggle{stoc}
\togglefalse{stoc}

\newcommand{\revise}[1]{{\draft{\color{blue}} #1}}
\newcommand{\draft}[1]{\iftoggle{final}{}{#1}}

\newcommand{\hlcom}[1]{\draft{\noindent{\textcolor{purple}{Holden: #1}}}}

\newcommand{\Pj}[0]{\mathbb{P}}

\newcommand{\cO}[0]{\mathcal{O}}

\newcommand{\bS}[0]{\mathbb{S}}

\newcommand{\al}[0]{\alpha}
\newcommand{\be}[0]{\beta}
\newcommand{\ga}[0]{\gamma}

\newcommand{\de}[0]{\delta}

\newcommand{\ep}[0]{\varepsilon}
\newcommand{\eph}[0]{\frac{\varepsilon}{2}}

\newcommand{\ka}[0]{\kappa}

\newcommand{\rh}[0]{\rho}

\newcommand{\Te}[0]{\Theta}

\newcommand{\Om}[0]{\Omega}
\newcommand{\si}[0]{\sigma}

\newcommand{\nin}[0]{\not\in}

\newcommand{\ot}[0]{\otimes}

\newcommand{\subeq}[0]{\subseteq}

\newcommand{\iy}[0]{\infty}

\newcommand{\rc}[1]{\frac{1}{#1}}
\newcommand{\prc}[1]{\pa{\rc{#1}}}

\newcommand{\fc}[2]{\frac{#1}{#2}}
\newcommand{\sfc}[2]{\sqrt{\frac{#1}{#2}}}
\newcommand{\pf}[2]{\pa{\frac{#1}{#2}}}

\newcommand{\dd}[2]{\frac{d #1}{d #2}}

\newcommand{\ddd}[1]{\frac{d}{d #1}}

\newcommand{\nb}[0]{\nabla}

\newcommand{\dx}{\ud x}
\newcommand{\rdx}[0]{\mathrm dx} 
\newcommand{\dt}{\ud t}
\newcommand{\dB}{\ud B}

\newcommand{\ab}[1]{\left| {#1} \right|}

\newcommand{\ba}[1]{\left[ {#1} \right]}
\newcommand{\bc}[1]{\left\{ {#1} \right\}}

\newcommand{\ce}[1]{\left\lceil {#1}\right\rceil}
\newcommand{\fl}[1]{\left\lfloor {#1}\right\rfloor}

\newcommand{\pa}[1]{\left( {#1} \right)}
\newcommand{\pat}[1]{\left( \text{#1} \right)}
\newcommand{\ve}[1]{\left\Vert {#1}\right\Vert}

\newcommand{\ved}[0]{\ve{\cdot}}
\newcommand{\set}[2]{\left\{{#1}:{#2}\right\}}

\newcommand{\ol}[1]{\overline{#1}}

\newcommand{\ub}[2]{\underbrace{#1}_{#2}}

\newcommand{\amin}{\operatorname{argmin}}

\newcommand{\Var}[0]{\operatorname{Var}}

\providecommand{\cal}[1]{\mathcal{#1}}
\renewcommand{\cal}[1]{\mathcal{#1}}

\newcommand{\pull}[9]{
#1\ar@/_/[ddr]_{#2} \ar@{.>}[rd]^{#3} \ar@/^/[rrd]^{#4} & &\\
& #5\ar[r]^{#6}\ar[d]^{#8} &#7\ar[d]^{#9} \\}

\newcommand{\cmp}[9]{
\xymatrix{
#1 \ar[r]^{#4}{#5} \ar@/_2pc/[rr]^{#8}_{#9} & #2 \ar[r]^{#6}_{#7} & #3
}
}

\newcommand{\ha}[1]{\ar@{^(->}[#1]}
\newcommand{\ls}[1]{\ar@{-}[#1]}
\newcommand{\sj}[1]{\ar@{->>}[#1]}
\newcommand{\aq}[1]{\ar@{=}[#1]}
\newcommand{\acir}[1]{\ar@{}[#1]|-{\textstyle{\circlearrowright}}}
\newcommand{\acil}[1]{\ar@{}[#1]|-{\textstyle{\circlearrowleft}}}
\newcommand{\ard}[1]{\ar@{.>}[#1]}
\newcommand{\mt}[1]{\ar@{|->}[#1]}
\newcommand{\inm}[1]{\ar@{}[#1]|-{\in}}
\newcommand{\inr}{\ar@{}[d]|-{\rotatebox[origin=c]{-90}{$\in$}}}
\newcommand{\inl}{\ar@{}[u]|-{\rotatebox[origin=c]{90}{$\in$}}}

\newcommand{\sumo}[2]{\sum_{#1=1}^{#2}}
\newcommand{\sumz}[2]{\sum_{#1=0}^{#2}}
\newcommand{\prodo}[2]{\prod_{#1=1}^{#2}}

\newcommand{\coltwo}[2]{
\begin{pmatrix}
{#1}\\
{#2}
\end{pmatrix}}

\newcommand{\matt}[4]{
\begin{pmatrix}
{#1}&{#2}\\
{#3}&{#4}
\end{pmatrix}
}

\newcommand{\beq}[1]{\begin{equation}\llabel{#1}}
\newcommand{\eeq}[0]{\end{equation}}
\newcommand{\bal}[0]{\begin{align*}}
\newcommand{\eal}[0]{\end{align*}}
\newcommand{\ban}[0]{\begin{align}}
\newcommand{\ean}[0]{\end{align}}

\newcommand{\llabel}[1]{\label{#1}}

\newcommand{\arxiv}[1]{\url{http://www.arxiv.org/abs/#1}}

\allowdisplaybreaks[2]

\DeclareFontFamily{U}{wncy}{}
    \DeclareFontShape{U}{wncy}{m}{n}{<->wncyr10}{}
    \DeclareSymbolFont{mcy}{U}{wncy}{m}{n}
    \DeclareMathSymbol{\Sh}{\mathord}{mcy}{"58} 

\title{Estimating Normalizing Constants for Log-Concave Distributions: Algorithms and Lower Bounds}
\author{Rong Ge\thanks{Duke University, Computer Science Department \texttt{rongge@cs.duke.edu}}, \quad Holden Lee\thanks{Duke University, Mathematics Department \texttt{holee@math.duke.edu}}, \quad Jianfeng Lu\thanks{Duke University, Mathematics Department \texttt{jianfeng@math.duke.edu}}}

\begin{document}
\maketitle
\draft{\pagenumbering{roman}}

\begin{abstract}
    Estimating the normalizing constant of an unnormalized probability distribution has important applications in computer science, statistical physics, machine learning, and statistics. In this work, we consider the problem of estimating the normalizing constant $Z=\int_{\mathbb{R}^d} e^{-f(x)}\,\mathrm{d}x$ to within a  multiplication factor of $1 \pm \varepsilon$ for a $\mu$-strongly convex and $L$-smooth function $f$, given query access to $f(x)$ and $\nabla f(x)$.
    We give both algorithms and lowerbounds for this problem. Using an annealing algorithm combined with a multilevel Monte Carlo method based on underdamped Langevin dynamics, we show that     $\widetilde{\mathcal{O}}\Bigl(\frac{d^{4/3}\kappa + d^{7/6}\kappa^{7/6}}{\varepsilon^2}\Bigr)$ 
    queries to $\nabla f$ are sufficient, where $\kappa= L / \mu$ is the condition number. Moreover, we provide an information theoretic lowerbound, showing that at least $\frac{d^{1-o(1)}}{\varepsilon^{2-o(1)}}$ queries are necessary. This provides a first nontrivial lowerbound for the problem. 
\end{abstract}
\pagebreak

\draft{\tableofcontents}
\iftoggle{arxiv}{\tableofcontents\pagebreak}{}

\IGNORE{
    \begin{itemize}
        \item ``$\mu$-strongly convex" (as opposed to ``$\mu$-convex")
        \item 
        $\mu$ for constant of strong convexity (change from $m$) \hlcom{Change in multilevel section}
        \item
        $\ve{x}$ for norm of vector (change from $|x|$)
        \item
        $\log$ (change from $\ln$)
        \item $\ep$ (change from $\epsilon$)
        \item 
        $\cO(\cdot)$ (change from $O(\cdot)$) \hlcom{Change in multilevel section}
    \end{itemize}
    \color{black}
    }

\draft{\pagebreak}

\draft{
\pagenumbering{arabic}
\setcounter{page}{1}}


\section{Introduction}

Given a distribution $\rh$ on a space $\Omega$ with base measure $\dx$, defined by $\rh(\rdx) \propto e^{-f(x)}\dx$, its normalizing constant is the integral $Z := \int_\Om e^{-f(x)}\dx$. Estimating the normalizing constant is a fundamental problem in theoretical computer science, statistical physics (where it is called the partition function~\citep{balian2007microphysics,stoltz2010free}), 
and Bayesian statistics~\citep{GelmanMeng:1998}. In high dimensional settings, even when the function $f(x)$ is convex (and the distribution $\rh$ is log-concave), computing the exact normalizing constant is \#P-hard \cite{dyer1988complexity}. 
Hence, the goal is to approximate the normalizing constant up to $1\pm \ep$ multiplicative accuracy. Approximating the normalizing constant is closely related to the problem of sampling from the distribution $\rh$ \citep{jerrum1986random,sinclair1989approximate,dyer1991random}.  



Many polynomial time algorithms, starting from the seminal work of \cite{dyer1991random}, were known for estimating normalizing constants in various settings when $f(x)$ is convex. In this paper, we consider the special case where $\Om=\R^d$ and $f(x)$ is a $L$-smooth and $\mu$-strongly convex function (see equation \eqref{eq:smooth_strongconvex}). Given query access to $f(x)$ and/or $\nabla f(x)$, our goal is to estimate the 
normalizing constant 
\begin{equation}
  Z = \int_{\R^d} e^{-f(x)} \ud x 
\end{equation}
within a multiplicative factor of $1 \pm \eps$ with probability greater 
than $3/4$\footnote{For any algorithm, the probability can be easily amplified to $1-\zeta$ by repeating the algorithm $\cO(\log (1/\zeta))$ times and finding the median.}. 

This is a classical setting with applications to Bayesian statistics and machine learning. It is simpler than some of the settings considered before (such as volume estimation) because of strong convexity. 
Indeed, many faster sampling algorithms are known when $f$ is strongly convex. However, there are very few results for estimating the normalizing constant and they give suboptimal dependencies. On the lowerbound side, although lowerbounds were considered in different settings (e.g., \cite{RademacherVempala:2008}), there are no non-trivial lowerbounds when $f$ is strongly convex. In this paper, we give a new algorithm that only requires $\wt \cO \Bigl(\fc{d^{\fc 43}\ka +d^{\fc 76}\ka^{\fc 76}}{\ep^2}\Bigr)$ queries to $\nabla f(x)$, as well as a lowerbound that shows shows no algorithm can succeed with $\fc{d^{1-o(1)}}{\ep^{2-o(1)}}$ queries.

In high dimensions, most existing works rely on combining sampling algorithms for log-concave distributions and an annealing procedure. Our algorithm follows a similar recipe. We can use several sampling algorithms including Metropolis-Adjusted Langevin Algorithm (MALA), Underdamped Langevin Diffusion (ULD) and randomized midpoint method for ULD (ULD-RMM). However, a na\"ive combination of ULD and ULD-RMM with standard annealing procedure results in high query complexity. We use an approach called multilevel Monte Carlo~\citep{giles2008multilevel,giles2016multilevel} to improve the query complexity and running time of the algorithm. 

\begin{thm}[Upper bound] \label{t:ub}
Suppose $f:\R^d\to \R$ is $\mu$-strongly convex and $L$-smooth, and let $\ka = \fc L\mu$. Consider the problem of estimating 
$\int_{\R^d} e^{-f(x)}\dx$ within $1\pm \ep$ with success probability $3/4$.
\begin{enumerate}
    \item Algorithm~\ifstoc{2}{\ref{alg:normmala}} (annealing with MALA) solves the problem with $\wt \cO\pa{\frac{d^2\ka }{\veps^2} \max \bigl\{ 1, \sqrt{\kappa / d} \bigr\}}$ queries (Theorem~\ifstoc{B.3}{\ref{thm:normmala}}). 
    \item Algorithm~\ifstoc{6}{\ref{a:ml-norm}} (annealing with multilevel Monte Carlo) run using Algorithm~\ifstoc{4}{\ref{a:uld-couple}} (ULD) solves the problem with $\wt \cO \Bigl(\fc{d^{\fc 32}\ka^{2}}{\ep^2}\Bigr)$ queries (Theorem~\ifstoc{C.12}{\ref{t:muld-norm}}).  
    \item Algorithm~\ifstoc{6}{\ref{a:ml-norm}} (annealing with multilevel Monte Carlo) run using Algorithm~\ifstoc{5}{\ref{a:uld-rmm-couple}} (ULD-RMM) solves the problem with  \iftoggle{stoc}{\\}{} $\wt \cO \Bigl(\fc{d^{\fc 43}\ka +d^{\fc 76}\ka^{\fc 76}}{\ep^2}\Bigr)$ queries (Theorem~\ifstoc{C.13}{\ref{t:muld-rmm-norm}}).  
\end{enumerate}
\end{thm}
Note that these algorithms are also computationally efficient: for all of these algorithms, the runtime (in terms of number of vector operations in $\R^d$) is comparable to the number of queries. 
On the way to proving this theorem, we establish improved rates for estimating an expected value of a function using multilevel ULD. This result may be of independent interest.
\begin{thm}[Multilevel ULD]\label{thm:expectation_main}
Let $\rh(\rdx)\propto e^{-f(x)}\dx$, where $f:\R^d\to \R$ is $\mu$-strongly convex and $L$-smooth. Let $g:\R^d\to \R$ be $L_g$-Lipschitz.
Suppose $0<\ep<\fc{L_g}{\sqrt \mu}$. 
Consider the problem of outputting $\wh R$ such that $|\wh R-\E_\rh[g(x)]|\le \ep$. With probability at least $\fc 34$, Algorithm~\ifstoc{3}{\ref{a:ml}} (Multilevel Monte Carlo) has the following guarantees:
\begin{enumerate}
    \item 
    When run using Algorithm~\ifstoc{4}{\ref{a:uld-couple}} (ULD), it succeeds using \iftoggle{stoc}{\\}{}
    $\wt\cO\Bigl(\fc{L_g^2 d^{\rc 2} \ka^2}{\mu\ep^2}\Bigr)$ queries
    (Theorem~\ifstoc{C.4}{\ref{t:muld}}). 
    \item 
    Using Algorithm~\ifstoc{5}{\ref{a:uld-rmm-couple}} (ULD-RMM), it succeeds using \iftoggle{stoc}{\\}{} $\wt\cO\Bigl(\fc{L_g^2 (d^{\rc 3}\ka + d^{\rc 6}\ka^{\fc 76})}{\mu \ep^2}\Bigr)$ queries (Theorem~\ifstoc{C.6}{\ref{t:muld-rmm}}). 
\end{enumerate}
\end{thm}

Intuitively, the multi-level Monte Carlo method is a way to reduce the variance of the final sample by coupling several different Markov chains at different step sizes, which reduces the number of queries when the running time of the sampling algorithm depends polynomially on the desired accuracy (see Section~\ref{sec:multilevel} for more details).

We also give the first lowerbound for the complexity of estimating the normalizing constant:

\begin{theorem}[Lower bound]
Even for an $L$-smooth and $\mu$-strongly convex function $f(x)$ with $\kappa = L/\mu$ being a constant, any algorithm that uses $\fc{d^{1-o(1)}}{\ep^{2-o(1)}}$ queries cannot estimate the normalizing constant of $f(x)$ with accuracy $(1\pm \ep)$ with probability more than $3/4$.
 \end{theorem}
 
 Our lowerbound matches the dependency on $\ep$ in high dimensions (note that this is impossible in low dimensions due to deterministic quadrature methods; see Appendix~\ifstoc{E}{\ref{s:quad}}). The lowerbound also shows that there is an inherent dependency on dimension $d$ even when the condition number is a constant, which makes the problem of estimating the normalizing constant different from 
 optimization. 
 The lowerbound is information theoretic. We construct a function with many independent cells with two types. The final normalizing constant depends on the relative fraction of the cells of type 2. Making one query to function $f$ can reveal the type of at most one cell; therefore a standard argument shows estimating the frequencies of cell-types requires a large number of queries.



\subsection{Notation and Assumptions}

For any function $f$, we let $\cO(f)$ and $\Om(f)$ denote the class of functions that are $\le Cf$ and $\ge Cf$, respectively, for some constant $C>0$.
Let $\wt \cO(f)$ denote the class $\cO(f)\cdot \log^{\cO(1)}(f)$, and $\wt \Om(f)$ denote the class $\Om(f) \cdot \log^{-\cO(1)}(f)$. Let $\Te(f)$ denote the class of functions that are both $\cO(f)$ and $\Om(f)$, and $\wt \Te(f)$ denote the class of functions that are both $\wt \cO(f)$ and $\wt \Om(f)$. 

For a vector $v\in \R^d$, let $\ve{v}$ denote its Euclidean norm; and for a matrix $A \in \R^{d \times d}$, $\norm{A}$ denotes its spectral norm. For $x,y\in \R$, let $x\wedge y=\min\{x,y\}$ and $x\vee y = \max\{x,y\}$.

The $p$th Wasserstein distance between two probability measures $\mu$ and $\nu$ is defined as 
\begin{align*}
    W_p(\mu,\nu) &=
    \pa{\inf_{(X,Y)\in \cal C(\mu,\nu)} \E[\ve{X-Y}^p]}^{\rc p}
\end{align*}
where $\cal C(\mu,\nu)$ denotes the set of couplings between $\mu$ and $\nu$. 
The TV-distance is defined as 
$d_{\mathrm{TV}}(\mu,\nu) = \sup_A |\mu(A)-\nu(A)|$, where the sup is over all measurable subsets.

Throughout this work, we consider a log-concave distribution $\rho(\rd x) = \frac{1}{Z} e^{-f(x)} \ud x$. We assume that the negative log-density function $f(x)$ is twice continuously differentiable, $\mu$-strongly convex and $L$-smooth: For all $x, y \in \R^d$,
\begin{align}
  \frac{\mu}{2} \ve{x - y}^2 \le f(y) - f(x) - \nabla f(x)^{\top} (y - x) & \leq \frac{L}{2} \ve{x - y}^2. \label{eq:smooth_strongconvex} 
\end{align}

As we are concerned about the relative error for estimating the normalizing constant $Z$, it does not matter if $f$ is shifted by a constant, and hence for simplicity of the presentation, we will assume that
$f$ achieves its global minimum at $x^{\ast}$ with $f(x^{\ast}) = 0$
and only consider the (most challenging) regime that
$\mu \ll 1 \ll L$. In fact, to further simplify the presentation, we will
also assume $x^{\ast} = 0$, \textit{i.e.}, $f$ achieves the minimum
at the origin. 
In practice, we do not know $x^{\ast}$ a priori, however, using a
first-order optimization method like gradient descent, we can obtain an approximate of $x^{\ast}$ within error $\eta$ using $\kappa \log (1/ \eta)$
gradient evaluations. Such cost is negligible compared with other parts of the algorithm. 


\subsection{Roadmap} First in Section~\ref{sec:prelim} we review existing works on sampling and estimating normalizing constant; in particular we recall guarantees for the sampling algorithms that we use in this paper. In Section~\ref{s:anneal} we describe the annealing strategy that we use, which is similar to but has different parameters with existing work. We describe the main idea of our algorithm (especially the idea of using the multilevel Monte Carlo method) in Section~\ref{sec:multilevel}. Then we give the main ideas for the lowerbound in Section~\ref{sec:lowerbound}. Detailed algorithms and proofs are deferred to the appendices.

\section{Related works}\label{sec:prelim}

Many methods have been developed over the years for estimating the normalizing constant (also known as the partition function), see e.g., \citep{GelmanMeng:1998, stoltz2010free} and references therein.
However, not many works have given
non-asymptotic 
rates for 
algorithms to estimate the normalizing
constant of a strongly log-concave distribution. 
The closest work to ours is the recent work
\cite{BrosseDurmusMoulines:2018}, which gives a
$\wt{\Or}( \kappa^3 d^3 \veps^{-4})$ upperbound. An upperbound with
a rather high power dependence on $d$ is also established in
\cite{AndrieuRidgwayWhiteley} for a different algorithm. \revise{The works \cite{lovasz2006simulated,Lovasz} give an algorithm for arbitrary logconcave densities using only function queries with complexity $\wt{\Or}(d^4 \veps^{-2})$. Compared with previous works, our algorithm and analysis yield  better dependence on $d$, but also depend on the condition number $\ka$.}

The estimation of the normalizing constant for a log-concave distribution is closely related to volume computation of a convex set $K$ \cites{dyer1991random,lovasz1993random,lovasz2006simulated} (which can be thought of as the special case where $f(x)=0$ on $K$ and $f(x)=\iy$ outside of $K$). This can be done in $\wt{\Or}\bigl(\frac{d^3}{\ep^2}\bigr)$ time~\cite{CousinsVempala:2018}  
using an annealing algorithm combined with the Metropolis ball walk.  %
While our setup is quite different, the overall annealing algorithm follows the same spirit, albeit with different parameter choices. 

To the best of our knowledge, no lowerbound is known for the problem
under consideration. For volume computation of convex set, the best
known query lowerbound is $\wt{\Omega}(d^2)$ given by
\cite{RademacherVempala:2008} when $\ep = \Theta(1)$. The results are not comparable as the volume of convex body corresponds to a function $f$ that is not strongly convex, and the query is of membership rather than gradient type.


Non-asymptotic error analysis for Monte Carlo sampling algorithms has
received a lot of research focus in recent years. One popular type of
sampling algorithm is based on the Langevin dynamics, either the
underdamped Langevin dynamics
\begin{align*}
  & \ud x_t = v_t \ud t; \\
  & \ud v_t = - \nabla f(x_t) \ud t - \gamma v_t \ud t + \sqrt{2 \gamma} \ud B_t, 
\end{align*}
where $\gamma>0$ is a friction parameter and each component of $B_t \in \RR^d$ is independent standard Brownian motion, or the overdamped version (which can be obtained by taking $\gamma \to \infty$ of the underdamped Langevin while rescaling time $t \mapsto  t / \gamma$):  
\begin{equation*}
  \ud x_t = - \nabla f(x_t) \ud t + \sqrt{2} \ud B_t. 
\end{equation*}
After discretization of the SDE by a numerical integration scheme, the overdamped
Langevin dynamics leads to the unadjusted Langevin algorithms, whose
explicit non-asymptotic error bounds have been established by recent
works \cite{dalalyan2017theoretical, durmus2017nonasymptotic,
  dalalyan2017user, durmus2018analysis, vempala2019rapid}, with
complexity $\wt{\Or}\bigl(\frac{\kappa d}{\mu \veps^2}\bigr)$ to
achieve Wasserstein-$2$ error $\veps$ \cite{durmus2018analysis}. The
dependence on $d$ and $\veps$ can be improved by sampling algorithms
based on discretizing the underdamped Langevin dynamics, which has
been recently pursued by
\cite{cheng2017underdamped,DalalyanRiou-Durand,ma2019there,mou2019improved,shen2019randomized}. In
particular, the very recent work \cite{shen2019randomized} gives an
upperbound of query complexity
$\wt{\Or}\Bigl( \max \Bigl\{
\frac{d^{1/3}\kappa }{\mu^{1/3}\veps^{2/3} }, \frac{d^{1/6}\kappa^{7/6}}{\mu^{1/6}\veps^{1/3}}  \Bigr\}
\Bigr)$ for the ULD-RMM algorithm, upon which we will base our algorithm for the normalizing constant.

Metropolis-Hastings acceptance/rejection can be applied on top of
the unadjusted Langevin algorithm. The resulting algorithm is known as
Metropolis-Adjusted Langevin algorithm (MALA)
\cite{RobertsTweedie:1996}, which was in fact first developed in the
chemistry literature known as the smart Monte Carlo algorithm
\cite{RosskyDollFriedman:1978}. The non-asymptotic error bound for
MALA for log-concave probability distribution was recently studied by
\cite{DwivediChenWainwrightYu, ChenDwivediWainwrightYu}. The result
indicates that $\cO(\kappa d \log(1 / \veps))$ queries to $f$ and $\nabla f$ are needed
to achieve error $\veps$ measured in total variation (TV) distance. Thus using
Metropolis-Hastings acceptance/rejection improves the sampling
efficiency exponentially in terms of the error $\veps$, but suffers a worse dependence on $d$.

Besides the Langevin dynamics, sampling algorithms based on
the deterministic Hamiltonian dynamics have been also quite popular,
known as the Hamiltonian Monte Carlo (HMC) algorithms or hybrid Monte
Carlo algorithms originally proposed in \cite{Duane:1987}; see also
the review \cite{BouRabeeSanz-Serna:2018}. The non-asymptotic error analysis
has been considered recently in \cite{mangoubi2017rapid, lee2018algorithmic, lee2018convergence, chen2019optimal} for log-concave
case and in \cite{BouRabeeEberleZimmer} for more general cases using
coupling arguments.

\section{Annealing for Estimating the Normalizing Constant}
\label{s:anneal}
For estimating the normalizing constant $Z$, we consider an annealing 
algorithm similar to 
previous algorithms for normalization constant estimation (see
e.g., \cite{lovasz2006simulated, CousinsVempala:2018, BrosseDurmusMoulines:2018}). Similar  annealing strategies are widely used in calculation of normalizing constants, such as the annealed importance sampling \cite{Neal:2001}
in the statistic literature and thermodynamic integration
\cite{Jarzynski:1997} in the statistical physics literature. 

We define a sequence of auxiliary distributions, given by adding a quadratic function to $f$,  for $i = 1, 2, \ldots, M$
\begin{equation}
  f_i(x) = \frac{1}{2} \frac{\norm{x}^2}{\sigma_i^2} + f(x), 
\end{equation}
where $\sigma_1 \leq \sigma_2 \leq \cdots \leq \sigma_{M}$; for
convenience of notation, we also define $\sigma_{M+1} = \infty$ so that
$f_{M+1} = f$. Correspondingly, we consider the sequence of distributions
\begin{equation}
  \rho_i(\rd x) = Z_i^{-1} e^{-f_i(x)} \ud x, 
\end{equation}
where $Z_i$ is the normalizing constant
\begin{equation}
  Z_i = \int_{\RR^d} e^{-f_i(x)} \ud x. 
\end{equation}
The estimation of $Z$ is based on the identity 
\begin{equation}\label{eq:estimateZ}
  Z = Z_{M+1} = Z_1 \prod_{i=1}^M \frac{Z_{i+1}}{Z_i}. 
\end{equation}
In \eqref{eq:estimateZ}, we will approximate $Z_1$ by the normalizing factor of the Gaussian distribution with variance $\sigma_1^2$. The ratio  $\frac{Z_{i+1}}{Z_i}$ for $i = 1, \ldots, M$ can estimated using sampling algorithms for the distribution $\rho_i$, since 
\begin{equation}
  \frac{Z_{i+1}}{Z_i} = \int  \exp\Biggl(\frac{1}{2}\Bigl(\frac{1}{\sigma_i^2} - \frac{1}{\sigma_{i+1}^2}\Bigr) \norm{x}^2 \Biggr)\, \rho_i(\rd x) = \EE_{\rho_i}(g_i) 
\end{equation}
where
\begin{equation}
  g_i := \exp\pa{\frac{1}{2}\pa{\frac{1}{\sigma_i^2} - \frac{1}{\sigma_{i+1}^2}} \norm{x}^2 }.
\end{equation}
Thus, if $X_i^{(1)}, \ldots, X_i^{(K)}$ are iid sample points generated according to the distribution $\rho_i$ (or its approximation), we can estimate 
\begin{equation}
  \frac{Z_{i+1}}{Z_i} \approx \frac{1}{K} \sum_{k = 1}^K g_i(X_i^{(k)}). 
\end{equation}

For the sequence of $\sigma_i^2$, we choose the following annealing strategy: 
We start with $\sigma_1^2 = \frac{\veps}{2dL}$ and increase as
\begin{equation}
  \sigma_{i+1}^2 = \sigma_i^2 \pa{ 1 + \frac{1}{\sqrt{d}} }
\end{equation}
until $\sigma_M^2$ is large enough, as specified below. We remark that a slower annealing procedure of $\sigma_{i+1}^2 = \sigma_i^2 ( 1 + 1/d)$ was \revise{considered in  \cite{CousinsVempala:2018} to maintain a warm start}, as it gives a smaller relative variance of $g_i$ for each stage (on the order of $d^{-1}$). We take a faster annealing procedure \revise{as in \cite{lovasz2006simulated}} to take advantage of variance reduction by the multilevel Monte Carlo method, cf.~Section~\ref{sec:multilevel}.  

\smallskip 

In the above sketch of the algorithm, the approximation of $Z_1$ is guaranteed by the following lemma. Proofs of this and other lemmas in this section are postponed to Appendix~\ifstoc{A}{\ref{sec:annealingproof}}. 
  
\begin{lemma}[Starting distribution]\label{l:start}
Letting $\sigma_1^2 = \frac{\veps}{2dL}$, we have 
  \begin{equation}
    \Bigl(1 - \frac{\veps}{2}\Bigr)
    \int_{\RR^d} e^{-\frac{1}{2} \frac{\norm{x}^2}{\sigma_1^2}} \ud x \leq Z_1 \leq \int_{\RR^d} e^{-\frac{1}{2} \frac{\norm{x}^2}{\sigma_1^2}} \ud x.
  \end{equation}
\end{lemma}

Next we consider the ratio $\frac{Z_{M+1}}{Z_M}$ in \eqref{eq:estimateZ}. We have 
\begin{equation}
  \frac{Z_{M+1}}{Z_M} = \int_{\R^d}  \exp\Biggl(\frac{\norm{x}^2}{2 \sigma_M^2}   \Biggr)\, \rho_M(\rd x) = \EE_{\rho_M}(g_M) 
\end{equation}
with $g_M =\exp\bigl(\frac{\norm{x}^2}{2 \sigma_M^2}   \bigr)$.
To control the accuracy of Monte Carlo estimation of $\EE_{\rho_M}(g_M)$, we bound  the  relative variance in  the following lemma. The idea of the proof (deferred to Appendix~\ifstoc{A}{\ref{sec:annealingproof}}) comes from
\cite[Section 7.1]{CousinsVempala:2018}, in particular the proof of
\cite[Lemma 7.6]{CousinsVempala:2018}. 

\begin{lemma}\label{lem:varboundM}
 For any $\sigma_M^2 \geq \frac{2}{\mu}$, we have
  \begin{equation*}
      \frac{\EE_{\rho_M}(g_M^2)}{\EE_{\rho_M} (g_M)^2} = 
    \EE_{\rho}
    \exp\Bigl(-\frac{1}{2} \frac{\norm{x}^2}{\sigma_{M}^2}\Bigr)
    \, \EE_{\rho} \exp\Bigl(\frac{1}{2} \frac{\norm{x}^2}{\sigma_{M}^2}  \Bigr)
    \leq \exp \Bigl( \frac{4d}{\mu\sigma_M^4}  \Bigr).
  \end{equation*}
\end{lemma}

Let us now consider the estimate for $\frac{Z_{i+1}}{Z_i} = \EE_{\rho_i}(g_i)$ in
\eqref{eq:estimateZ}. 
To bound the variance of $g_i = \exp\bigl(\frac{1}{2} (\sigma_i^{-2} - \sigma_{i+1}^{-2}) \norm{x}^2 \bigr)$ under the distribution $\rho_i$, let $\sigma^2 = \sigma_{i+1}^2$ and $\sigma_i^2 = \sigma^2 / ( 1 + \alpha)$, and calculate 
\begin{equation}\label{e:vari}
    \frac{\EE_{\rho_i}(g_i^2)}{\EE_{\rho_i} (g_i)^2} = 
    \dfrac{\EE_{\rho} \exp\Bigl(-\frac{1+\alpha}{2} \frac{\norm{x}^2}{\sigma^2}\Bigr)
  \, \EE_{\rho} \exp\Bigl(-\frac{1-\alpha}{2} \frac{\norm{x}^2}{\sigma^2}\Bigr)}{\biggl(\EE_{\rho}  \exp\Bigl(-\frac{1}{2} \frac{\norm{x}^2}{\sigma^2}\Bigr)\biggr)^2}. 
\end{equation}
The next lemma gives an upper bound for the right hand side as
$\exp(4 \alpha^2 d)$. This suggests the choice $\alpha = \frac{1}{\sqrt{d}}$ used in our annealing strategy to give an $\Or(1)$ relative variance. The proof follows along similar lines as the previous
  lemma. 

\begin{lemma}\label{lem:varbound}
  Let $\rho$ be a logconcave distribution,  for $\alpha \leq \frac{1}{2}$, we have
  \begin{equation}\label{eq:varbound} 
    \dfrac{\EE_{\rho} \exp\Bigl(-\frac{1+\alpha}{2} \frac{\norm{x}^2}{\sigma^2}\Bigr)
      \, \EE_{\rho} \exp\Bigl(-\frac{1-\alpha}{2} \frac{\norm{x}^2}{\sigma^2}\Bigr)}{\biggl(\EE_{\rho}  \exp\Bigl(-\frac{1}{2} \frac{\norm{x}^2}{\sigma^2}\Bigr)\biggr)^2}
    \leq \exp \bigl( 4 \alpha^2 d \bigr) 
  \end{equation}
\end{lemma}

With these lemmas, it remains to choose a suitable sampling scheme to estimate $\EE_{\rho_i} g_i$ for each $i$. One possible approach is to use the Metropolis-Adjusted Langevin Algorithm (MALA) to generate independent samples with respect to $\rho_i$. Using the theoretical guarantees of MALA for strongly log-concave distributions recently established in \cite{DwivediChenWainwrightYu, ChenDwivediWainwrightYu}, and the choice of $\sigma_M^2 = \Theta(\frac{\sqrt{d}}{\mu})$, we arrive at an algorithm with total query complexity $\wt{\Or}\Bigl(\frac{d^2\kappa}{\veps^2} \max \bigl\{ 1, \sqrt{\kappa/d} \bigr\}\Bigr)$. 
This follows from the fact that MALA needs $\wt \cO\bigl(d\ka \max\bigl\{1,\sqrt{\ka/d}\bigr\}\bigr)$ queries to achieve $\ep$ error in TV distance, and we need $\fc{\sqrt d}{\ep^2}$ samples at each annealing stage to achieve relative variance $\wt \cO\pf{\ep^2}{\sqrt d}$, which leads to relative variance $\cO(\ep^2)$ for the product, and thus $\cO(\ep)$ relative error. 
See Appendix~\ifstoc{B}{\ref{s:mala}} for details. The dimension dependence can however be improved by exploiting the multilevel Monte Carlo algorithm, as we discuss in the following section.

\section{Estimating the Normalizing Constant using Multilevel ULD} \label{sec:multilevel}

Without making additional smoothness assumptions, for guarantees in KL or TV error, the best dependence on $d$ known is the $\wt \cO(d)$ dependence given by MALA.
However, for guarantees in Wasserstein ($W_2$) error, algorithms based on underdamped Langevin diffusion are known to give better dependence: \cite{cheng2017underdamped} show that to achieve $W_2$ error $\ep$, underdamped Langevin dynamics (ULD) has query complexity $\wt\cO \pa{\fc{d^{\rc 2}\ka^2 }{\mu^{\rc 2}\ep}}$, 
and \cite{DalalyanRiou-Durand} improves the dependence on $\ka$ to $\ka^{\fc 32}$. \cite{shen2019randomized} propose the Randomized Midpoint Method (RMM) to estimate the integral in ULD, and obtain query complexity $\wt \cO\pa{\fc{d^{\fc 13}\ka}{\ep^{\fc 23}\mu^{\rc 3}} +\fc{d^{\rc 6}\ka^{\fc 76} }{\mu^{\rc 6}\ep^{\rc 3} }}$.

Focusing on the dependence on $d$ and $\ep$, one may hope that a method which obtains $W_2$ error using $\cO\pf{d^\ga}{\ep^\de}$ queries can be used to compute the normalizing constant in time $\cO\pf{d^{1+\ga}}{\ep^{2+\de}}$. However, we show below that a naive substitution of the algorithm in the annealing procedure described in Section~\ref{s:anneal} fails. The key ingredient we need to obtain this $d^{1+\ga}$ dependence is multilevel Monte Carlo, which additionally achieves $\rc{\ep^2}$ dependence in $\ep$. This allows us to obtain the $\wt \cO \Bigl(\fc{d^{\fc 32}\ka^{2}}{\ep^2}\Bigr)$ and $\wt \cO \Bigl(\fc{ d^{\fc 43}\ka+d^{\fc 76}\ka^{\fc 76}}{\ep^2}\Bigr)$ rates in Theorem~\ref{t:ub}.

For simplicity, in the proof sketch below we assume the condition number and strong convexity are order 1 ($\ka =\cO(1)$, $\mu=\Te(1)$), and focus on just the dependence on $d$ and $\ep$. In our main theorem we do work out the dependence on $\ka$. We describe the guarantees that we would obtain by using ULD, but the same story holds for ULD-RMM with improved rates. For details, see Appendix~\ifstoc{C}{\ref{s:ml}}.

\subsection{Insufficiency of ULD}

Underdamped Langevin dynamics  has the following error guarantee: to estimate the distribution up to $W_2$-error $\ep$, we can take step size $\eta=\cO\pf{\ep}{\sqrt{d}}$ and number of steps $\fc{T}{\eta}=\wt \cO\pf{\sqrt{d}}{\ep}$. 

Suppose we use $\wt \cO(\sqrt d)$ temperatures, differing by factors of $1+\rc{\sqrt d}$. We chose the fewest number of temperatures such that the variance of $g_i(x)$ over $\rh_i$ is $\cO(1)$. (Using more temperatures, we need improved accuracy for estimating $R_i:=\E_{x\sim \rh_i} g_i(x)$ for each temperature, which results in the same running time per temperature.)
Then to estimate the normalizing constant within $1\pm\cO(\ep)$, we need to estimate the ratio $R_i$ at each step with relative accuracy $\fc{\ep}{\sqrt d}$. We can check that $g_i(x) = \exp\pa{\fc{\ve{x}^2}{\si_i^2(1+\sqrt d)}}$ is 
$\cO\pf{R_i}{\si_i}$-Lipschitz
around where $\rh_i$ is concentrated, that is, for $x$ such that $\ve{x}=\cO(\si_i\sqrt d)$. 
To estimate the product with $\ep$ relative accuracy, we need to estimate each $R_i$ with $\cO\pf{\ep R_i}{\sqrt d}$ accuracy, so we need to sample from $\wt \rh_i$ with $W_2(\wt\rh_i , \rh_i)\le \cO\pf{\ep \si_i}{\sqrt d}$. This requires us to choose a step size of $\eta = \cO\pf{\ep/\sqrt d}{\sqrt d} = \cO\pf{\ep}{d}$, so each sample takes $\wt\cO\pf{d}{\ep}$ queries to obtain.
In order to reduce the variance to $\fc{\ep^2}{\sqrt d}$, we need $\fc{\sqrt d}{\ep^2}$ samples at each temperature, for a total of $\wt \cO\pa{\sqrt d \cdot \fc{d}{\ep} \cdot \fc{\sqrt d}{\ep^2}} = \cO\pf{d^2}{\ep^3}$ steps.

\subsection{Multilevel ULD}


Multilevel Monte Carlo~\cite{giles2008multilevel} is a generic way to improve rates for estimating $\E Y$ for a random variable $Y$, when 
there are biased estimators $Y^\eta$ such that (1) as $\eta\to 0$, $\E Y^\eta \to \E Y$ and the cost to evaluate $Y^\eta$ increases, and (2) there is a way to couple $Y^\eta$ and $Y^{\eta'}$ when $\eta'<\eta$ that significantly reduces the variance, $\Var(Y^\eta - Y^{\eta'})\ll \Var(Y^\eta)$.

This is the case when we wish to estimate  $\E_{x\sim \rh} g(x)$, when $\rh$ can be (approximately) obtained from simulating a stochastic differential equation (SDE) for some time $T$. In this setting, $Y^\eta=g(X^\eta)$ and $X^\eta=x^\eta_T$, where $x^\eta_T\sim \rh^\eta$ is the point obtained by simulating the SDE with some discretization algorithm $\cal A$ for time $T$ and step size $\eta$. Using the same Brownian motion for simulating $x_t^\eta$ and $x_t^{\eta'}$ naturally defines a coupling. If $g$ is $L_g$-Lipschitz, $\Var(g(X^{\eta})-g(X^{\eta'})) \le L_g^2 \E[||X^{\eta}-X^{\eta'}||^2]$. The average distance $\E[||X^{\eta}-X^{\eta'}||^2]$ will be comparable to the Wasserstein error $W_2(\rh^{\eta},\rh)$. This is much smaller than the variance of $X^{\eta}$, which is comparable to the variance of $X\sim \rh$.

\iftoggle{stoc}{\begin{algorithm}[h!]
\caption*{Multilevel Monte Carlo (Algorithm~\ifstoc{3}{\ref{a:ml}})}
\begin{algorithmic}[1]
\REQUIRE Initial point $x_0$, time $T$, largest step size $\eta_0$, 
number of levels $k$, number of samples $N_0,\ldots, N_k$, function $f:\R^d\to \R$, function $g:\R^d\to \R$.
\REQUIRE Sampling algorithm $\cal A(x_0,f,\eta,T)$ which can give coupled samples $(x^\eta,x^{\eta/2})$ (or individual samples $x^\eta$).
\ENSURE Estimate of $\E_{x\sim \rho} g(x)$ where $\rho(\ud x)\propto e^{-f(x)}\dx$
\FOR{$1\le i\le N_0$}
  \STATE Run $\cal A$ with initial point $x_0$, function $f$, step size $\eta_0$, and time $T$ to obtain $X^{\eta_0}_i$.
\ENDFOR
\FOR{$1\le j\le k$}
	\FOR{$1\le i\le N_j$}
	  \STATE Run coupled $\cal A$ with initial point $x_0$, function $f$, step size $\eta=\eta_0/2^{j-1}$, and time $T$, to obtain $(X^{\eta-}_i,X^{\eta/2+}_i)$.
	\ENDFOR
\ENDFOR
\RETURN $\rc{N_0} \sumo i{N_0} g(X_i^{\eta_0})
+\sumo jk \rc{N_j} \sumo i{N_j} [g(X_i^{\eta_0/2^{j}-}) - g(X_i^{\eta_0/2^{j-1}+})]$.
\end{algorithmic}
\end{algorithm}}{}

The idea of Multilevel Monte Carlo (Algorithm~\ifstoc{3}{\ref{a:ml}}) is to choose decreasing step sizes $\eta_0,\ldots, \eta_k$ (e.g. with $\eta_j = \fc{\eta_0}{2^j}$), and write $g(X^{\eta_k})$ as
\begin{align}
\llabel{e:ml}
g(X^{\eta_k})=
g(X^{\eta_0}) + \sumo jk [g(X^{\eta_{j}})-g(X^{\eta_{j-1}})]
\end{align}
We estimate each of these terms by taking $N_0$ samples at the highest level $X_i^{\eta_0}$, and $N_j$ coupled samples $(X_i^{\eta_j-},X_i^{\eta_{j-1}+})$, to obtain the estimate
\begin{align}
\llabel{e:ml-est-intro}
    \wh R := \rc{N_0} \sumo i{N_0} g(X_i^{\eta_0})
+\sumo jk \rc{N_j} \sumo i{N_j} [g(X_i^{\eta_{j}-}) - g(X_i^{\eta_{j-1}+})].
\end{align}
Suppose we would like to give an estimate with bias $\ep_b$ and variance $\ep_\si^2$. 
The expected value of $\wh R$ is simply $\E_{X^{\eta_k}\sim \rh^{\eta_k}}g(X^{\eta_k})$, so to ensure bias $\le \ep_b$, it suffices to choose $\eta_k$ small enough. Supposing the variance of $g(X_i^{\eta_{j}-}) - g(X_i^{\eta_{j-1}+})$ is $F(\eta_j)$, the total variance is $\fc{\Var(g(X^{\eta_0}))}{N_0}
+ \sumo jk \fc{F(\eta_j)}{N_j}$.
For smaller step size, because the variance $F(\eta_j)$ is smaller, it suffices to choose a smaller number of samples $N_j$, which offsets the increased number of steps $\fc{T}{\eta_j}$.
Optimally choosing $N_j$ to balance this with the total time necessary, $\sumo jk \fc{TN_j}{\eta_j}$, gives the following.

\begin{lem*}[Lemma~\ifstoc{C.2}{\ref{l:ml2}} with $L_g=\sqrt\mu$, $F(\eta)=C\eta^\be$]
Suppose that $\rh(\rdx)\propto e^{-f(x)}$, $f$ is $\mu$-strongly convex and $g:\R^d\to \R$ is $\sqrt{\mu}$-Lipschitz. Suppose algorithm $\cal A$ with step size $\eta$ takes $\fc{T}{\eta}$ gradient queries to generate the random variable $X^{\eta}$. 
Let $X^0$ denote the corresponding continuous process. Suppose there is a coupling between $X^\eta$ and $X^0$ such that $\E[||X^\eta-X^0||^2]\le F(\eta):=C\eta^\be$ (for some $\be>1$), and $T(\cdot)$ is a function such that $W_2(\rh^\eta,\rh)^2 \le F(\eta) \wedge \ep^2$ whenever $T\ge T(\ep)$.
Let $\eta_0$ be such that $F(\eta_0)=\rc{\mu}$ and $F(\eta_k) \le \fc{\ep_b^2}{\mu}$. For $T\ge T\pf{\ep}{\sqrt\mu}$ and appropriate number of samples $N_j$, multilevel Monte Carlo (Algorithm~\ifstoc{3}{\ref{a:ml}}) run using $\cal A$ returns an estimate $\wh R$ of $\E_\rh g$ satisfying $|\E \wh R - \E_\rh g| \le \ep_b$ and $\Var(\wh R)\le \ep_\si^2$ using $\cO\pa{T\pa{\fc{1}{\ep_\si^2\eta_0} + \fc{1}{\eta_k}}}$ gradient queries.
\end{lem*}
Note the scaling above is so that the variance of $g$ over $\rh$ is at most 1. Without multilevel Monte Carlo, the number of gradient queries would be significantly worse: $\cO\pa{\fc{T}{\eta_k} \cdot \fc{1}{\ep_\si^2}}$, because we need to take a step size of $\eta_k$, and the number of samples to reduce the variance from 1 to $\ep_\si^2$ is $\rc{\ep_\si^2}$. Using multilevel MC, we only need to pay $\cO\prc{\ep_\si^2}$ samples at the highest level $k=0$, and  
we only need to take $\eta_0$ small enough so that $F(\eta_0)=\rc{\mu}$ (which makes $\Var(g(X^{\eta_0}))\le 1$).  

We use this result to give a non-asymptotic analysis of the rate for multilevel ULD (Theorem~\ifstoc{C.4}{\ref{t:muld}}) and ULD-RMM (Theorem~\ifstoc{C.6}{\ref{t:muld-rmm}}).
The results of \cite{cheng2017underdamped,DalalyanRiou-Durand} show that for underdamped Langevin dynamics, the hypotheses of the lemma hold with $F(\eta)=\cO\pa{\fc d\mu \eta^2}$, which suggests we take the largest step size to be $\eta_0=\cO(d^{-\rc 2})$. For ULD with the randomized midpoint method,~\cite{shen2019randomized} show that the hypotheses hold with $F(\eta) = \cO \pa{\fc d\mu \eta^3}$, which suggests we take $\eta_0 = \cO\pa{d^{-\rc 3}}$.



For the problem of estimating the normalizing constant, for each temperature $i$ we apply Lemma~\ifstoc{C.2}{\ref{l:ml2}} with $g\mapsfrom \fc{g_i}{R_i}$, which has Lipschitz constant $\cO\prc{\si_i}=\cO(\sqrt{\mu_i})$ around where it is concentrated, where $\mu_i$ is the strong convexity constant of $f_i$.
Then, to obtain bias $\ep_b = \cO\pf{\ep}{\sqrt d}$ and variance $\ep_\si^2 = \cO\pf{\ep^2}{\sqrt d}$, we need $\eta_k = \cO\pf{\ep_b}{\sqrt d}$ and so $\cO\pa{T \pa{\fc{\sqrt d\sqrt d}{\ep^2} + \fc{\sqrt d\sqrt d}{\ep}}} = \wt\cO\pf{d}{\ep^2}$ queries. 
Since there are $\wt\cO(\sqrt d)$ temperatures, the total number of queries over all temperatures is $\wt \cO\pf{d^{3/2}}{\ep^2}$.
Similarly for ULD-RMM, we need $\eta_k =\cO \pf{\ep_b^{2/3}}{d^{1/3}}$ and so $\cO\pa{T\pa{\fc{d^{\rc 3}\sqrt d}{\ep^2} + \fc{d^{\rc 3}d^{\rc3}}{\ep^{\fc 23}}}}$ queries per temperature, and $\wt \cO\pf{d^{4/3}}{\ep^2}$ queries in total.



Note that it is important to keep track of $\ep_\si$ and $\ep_b$ separately when computing the rates for multilevel MC. In our application, we can tolerate a larger $\ep_\si$ than $\ep_b$ at each temperature. This is because when there are $M$ temperatures, when adding up the contributions from the different temperatures, the standard deviation will only be multiplied by $\sqrt{M}$, while the bias will be multiplied by $M$. This allowed us to take $\ep_\si = \Te\pf{\ep}{d^{1/4}} \gg \ep_b = \Te\pf{\ep}{\sqrt d}$. If we lowered $\ep_\si$ to make it equal to $\ep_b$, then we need a factor of $\sqrt d$ more samples for each temperature.

Compared to existing theoretical analysis of multilevel Monte Carlo~\cite{giles2008multilevel}, we only consider the case where  the variance is decreasing quickly enough as step size ($\be>1$ in the lemma), while~\cite[Theorem 3.1]{giles2008multilevel} gives non-asymptotic bounds for the regimes $\be>1$, $\be=1$, $0<\be<1$. While our proof follows the same argument, we give a more flexible version of the bound. Firstly, in Lemma~\ifstoc{C.2}{\ref{l:ml2}}, we consider any $F(\eta)$ that decays quickly enough as $\eta\to 0$ rather than just a power function; we need this for technical reasons. Secondly, rather than only bounding the mean squared error, we allow bounding the bias and variance separately, as noted above.

\subsection{Technical Issues}
\label{subsec:technical}
We glossed over several technical issues in the above proof sketch. First, we wish to estimate $\E_{x\sim \rh_i}g_i(x)$ where $\rh_i$ is the distribution at the $i$th temperature and $g_i$ is the ratio, but $g_i(x)=\exp\pa{\fc{\ve{x}^2}{\si_i^2(1+\al^{-1})}}$ is not Lipschitz. Instead, we truncate it for large $x$, and using concentration of $\ve{x}$ on the log-concave distribution $\rh_{i+1}$ to show that the bias introduced is small (Section~\ifstoc{C.4}{\ref{s:bias}}, Lemmas~\ifstoc{C.7}{\ref{l:g-tail}} and~\ifstoc{C.8}{\ref{l:bias}}). 
More precisely, let $h_i(x) = g_i(x) \wedge \exp\pf{r_i^{+2}}{\si_i^2 (1+\al^{-1})}$. We show that for some choice of \begin{align*}
    \al&=\cO\prc{\sqrt{d}\log\prc{\ep}}\\
    r_i^+ &= \E_{\rh_{i+1}}\ve{x} + \Te\pa{\si_i \sqrt{(1+\al)\log\prc{\ep}}},
\end{align*} we have (1) $\fc{h_i}{\E_{\rh_i} g_i}$ is $\cO\prc{\si_i}$-Lipschitz, and (2) the bias introduced is small, $|\E_{\rh_i}(h-g)|\le \ep$.

We need to know at what radius $r_i^+$ to truncate $g_i$; we can do this by estimating $\E_{x\sim \rh_{i+1}} \ve{x}$ using samples and then adding a suitable multiple of $\si_i$ (Lemma~\ifstoc{C.11}{\ref{l:r}}). Finally, we put all the bounds together to prove the main Theorem~\ifstoc{C.12}{\ref{t:muld-norm}} for ULD and Theorem~\ifstoc{C.13}{\ref{t:muld-rmm-norm}} for ULD-RMM.

\section{Lowerbound on Number of Queries}
\label{sec:lowerbound}
In this section, we give a lowerbound on the number of queries required to estimate the normalizing constant $\int e^{-f(x)} dx$. More precisely, we prove the following theorem:

\begin{theorem} \label{thm:lowerbound} For any fixed constant $\gamma > 0$, 
for large enough $d$, 
given query access to gradient or function value of a function $f:\R^d \to \R$ that is 1.5-smooth and $0.5$-strongly convex, any algorithm that makes $o\left(d^{1-\gamma}\ep^{-(2-\gamma)}\right)$ queries cannot estimate the normalizing constant $Z = \int_{\R^d} e^{-f(x)} dx$ within a multiplicative factor of $1\pm\ep$ with probability more than $3/4$.
\end{theorem}


In fact, even if the algorithm is allowed to query any local information (such as the Hessian of $f$ at $x$), our lowerbound still holds. Our construction also satisfies the Hessian Lipschitz property, which was used in some of the sampling results, see e.g., \cite{DalalyanRiou-Durand, BrosseDurmusMoulines:2018,mangoubi2018dimensionally,LiWuMackeyErdogdu}. 
Note that the bound hides constants that depend on $\gamma$, and $d$ needs to be at least as large as $\Omega(1/\gamma)$. One might hope that $d\ep^{-2}$ can be a lowerbound for every dimension $d$. However, this is impossible as when $d\leq3$ quadrature methods give better dependency in terms of $\ep$ (see Appendix~\ifstoc{E}{\ref{s:quad}}).

To prove Theorem~\ref{thm:lowerbound}, we first construct a $k$-dimensional function (where $k = \Theta(1/\gamma)$), and show that any algorithm that estimates its normalizing constant requires at least $\Omega\left(\ep^{-(2-\gamma)}\right)$ queries. 
Then we construct the function $f:\R^d \to \R$ in Theorem~\ref{thm:lowerbound} by partitioning the $d$ dimensions into $d/k$ groups of size $k$, and use a product distribution whose marginal on each group corresponds to the function that we construct for the low-dimensional regime.



\paragraph{Lowerbound for low dimensions}
In low dimensions, our goal is to give a lowerbound that depends on the accuracy $\ep$:

\begin{theorem} \label{thm:lowerbound_constant} For any fixed integer $k > 0$, given query access to gradient or function value of a function $f:\R^k \to \R$ that is 1.5-smooth and $0.5$-strongly convex, any algorithm that makes $o(\ep^{-\frac{2}{1+4/k}})$ queries cannot estimate the normalizing constant $Z = \int_{\R^k} e^{-f(x)} dx$ within a multiplicative factor of $1\pm\ep$ with probability more than $3/4$.
\end{theorem}

Note that if we would like to get guarantee in terms of $\ep$ similar to Theorem~\ref{thm:lowerbound} we only need to choose $k$ such that $-\frac{2}{1+4/k} = - (2-\gamma)$. It suffices to choose $k = \Theta(1/\gamma)$.

The main idea of proving this theorem is that we will construct a large number of independent ``cells'' in the space $\R^k$, where each cell can be one of two types. The final normalizing constant will depend on how many cells are of type 1. We will then pick a value $\delta$ (closely related to the accuracy $\ep$) and consider two distributions of functions: 
in the first distribution, each cell is of type 1 with probability $1/2+\delta$; in the second distribution, each cell is of type 1 with probability $1/2-\delta$. When the number of cells is large enough (much more than $1/\delta^2$), the functions from these two distributions will have different normalizing constants (with large constant probability). However, making one query to the function at best gives information about a single cell. By a standard argument (see Claim~\ifstoc{D.1}{\ref{clm:biasedcoin}}) 
we know in order to distinguish between two Bernoulli random variables with bias $\delta$ with better than $1/2$ probability, one needs at least $\Omega(1/\delta^2)$ queries. Any algorithm that uses fewer queries will not be able to distinguish the two distributions, and thus cannot estimate the normalizing constant accurately.

To construct these two distributions, we will start from a basic function $f_0(x) = \frac{\norm{x}^2}{2}$. The normalizing constant for this function is well-known:
\[
\int_{\R^k} e^{-f_0(x)}dx = (2\pi)^{k/2}.
\]

To construct $n$ cells, let $l = 1/(\sqrt{k}n^{1/k})$ (wlog we assume $n^{1/k}$ is an integer), and partition $[-1/\sqrt{k}, 1/\sqrt{k}]$ into $n^{1/k}$ intervals each of length $2l$. Let $I_i (i = 1,2,...,n^{1/k})$ be the $i$-th interval. Each cell $\tau$ will be indicated by a $k$-tuple $(i_1,i_2,...,i_k) \in \{1,2,...,n^{1/k}\}^k$, and the cell $\tau$ corresponds to $I_{i_1}\times I_{i_2}\times \cdots\times I_{i_k}$ in $\R^k$.

Next we will discuss how to modify the function within the cells. For cell $\tau$, we will modify the function to be $f_0(x) + c q(\frac{1}{l}(x-v_\tau))$ for $x$ in the cell, where $v_\tau$ is the center of cell $\tau$. Note that here the input $\frac{1}{l}(x-v_\tau)$ of $q$ ranges in $[-1,1]^k$. There are two major constraints for designing the function $q$: (1) it is possible to modify adjacent cells independently without violating the smoothness and strongly convex constraints; (2) it is possible to choose a large enough $c$ such that $\int_{x\in \tau} \exp(- (f_0(x) + c q(\frac{1}{l}(x-v_\tau))))dx$ is significantly smaller. The exact property of the $q$ function and the construction is deferred to Lemma~\ifstoc{D.1}{\ref{lem:qproperty}} in Section~\ifstoc{D}{\ref{sec:lowerbound_appendix}}.

Now, we modify the functions within each cell by adding in a scaled version of $q$, as in the following lemma:

\begin{lemma} \label{lem:function_construct}
For any $n$ where $n^{1/k}$ is an integer, let $l = 1/(\sqrt{k}n^{1/k})$. For each cell $\tau = (i_1,...,i_k)$, let $v_\tau$ be its center. Construct the function $f(x)$ as
\[
f(x) = \left\{\begin{array}{cl}f_0(x), & \mbox{cell $\tau$ is of type 1} \\ f_0(x)+c_\tau q\left(\frac{1}{l}(x-v_\tau)\right), &\mbox{cell $\tau$ is of type 2.}\end{array}\right.
\]
Here $q$ is the function constructed in Lemma~\ifstoc{D.1}{\ref{lem:qproperty}}.
There exists a way to choose $c_\tau$'s such that no matter what types each cell has, the family of functions satisfies the following properties:
\begin{enumerate}
    \item $f(x)$ is $1.5$-smooth and $0.5$-strongly convex.
    \item The normalizing constant $Z_f = \int_{\R^k} e^{-f(x)}dx = (2\pi)^{k/2} - C\frac{n_2}{n}$, where $n_2$ is the number of type-2 cells, and $C$ is at least $\Omega\left(l^2\right)$.
\end{enumerate}
\end{lemma}

With this lemma, one can construct two distributions of functions as follows: choose $\delta$ such that $\ep = \Theta(\delta^{1+4/k})$, $n \approx 1/\delta^2$, and let each cell be of type 1 with probability $1/2\pm \delta$ for the two classes. Claim~\ifstoc{D.1}{\ref{clm:biasedcoin}} 
shows that any algorithm that makes fewer than $o(1/\delta^2)$ queries cannot distinguish the two distributions, while Lemma~\ref{lem:function_construct} shows that the normalizing constant for two distributions differ by at least $1+\Omega(l^2\delta)$ factor where $l = \Theta(n^{1/k}) = \Theta(\delta^{-2/k})$. This gives the desired trade-off in Theorem~\ref{thm:lowerbound_constant}. A more detailed proof is given in Appendix~\ifstoc{D}{\ref{sec:lowerbound_appendix}}.

\paragraph{Lowerbound for high dimensions}

To generalize Theorem~\ref{thm:lowerbound_constant}, as we mentioned earlier, we partition the $d$ dimensions into $d/k$ groups of size $k$, and use a product distribution. If we use $S_i$ to denote the set of coordinates for the $i$-th group, we can write $f(x) = \sum_{i=1}^{d/k} f_i(x_{S_i})$. In particular, for the two distributions of functions that the algorithm is trying to distinguish, the $f_i(x_{S_i})$ are sampled from the two distributions of functions we defined for Theorem~\ref{thm:lowerbound_constant}. 
Since the normalizing constant of $f(x)$ is equal to the product of normalizing constants for $f_i$'s, the gap between the two distributions is amplified by a power of $d/k = \Omega(d)$. Therefore, in order to achieve accuracy $1\pm \ep$ for function $f$, one would need to achieve an accuracy of $1\pm \ep k/d$ for functions $f_i$. On the other hand, one query in $f$ can simultaneously give information on $d/k$ of the functions $f_i$'s. Intuitively, if the lowerbound for the $k$ dimensional case is $L(\ep)$, the new lowerbound should be $L(\ep k/d)/(d/k)$. Together with Theorem~\ref{thm:lowerbound_constant} and the choice $k = \Theta(1/\gamma)$, this gives the guarantee in Theorem~\ref{thm:lowerbound}. The detailed proof is given in Appendix~\ifstoc{D}{\ref{sec:lowerbound_appendix}}. 

\section{Conclusion and Future Work}

In this paper, using multilevel Monte Carlo method we give a better algorithm for estimating the normalizing constant that only uses $\widetilde{\mathcal{O}}\pf{d^{4/3}\ka + d^{7/6}\kappa^{7/6}}{\varepsilon^2}$ queries to the gradient. We also give the first lowerbound that no algorithm can estimate the normalizing constant up to $1\pm \ep$ accuracy with $\frac{d^{1-o(1)}}{\varepsilon^{2-o(1)}}$ queries. For well-conditioned functions, the two bounds differ by $\cO(d^{1/3+o(1)}\ep^{-o(1)})$. Closing the gap is an immediate open problem, however we are not sure which side (if any) is tight. Any better rate for Langevin dynamics or related methods can give a better running time when combined with the multilevel Monte Carlo framework. On the other hand, improving our lowerbound might involve giving a lowerbound for sampling problems that depends on the dimension $d$.





There are many other settings where the idea of multilevel Monte Carlo may help improving the upperbound. This includes when only stochastic gradient queries are available (or when $f$ is a sum of simpler functions). We note that multilevel methods can work with stochastic gradients as well~\cite{giles2016multilevel}, and variance reduction techniques are available~\cite{chatterji2018theory}.
It is an interesting question whether multilevel Langevin dynamics or multilevel hybrid Monte Carlo can improve running times for volume estimation of convex sets (like polytopes)~\cite{lee2017geodesic,lee2018convergence}, or smooth log-concave distributions restricted to convex sets.


\section*{Acknowledgements}

HL would like to thank Oren Mangoubi for introducing multilevel methods. RG and HL would like to thank Ruoqi Shen for explaining the paper~\cite{shen2019randomized}. RG acknowledges funding from NSF CCF-1704656, NSF CCF-1845171 (CAREER), Sloan Fellowship and Google Faculty Research Award. Part of the work was done while RG was visiting the Institute for Advanced Study. The work of JL is supported in part by National Science Foundation via grants DMS-1454939 and CCF-1934964. 

\bibliographystyle{amsxport}
\bibliography{sampling}

\begin{bibdiv}
\begin{biblist}

\bib{AndrieuRidgwayWhiteley}{misc}{
      author={Andrieu, C.},
      author={Ridgway, J.},
      author={Whiteley, N.},
       title={Sampling normalizing constants in high dimensions using
  inhomogeneous diffusions},
        date={2016},
        note={preprint, arXiv:1612.07583},
}

\bib{balian2007microphysics}{book}{
      author={Balian, Roger},
       title={From microphysics to macrophysics: methods and applications of
  statistical physics},
   publisher={Springer Science \& Business Media},
        date={2007},
      volume={2},
}

\bib{BrosseDurmusMoulines:2018}{article}{
      author={Brosse, N.},
      author={Durmus, A.},
      author={Moulines, E.},
       title={Normalizing constants of log-concave densities},
        date={2018},
     journal={Electron. J. Statist.},
      volume={12},
       pages={851\ndash 889},
}

\bib{bakry1985diffusions}{incollection}{
      author={Bakry, Dominique},
      author={{\'E}mery, Michel},
       title={Diffusions hypercontractives},
        date={1985},
   booktitle={S{\'e}minaire de probabilit{\'e}s xix 1983/84},
   publisher={Springer},
       pages={177\ndash 206},
}

\bib{bakry2013analysis}{book}{
      author={Bakry, Dominique},
      author={Gentil, Ivan},
      author={Ledoux, Michel},
       title={Analysis and geometry of {M}arkov diffusion operators},
   publisher={Springer Science \& Business Media},
        date={2013},
      volume={348},
}

\bib{BouRabeeEberleZimmer}{misc}{
      author={Bou-Rabee, N.},
      author={Eberle, A.},
      author={Zimmer, R.},
       title={Coupling and convergence for {H}amiltonian {M}onte {C}arlo},
        date={2018},
        note={preprint, arXiv:1805.00452},
}

\bib{BouRabeeSanz-Serna:2018}{article}{
      author={Bou-Rabee, N.},
      author={Sanz-Serna, J.~M.},
       title={Geometric integrators and the {H}amiltonian {M}onte {C}arlo
  method},
        date={2018},
     journal={Acta Numer.},
      volume={27},
       pages={113\ndash 206},
}

\bib{cheng2017underdamped}{article}{
      author={Cheng, Xiang},
      author={Chatterji, Niladri~S},
      author={Bartlett, Peter~L},
      author={Jordan, Michael~I},
       title={Underdamped langevin mcmc: A non-asymptotic analysis},
        date={2017},
     journal={arXiv preprint arXiv:1707.03663},
}

\bib{ChenDwivediWainwrightYu}{misc}{
      author={Chen, Y.},
      author={Dwivedi, R.},
      author={Wainwright, M.~J.},
      author={Yu, B.},
       title={Fast mixing of {M}etropolized {H}amiltonian {M}onte {C}arlo:
  {B}enefits of multi-step gradients},
        date={2019},
        note={preprint, arXiv:1905.12247},
}

\bib{chatterji2018theory}{inproceedings}{
      author={Chatterji, Niladri},
      author={Flammarion, Nicolas},
      author={Ma, Yian},
      author={Bartlett, Peter},
      author={Jordan, Michael},
       title={On the theory of variance reduction for stochastic gradient
  {M}onte {C}arlo},
        date={2018},
   booktitle={Proceedings of the 35th international conference on machine
  learning},
      editor={Dy, Jennifer},
      editor={Krause, Andreas},
      series={Proceedings of Machine Learning Research},
      volume={80},
   publisher={PMLR},
     address={Stockholmsm\"assan, Stockholm Sweden},
       pages={764\ndash 773},
         url={http://proceedings.mlr.press/v80/chatterji18a.html},
}

\bib{CousinsVempala:2018}{article}{
      author={Cousins, B.},
      author={Vempala, S.},
       title={Gaussian cooling and {$O^{\ast}(n^3)$} algorithm for volume and
  {G}aussian volume},
        date={2018},
     journal={SIAM J. Comput.},
      volume={47},
       pages={1237\ndash 1273},
        note={preprint, arXiv:1409.6011v3},
}

\bib{chen2019optimal}{article}{
      author={Chen, Zongchen},
      author={Vempala, Santosh~S},
       title={Optimal convergence rate of hamiltonian monte carlo for strongly
  logconcave distributions},
        date={2019},
     journal={arXiv preprint arXiv:1905.02313},
}

\bib{dalalyan2017theoretical}{article}{
      author={Dalalyan, Arnak~S},
       title={Theoretical guarantees for approximate sampling from smooth and
  log-concave densities},
        date={2017},
     journal={Journal of the Royal Statistical Society: Series B (Statistical
  Methodology)},
      volume={79},
      number={3},
       pages={651\ndash 676},
}

\bib{DwivediChenWainwrightYu}{misc}{
      author={Dwivedi, R.},
      author={Chen, Y.},
      author={Wainwright, M.~J.},
      author={Yu, B.},
       title={Log-concave sampling: {M}etropolis-{H}astings algorithms are
  fast},
        date={2018},
        note={preprint, arXiv:1801.02309},
}

\bib{dyer1988complexity}{article}{
      author={Dyer, Martin~E.},
      author={Frieze, Alan~M.},
       title={On the complexity of computing the volume of a polyhedron},
        date={1988},
     journal={SIAM Journal on Computing},
      volume={17},
      number={5},
       pages={967\ndash 974},
}

\bib{dyer1991random}{article}{
      author={Dyer, Martin},
      author={Frieze, Alan},
      author={Kannan, Ravi},
       title={A random polynomial-time algorithm for approximating the volume
  of convex bodies},
        date={1991},
     journal={Journal of the ACM (JACM)},
      volume={38},
      number={1},
       pages={1\ndash 17},
}

\bib{dalalyan2017user}{article}{
      author={Dalalyan, Arnak~S},
      author={Karagulyan, Avetik~G},
       title={User-friendly guarantees for the langevin monte carlo with
  inaccurate gradient},
        date={2017},
     journal={arXiv preprint arXiv:1710.00095},
}

\bib{Duane:1987}{article}{
      author={Duane, S.},
      author={Kennedy, A.~D.},
      author={Pendleton, B.~J.},
      author={Roweth, D.},
       title={Hybrid {M}onte {C}arlo},
        date={1987},
     journal={Phys. Lett. B},
      volume={195},
       pages={216\ndash 222},
}

\bib{durmus2018analysis}{article}{
      author={Durmus, Alain},
      author={Majewski, Szymon},
      author={Miasojedow, B{\l}a{\.z}ej},
       title={Analysis of langevin monte carlo via convex optimization},
        date={2019},
     journal={Journal of Machine Learning Research},
      volume={20},
       pages={1\ndash 46},
}

\bib{durmus2017nonasymptotic}{article}{
      author={Durmus, Alain},
      author={Moulines, Eric},
      author={others},
       title={Nonasymptotic convergence analysis for the unadjusted langevin
  algorithm},
        date={2017},
     journal={The Annals of Applied Probability},
      volume={27},
      number={3},
       pages={1551\ndash 1587},
}

\bib{DalalyanRiou-Durand}{misc}{
      author={Dalalyan, A.~S.},
      author={Riou-Durand, L.},
       title={On sampling from a log-concave density using kinetic {L}angevin
  diffusions},
        date={2018},
        note={preprint, arXiv:1807.09382},
}

\bib{giles2008multilevel}{article}{
      author={Giles, Michael~B},
       title={Multilevel monte carlo path simulation},
        date={2008},
     journal={Operations Research},
      volume={56},
      number={3},
       pages={607\ndash 617},
}

\bib{GelmanMeng:1998}{article}{
      author={Gelman, A.},
      author={Meng, X.-L.},
       title={Simulating normalizing constants: {F}rom importance sampling to
  bridge sampling to path sampling},
        date={1998},
     journal={Stat. Sci.},
      volume={13},
       pages={163\ndash 185},
}

\bib{giles2016multilevel}{article}{
      author={Giles, Mike},
      author={Nagapetyan, Tigran},
      author={Szpruch, Lukasz},
      author={Vollmer, Sebastian},
      author={Zygalakis, Konstantinos},
       title={Multilevel monte carlo for scalable bayesian computations},
        date={2016},
     journal={arXiv preprint arXiv:1609.06144},
}

\bib{Jarzynski:1997}{article}{
      author={Jarzynski, C.},
       title={Nonequilibrium equality for free energy differences},
        date={1997},
     journal={Phys. Rev. Lett.},
      volume={78},
       pages={2690\ndash 2693},
}

\bib{jerrum1986random}{article}{
      author={Jerrum, Mark~R},
      author={Valiant, Leslie~G},
      author={Vazirani, Vijay~V},
       title={Random generation of combinatorial structures from a uniform
  distribution},
        date={1986},
     journal={Theoretical Computer Science},
      volume={43},
       pages={169\ndash 188},
}

\bib{laurent2000adaptive}{article}{
      author={Laurent, Beatrice},
      author={Massart, Pascal},
       title={Adaptive estimation of a quadratic functional by model
  selection},
        date={2000},
     journal={Annals of Statistics},
       pages={1302\ndash 1338},
}

\bib{lovasz1993random}{article}{
      author={Lov{\'a}sz, L{\'a}szl{\'o}},
      author={Simonovits, Mikl{\'o}s},
       title={Random walks in a convex body and an improved volume algorithm},
        date={1993},
     journal={Random structures \& algorithms},
      volume={4},
      number={4},
       pages={359\ndash 412},
}

\bib{lee2018algorithmic}{article}{
      author={Lee, Yin~Tat},
      author={Song, Zhao},
      author={Vempala, Santosh~S},
       title={Algorithmic theory of odes and sampling from well-conditioned
  logconcave densities},
        date={2018},
     journal={arXiv preprint arXiv:1812.06243},
}

\bib{Lovasz}{inproceedings}{
      author={Lovasz, Laszlo},
      author={Vempala, Santosh},
       title={Fast algorithms for logconcave functions: Sampling, rounding,
  integration and optimization},
        date={2006},
   booktitle={Proceedings of the 47th annual ieee symposium on foundations of
  computer science},
      series={FOCS '06},
   publisher={IEEE Computer Society},
     address={Washington, DC, USA},
       pages={57\ndash 68},
         url={http://dx.doi.org/10.1109/FOCS.2006.28},
}

\bib{lovasz2006simulated}{article}{
      author={Lov{\'a}sz, L{\'a}szl{\'o}},
      author={Vempala, Santosh},
       title={Simulated annealing in convex bodies and an o*(n4) volume
  algorithm},
        date={2006},
     journal={Journal of Computer and System Sciences},
      volume={72},
      number={2},
       pages={392\ndash 417},
}

\bib{lee2017geodesic}{inproceedings}{
      author={Lee, Yin~Tat},
      author={Vempala, Santosh~S},
       title={Geodesic walks in polytopes},
organization={ACM},
        date={2017},
   booktitle={Proceedings of the 49th annual acm sigact symposium on theory of
  computing},
       pages={927\ndash 940},
}

\bib{lee2018convergence}{inproceedings}{
      author={Lee, Yin~Tat},
      author={Vempala, Santosh~S},
       title={Convergence rate of riemannian hamiltonian monte carlo and faster
  polytope volume computation},
organization={ACM},
        date={2018},
   booktitle={Proceedings of the 50th annual acm sigact symposium on theory of
  computing},
       pages={1115\ndash 1121},
}

\bib{LiWuMackeyErdogdu}{misc}{
      author={Li, Xuechen},
      author={Wu, Denny},
      author={Mackey, Lester},
      author={Erdogdu, Murat~A.},
       title={Stochastic {R}unge-{K}utta accelerates {L}angevin {M}onte {C}arlo
  and beyond},
        date={2019},
        note={preprint, arXiv:1906.07868},
}

\bib{ma2019there}{article}{
      author={Ma, Yi-An},
      author={Chatterji, Niladri},
      author={Cheng, Xiang},
      author={Flammarion, Nicolas},
      author={Bartlett, Peter},
      author={Jordan, Michael~I},
       title={Is there an analog of nesterov acceleration for mcmc?},
        date={2019},
     journal={arXiv preprint arXiv:1902.00996},
}

\bib{mou2019improved}{article}{
      author={Mou, Wenlong},
      author={Flammarion, Nicolas},
      author={Wainwright, Martin~J},
      author={Bartlett, Peter~L},
       title={Improved bounds for discretization of langevin diffusions:
  Near-optimal rates without convexity},
        date={2019},
     journal={arXiv preprint arXiv:1907.11331},
}

\bib{mangoubi2017rapid}{article}{
      author={Mangoubi, Oren},
      author={Smith, Aaron},
       title={Rapid mixing of {Hamiltonian} {Monte} {Carlo} on strongly
  log-concave distributions},
        date={2017},
     journal={arXiv preprint arXiv:1708.07114},
}

\bib{mangoubi2018dimensionally}{inproceedings}{
      author={Mangoubi, Oren},
      author={Vishnoi, Nisheeth},
       title={Dimensionally tight bounds for second-order hamiltonian monte
  carlo},
        date={2018},
   booktitle={Advances in neural information processing systems},
       pages={6028\ndash 6038},
}

\bib{Neal:2001}{article}{
      author={Neal, R.~M.},
       title={Annealed importance sampling},
        date={2001},
     journal={Statist. Comput.},
      volume={11},
       pages={125\ndash 139},
}

\bib{RosskyDollFriedman:1978}{article}{
      author={Rossky, P.~J.},
      author={Doll, J.~D.},
      author={Friedman, H.~L.},
       title={Brownian dynamics as smart {M}onte {C}arlo simulation},
        date={1978},
     journal={J. Chem. Phys.},
      volume={69},
       pages={4628},
}

\bib{RobertsTweedie:1996}{article}{
      author={Roberts, G.~O.},
      author={Tweedie, R.~L.},
       title={Exponential convergence of {L}angevin distributions and their
  discrete approximations},
        date={1996},
     journal={Bernoulli},
      volume={2},
      number={4},
       pages={341\ndash 363},
}

\bib{RademacherVempala:2008}{article}{
      author={Rademacher, L.},
      author={Vempala, S.},
       title={Disperson of mass and the complexity of randomized geometric
  algorithms},
        date={2008},
     journal={Adv. Math.},
      volume={219},
       pages={1037\ndash 1069},
}

\bib{sinclair1989approximate}{article}{
      author={Sinclair, Alistair},
      author={Jerrum, Mark},
       title={Approximate counting, uniform generation and rapidly mixing
  markov chains},
        date={1989},
     journal={Information and Computation},
      volume={82},
      number={1},
       pages={93\ndash 133},
}

\bib{shen2019randomized}{article}{
      author={Shen, Ruoqi},
      author={Lee, Yin~Tat},
       title={The randomized midpoint method for log-concave sampling},
        date={2019},
     journal={arXiv preprint arXiv:1909.05503},
}

\bib{stoltz2010free}{book}{
      author={Stoltz, Gabriel},
      author={Rousset, Mathias},
      author={others},
       title={Free energy computations: A mathematical perspective},
   publisher={World Scientific},
        date={2010},
}

\bib{vempala2019rapid}{article}{
      author={Vempala, Santosh~S},
      author={Wibisono, Andre},
       title={Rapid convergence of the unadjusted langevin algorithm:
  Log-sobolev suffices},
        date={2019},
     journal={arXiv preprint arXiv:1903.08568},
}

\end{biblist}
\end{bibdiv}
\appendix
\section{Proofs for Annealing Strategy}\label{sec:annealingproof}

We provide proofs here for Lemmas in Section~\ref{s:anneal}. 

\begin{proof}[Proof of Lemma~\ref{l:start}]
  Without loss of generality, we assume $x^{\ast} = 0$ (as it amounts
  to a change of variable $x \to x - x^{\ast}$ which does not affect
  the normalizing constant). The upper bound is obvious since
  $f(x) \geq 0$ by our assumption (recall that we only concern about the relative error for normalizing constant, so that shifting $f$ by a constant has no impact). For the lower bound of $Z_1$, note
  that $f(x) \leq \frac{1}{2} L \norm{x}^2$, we have
  \begin{equation}
    \begin{aligned}
      Z_1 & = \int_{\RR^d} e^{-f(x) - \frac{1}{2} \frac{\norm{x}^2}{\sigma_1^2}} \ud x \\
      & \geq \int_{\RR^d} e^{ - \frac{1}{2} \bigl(L + \sigma_1^{-2} \bigr) \norm{x}^2} \ud x = \Bigl(2\pi \bigl( L + \sigma_1^{-2} \bigr)^{-1} \Bigr)^{d/2}
    \end{aligned}
  \end{equation}
  Thus, 
  \begin{equation}
    \frac{Z_1}{(2\pi \sigma_1^2)^{d/2}} = \Bigl( 1 + \sigma_1^2 L  \Bigr)^{-d/2} \geq e^{-d\sigma_1^2 L / 2} 
  \end{equation}
  which is larger than $1 - \frac{\veps}{2}$ for $\sigma_1^2 = \frac{\veps}{2dL}$.
\end{proof}

\begin{proof}[Proof of Lemma~\ref{lem:varboundM}]
  Define
  \begin{equation*}
    h(t) := \EE_{\rho} e^{-t\norm{x}^2} \EE_{\rho} e^{t\norm{x}^2}. 
  \end{equation*}
  We have
  \begin{equation*}
    \begin{aligned}
      \frac{h'(t)}{h(t)} & = \dfrac{\EE_{\rho} \Bigl( \norm{x}^2
        e^{t\norm{x}^2} \Bigr)} {\EE_{\rho} e^{t\norm{x}^2}} -
      \dfrac{\EE_{\rho} \Bigl( \norm{x}^2 e^{-t\norm{x}^2} \Bigr)}
      {\EE_{\rho} e^{-t\norm{x}^2}}  \\
      & = \int_{-t}^t v'(s)\ud s, 
    \end{aligned}
  \end{equation*}
  where
  \begin{equation*}
    v(s) := \dfrac{\EE_{\rho} \Bigl( \norm{x}^2
      e^{s\norm{x}^2} \Bigr)} {\EE_{\rho} e^{s\norm{x}^2}},. 
  \end{equation*}
  Thus
  \begin{equation*}
    \begin{aligned}
      v'(s) & = \dfrac{\EE_{\rho} \Bigl( \norm{x}^4 e^{s\norm{x}^2}
        \Bigr)\EE_{\rho} e^{s\norm{x}^2} - \biggl(\EE_{\rho} \Bigl(
        \norm{x}^2
        e^{s\norm{x}^2} \Bigr)\biggr)^2 } {\biggl(\EE_{\rho} e^{s\norm{x}^2}\biggr)^2}\\
      & = \var_{\rho_s} \bigl( \norm{x}^2 \bigr), 
    \end{aligned}
  \end{equation*}
  where $\rho_s$ is a distribution with
  $\frac{\ud\rho_s}{\ud\rho} \propto e^{s\norm{x}^2}$. Since $\rho$ is
  strongly log-concave with convexity parameter $\mu$, $\rho_s$
  satisfies the Poincar\'e inequality with constant
  $1/ (\mu - 2s) \leq 2/\mu$ for $s \leq \frac{1}{4} \mu$, thus
  \begin{equation}\label{eq:varrhos}
    \var_{\rho_s} \bigl( \norm{x}^2 \bigr) \leq \frac{8}{\mu} \EE_{\rho_s} \bigl( \norm{x}^2 \bigr) \leq \frac{16}{\mu} d,
  \end{equation}
  where the last inequality follows from the concentration property of log-concave distribution.

  Therefore,
  \begin{equation*}
    \begin{aligned}
      \ln h\bigl(\frac{1}{2\sigma_M^2}\bigr) & = \ln h(0) + \int_0^{1/(2\sigma_M^{2})} \frac{h'(t)}{h(t)} \ud t \\
      & \leq \int_0^{1/(2\sigma_M^{2})} \int_{-t}^t \var_{\rho_s} \bigl( \norm{x}^2 \bigr) \ud s
      \ud t \\
      & \leq \int_0^{1/(2\sigma_M^{2})} \int_{-t}^t \frac{16d}{\mu} \ud t \\
      & = \frac{4d}{\mu\sigma_M^4} 
    \end{aligned}
  \end{equation*}
  where in the last inequality we have used \eqref{eq:varrhos} that $s \leq 1/(2\sigma_M^2) \leq \frac{\mu}{4}$ by our assumption on $\sigma_M^2$. 
  Thus we arrive at
  \begin{equation*}
    \EE_{\rho}
    \exp\Bigl(-\frac{1}{2} \frac{\norm{x}^2}{\sigma_{M}^2}\Bigr)
    \, \EE_{\rho} \exp\Bigl(\frac{1}{2} \frac{\norm{x}^2}{\sigma_{M}^2}  \Bigr)
    \leq \exp \Bigl( \frac{4d}{\mu\sigma_M^4}  \Bigr).\qedhere
  \end{equation*}
\end{proof}

\begin{proof}[Proof of Lemma~\ref{lem:varbound}]  Define
  \begin{equation*}
    h(\alpha) :=     \EE_{\rho} \exp\Bigl(-\frac{1+\alpha}{2} \frac{\norm{x}^2}{\sigma^2}\Bigr)
      \, \EE_{\rho} \exp\Bigl(-\frac{1-\alpha}{2} \frac{\norm{x}^2}{\sigma^2}\Bigr).
  \end{equation*}
  It follows then 
  \begin{equation*}
    \begin{aligned}
      \frac{h'(\alpha)}{h(\alpha)} & = - \frac{1}{2\sigma^2} \left( \dfrac{\EE_{\rho} \Bigl( \norm{x}^2 \exp\Bigl(-\frac{1+\alpha}{2} \frac{\norm{x}^2}{\sigma^2}\Bigr) \Bigr)}{\EE_{\rho} \exp\Bigl(-\frac{1+\alpha}{2} \frac{\norm{x}^2}{\sigma^2}\Bigr)} - \dfrac{\EE_{\rho} \Bigl(\norm{x}^2 \exp\Bigl(-\frac{1-\alpha}{2} \frac{\norm{x}^2}{\sigma^2}\Bigr)\Bigr)}{\EE_{\rho} \exp\Bigl(-\frac{1-\alpha}{2} \frac{\norm{x}^2}{\sigma^2}\Bigr)} \right) \\
      & = - \frac{1}{2\sigma^2} \int_{1-\alpha}^{1+\alpha} v'(t) \ud t, 
    \end{aligned}
  \end{equation*}
  where $v(t)$ is defined as 
  \begin{equation*}
    v(t) := \dfrac{\EE_{\rho} \Bigl( \norm{x}^2 \exp\Bigl(-\frac{t}{2} \frac{\norm{x}^2}{\sigma^2}\Bigr) \Bigr)}{\EE_{\rho}  \exp\Bigl(-\frac{t}{2} \frac{\norm{x}^2}{\sigma^2}\Bigr)}. 
  \end{equation*}
  Explicit calculation gives 
  \begin{equation*}
    \begin{aligned}
      v'(t) & = - \frac{1}{2\sigma^2} \dfrac{\EE_{\rho} \Bigl(
        \norm{x}^4 \exp\Bigl(-\frac{t}{2}
        \frac{\norm{x}^2}{\sigma^2}\Bigr) \Bigr)
        \EE_{\rho}  \exp\Bigl(-\frac{t}{2} \frac{\norm{x}^2}{\sigma^2}\Bigr) - \biggl(\EE_{\rho} \Bigl( \norm{x}^2 \exp\Bigl(-\frac{t}{2} \frac{\norm{x}^2}{\sigma^2}\Bigr) \Bigr) \biggr)^2}{\biggl(\EE_{\rho}  \exp\Bigl(-\frac{t}{2} \frac{\norm{x}^2}{\sigma^2}\Bigr)\biggr)^2} \\
      & = - \frac{1}{2\sigma^2} \var_{\rho_t}(\norm{x}^2). 
    \end{aligned}
  \end{equation*}
  Here $\rho_t$ is the distribution given by 
  \begin{equation*}
    \frac{\ud \rho_t}{\ud \rho} \propto \exp\Bigl(-\frac{t}{2} \frac{\norm{x}^2}{\sigma^2}\Bigr). 
  \end{equation*}
  By the Poincar\`e inequality and concentration property of strongly log-concave measure 
  \begin{equation*}
    \var_{\rho_t}(\norm{x}^2) \leq \frac{4\sigma^2}{t} \EE_{\rho_t} \norm{x}^2 
    \leq 8\frac{\sigma^4}{t^2} d. 
  \end{equation*}
  Therefore, we arrive at the inequality 
  \begin{equation*}
    \begin{aligned}
      \dfrac{h'(\alpha)}{h(\alpha)} & = \frac{1}{4\sigma^4} \int_{1-\alpha}^{1+\alpha} \var_{\rho_t} (\norm{x}^2)  \ud t \\
      & \leq 2d \int_{1-\alpha}^{1+\alpha} \frac{1}{t^2} \ud t \\
      & = 2d \Bigl( \frac{1}{1-\alpha} - \frac{1}{1+\alpha} \Bigr) \\
      & \leq 8 d \alpha.
    \end{aligned}
  \end{equation*}
  This gives 
  \begin{equation*}
    \ln h(\alpha) - \ln h(0) = \int_0^{\alpha} \dfrac{h'(\alpha)}{h(\alpha)} \ud \alpha  \leq 4 d \alpha^2. 
  \end{equation*}
  Thus, we arrive that 
  \begin{equation*}
    \frac{h(\alpha)}{h(0)} \leq e^{4d \alpha^2},  
  \end{equation*}
  which is the desired inequality by the definition of $h$.
\end{proof}

\section{Estimating the Normalizing Constant using MALA and Annealing}
\label{s:mala}

Let us first recall the Metropolis adjusted Langevin algorithm (MALA)
\cite{RobertsTweedie:1996}, Algorithm~\ref{alg:mala}, which is a Metropolis-Hasting algorithm with the proposal step given by  discretized overdamped Langevin diffusion. 
\begin{algorithm} 
  \caption{Metropolis adjusted Langevin algorithm (MALA)}\label{alg:mala} 
  \begin{algorithmic}[1]
    \REQUIRE{Step size $h$ and a sample $x_0$ from a starting distribution $\mu_0$} 
    \ENSURE{Sequence $x_1, x_2, \ldots$}
    \FOR{$i = 0, 1, \ldots$}
    \STATE Draw $z_{i+1} \sim \mathcal{N}(x_i - h \nabla f(x_i), 2 h I)$
    \STATE Compute $\alpha_{i+1} \gets \min \Biggl\{1, \dfrac{\exp\bigl(-f(z_{i+1}) - \abs{x_i - z_{i+1} + h \nabla f(z_{i+1})}_2^2 / (4h) \bigr)}{\exp\bigl(-f(x_{i}) - \abs{z_{i+1} - x_i + h \nabla f(x_i)}_2^2 / (4h)\bigr)} \Biggr\}$
    \STATE With probability $\alpha_{i+1}$ accept the proposal $x_{i+1} \gets z_{i+1}$
    \STATE With probability $1 - \alpha_{i+1}$ reject the proposal $x_{i+1} \gets x_i$
    \ENDFOR
  \end{algorithmic}
\end{algorithm} 

Following the recent theoretical analysis for MALA
\cite{DwivediChenWainwrightYu, ChenDwivediWainwrightYu}, we consider
the $\tfrac{1}{2}$-lazy version of MALA, namely, for each step, for
probability $\tfrac{1}{2}$ one stays at the previous iterate and for
probability $\tfrac{1}{2}$ one takes a MALA step. The laziness
guarantees that the Markov chain is aperiodic and hence has a unique
invariant measure, given by the target distribution thanks to the
Metropolis acceptance-rejection step. The convergence of the empirical measure to the target measure has been established in \cite{DwivediChenWainwrightYu, ChenDwivediWainwrightYu}, which we recall here:
\begin{theorem}[{\cite[Theorem 2]{ChenDwivediWainwrightYu}}] \label{thm:MALA}
  Assume the target distribution $\rho$ is strongly log-concave with
  $L$-smooth and $\mu$-strongly convex negative log-density. Then
  given the initial distribution
  $\rho_0 = \mathcal{N}(x^{\ast}, \frac{1}{L} I)$, the $\tfrac{1}{2}$-lazy
  version of MALA with step size
  $h = c \bigl( L d \max\{1, \sqrt{\kappa / d}\}\bigr)^{-1}$
  achieves
  \begin{equation*}
    d_{\mathrm{TV}}(\rho_n, \rho) \leq \delta
  \end{equation*}
  for steps 
  \begin{equation*}
    n \geq C  d \kappa \log \frac{d}{\delta} \max \bigl\{ 1, \sqrt{\kappa / d} \bigr\}, 
  \end{equation*}
  where $c$ and $C$ above are universal constants.
\end{theorem}

The above Theorem assumes $x^{\ast}$, the minimum of $f(x)$.  In
practice, we do not know $x^{\ast}$ a priori, however, using a
first-order method like gradient descent, we can obtain an
$\eta$-approximate mode $\wt{x}$ using $\kappa \log (1/ \eta)$
gradient evaluations. If we instead take the initial distribution
$\wt{\rho}_0 = \mathcal{N}(\wt{x}, \frac{1}{2L} I)$, the warmness
parameter with respect to the target distribution becomes
$\exp\bigl(\frac{d}{2} \log(2\kappa) + L\eta^2 \bigr)$ instead of
$\kappa^{d/2}$ for $\rho_0 = \mathcal{N}(x^{\ast}, \frac{1}{L} I)$. As
discussed in \cite{DwivediChenWainwrightYu}*{Section 3.2}, with a
slightly modified step size, the MALA sampling then requires
\begin{equation*}
  n \geq C d \kappa \log \frac{d}{\delta} \max \bigl\{ 1, \sqrt{\kappa / d} \bigr\} \Bigl(2 + \frac{2L \eta^2}{d \log \kappa}\Bigr)
\end{equation*}
steps to achieve TV error less than $\delta$. Thus with a negligible
amount of increased cost for finding $\wt{x}$ that is $1/\sqrt{L}$
accurate: $\norm{\wt{x} - x^{\ast}} \leq 1/\sqrt{L}$, we have the
number of steps of MALA for achieving $\delta$ error in TV norm
remains
$\Or \bigl( d \kappa \log \frac{d}{\delta} \max \bigl\{ 1,
\sqrt{\kappa / d} \bigr\} \bigr)$.

\medskip 

Come back to the problem of estimating the normalizing constant. We will estimate the normalizing constant based on the annealing algorithm. The Lemma~\ref{lem:varboundM} suggests the choice of $\sigma_M^2$ to be larger than $\frac{2\sqrt{d}}{\mu}$ so that it satisfies the assumption of the Lemma the last stage has the same $\Or(1)$ relative variance as the previous steps, guaranteed by Lemma~\ref{lem:varbound}. This implies that the number of stages 
\begin{equation}
  M \leq C \sqrt{d} \Bigl( \log \frac{\kappa d}{\veps} + 1 \Bigr) = \wt{\Or}(\sqrt{d}). 
\end{equation}

Given the annealing sequence, we approximate  $Z_1$ by the normalizing constant of Gaussian with variance $\sigma_1^2$. Lemma~\ref{l:start} guarantees that this would only introduce at most $\veps/2$ relative error. Thus the task remains to estimate the ratio $Z_{i+1}/Z_i$ for $i = 1, \ldots, M$, or equivalently to estimate the expectation of
\begin{equation}
    g_i = \exp \Bigl( \frac{1}{2} \Bigl(\frac{1}{\sigma_{i}^2} - \frac{1}{\sigma_{i+1}^2} \Bigr) \norm{x}^2 \Bigr)
\end{equation}
under the distribution $\rho_i$, proportional to $\exp\bigl( - \frac{1}{2} \frac{\norm{x}^2}{\sigma_{i+1}^2} - f(x)  \bigr) \ud x$. Suppose we generate $K$ iid samples $X_i^{(1)}, \cdots, X_i^{(K)}$ according to $\rho_i$, we estimate the ratio $Z_{i+1}/Z_i$ by 
\begin{equation}
  \wh{g}_i = \frac{1}{K} \sum_{k=1}^K g_i(X_i^{(k)}). 
\end{equation}
Denote the short hand $\wb{g}_i = \EE_{\rho_i} g_i$, we use the relative variance bounds shown in Lemma~\ref{lem:varbound} and Lemma~\ref{lem:varboundM} to upper bound
\begin{equation}\label{eq:gvarbound}
    \begin{aligned}
    \EE(\wh{g}_i^2) & = \frac{1}{K^2} \Bigl( \sum_{k=1}^K \EE \bigl(g_i(X_i^{(k)}) \bigr)^2   + K(K-1) \wb{g}_i^2 \Bigr) \\
    & \leq \frac{1}{K^2} \bigl( e^4 K + K(K-1) \bigr) \wb{g}_i^2 \\
    & \leq \Bigl(1 + \frac{60}{K} \Bigr) \wb{g}_i^2
    \end{aligned}
\end{equation}

\begin{lemma}\label{lem:prodvar}\label{l:prod}
    Let $Y_i$, $i = 1, \ldots, M$ be independent variables and let $\bar{Y}_i = \EE Y_i$. Assume there exists $\eta > 0$  such that $\eta M \leq \frac{1}{5}$ and
    \begin{equation*}
        \EE Y_i^2 \leq (1 + \eta) \bar{Y}_i^2,
    \end{equation*}
    then for any $\veps > 0$
    \begin{equation*}
        \mathbb{P}\Biggl( \frac{ \abs{ Y_1 \cdots Y_M - \bar{Y}_1 \cdots \bar{Y}_M}}{\bar{Y}_1 \cdots \bar{Y}_M} \geq \frac{\veps}{2} \Biggr) \leq \frac{5 \eta M}{\veps^2}.
    \end{equation*}
\end{lemma}
\begin{proof}
  The proof follows the Chebyshev's inequality: 
  \begin{align*}
      \mathbb{P}\Biggl( \frac{ \abs{ Y_1 \cdots Y_M - \bar{Y}_1 \cdots \bar{Y}_M}}{\bar{Y}_1 \cdots \bar{Y}_M} \geq \frac{\veps}{2} \Biggr) & \leq \frac{4}{\veps^2} \frac{\var(Y_1 \cdots Y_M)}{\bar{Y}_1^2 \cdots \bar{Y}_M^2} \\
      & = \frac{4}{\veps^2} \Biggl( \frac{\EE (Y_1^2 \cdots Y_M^2)}{\bar{Y}_1^2 \cdots \bar{Y}_M^2} - 1 \Biggr) \\
      & \leq \frac{4}{\veps^2} \bigl( (1 + \eta)^M - 1) \\
      & \leq \frac{4}{\veps^2} (e^{\eta M} - 1) \\
      & \leq \frac{5 \eta M}{\veps^2},
  \end{align*}
  where the last inequality follows from $e^{\eta M} - 1 \leq \frac{5}{4} \eta M$ for $\eta M \leq \frac{1}{5}$.
\end{proof}

Applying Lemma~\ref{lem:prodvar} by taking $Y_i = \wh{g}_i$ and $\eta = \frac{60}{K}$, we obtain 
\begin{equation}
   \mathbb{P}\Biggl( \frac{\abs{\wh{g}_1 \cdots \wh{g}_M - \wb{g}_1 \cdots \wb{g}_M}}{\wb{g}_1 \cdots \wb{g}_M} \geq \frac{\veps}{2} \Biggr)  \leq \frac{300 M}{\veps^2 K}.
\end{equation}
This suggests us to take the number of samples $K = \frac{1200 M}{\veps^2}$, so that the right hand side of above is bounded by $\frac{1}{4}$. Since we have $M$ stages in total, the total number of samples we need in the whole algorithm is 
\begin{equation}
  N_{\mathrm{tot}} = MK = \Or\bigl( \frac{M^2}{\veps^2} \bigr) = \wt{\Or}\bigl( \frac{d}{\veps^2} \bigr).
\end{equation}

To generate the iid samples $X_i^{(1)}, \ldots, X_i^{(K)}$, $i = 1, \ldots, M$, we will use the $\tfrac{1}{2}$-lazy version of MALA algorithm, and choose parameter
$\delta = \frac{1}{4} \frac{1}{N_{\mathrm{tot}}}$, so that for
probability at least $\frac{3}{4}$, every sample in our algorithm is
guaranteed to follow the desired distribution, since we have in total
$N_{\mathrm{tot}}$ samples.

Note that we have a uniform bound over the condition number of
$\rho_i, i = 1, \ldots, M$ by $\kappa = L / \mu$ thanks to the
strongly log-concave assumption on $\rho$. Thus, for each sample, the number of steps it
takes is bounded by 
$\Or \bigl( d \kappa \log (d N_{\mathrm{tot}}) \max \bigl\{ 1, \sqrt{\kappa / d} \bigr\}
\bigr)$ by Theorem~\ref{thm:MALA}. 

We summarize the procedure of estimating the normalizing constant based on the MALA sampling below. 
\begin{algorithm}[H]
  \caption{Annealing algorithm for normalizing constant based on MALA}\label{alg:normmala} 
  \begin{algorithmic}[1]
    \REQUIRE{$\mu$-strongly convex and $L$-smooth function $f$, error threshold $\veps$} 
    \ENSURE{An estimate $\wh{Z}$ for the normalizing constant $Z = \int e^{-f(x)} \ud x$ within relative error $\Or(\veps)$}
    
    \STATE $\sigma_1^2 \gets \frac{\veps}{2dL}$
    \STATE $M \gets \Bigl\lceil \log \bigl(\frac{2 d^{3/2} \kappa}{\veps} \bigr) / \log\bigl(1 + \frac{1}{\sqrt{d}}\bigr) \Bigr\rceil$
    \STATE $K \gets \frac{1200 M}{\veps^2}$
    \STATE $\wh{Z} \gets \bigl( 2\pi \sigma_1^2 \bigr)^{d/2}$
    \FOR{$i = 1, 2, \ldots, M$}
    \IF{$i < M$} 
        \STATE {$\sigma_{i+1}^2 \gets \sigma_i^2 \Bigl( 1 + \frac{1}{\sqrt{d}} \Bigr)$} 
    \ELSE 
        \STATE{$\sigma_{i+1}^2 \gets \infty$}
    \ENDIF
    \STATE Use $\frac{1}{2}$-lazy MALA to generate random variables $X_i^{(1)}, \ldots, X_i^{(K)}$ iid wrt $\rho_i$ with TV error guarantee $\delta = \frac{1}{4MK}$. 
    \STATE $\wh{g}_i \gets \frac{1}{K} \sum_{k=1}^K \exp\Bigl(\frac{1}{2} \Bigl(\frac{1}{\sigma_i^2} - \frac{1}{\sigma_{i+1}^2} \Bigr) \norm{X_i^{(k)}}^2 \Bigr)$ 
    \STATE $\wh{Z} \gets \wh{Z}\, \wh{g}_i$
    \ENDFOR
    \RETURN $\wh{Z}$
  \end{algorithmic}
\end{algorithm} 
Putting together all the above estimates, we arrive at the following guarantee for the Algorithm~\ref{alg:normmala}. 

\begin{theorem}\label{thm:normmala} 
Let $f:\RR^d \to \RR$ be a $\mu$-strongly convex and $L$-smooth function. 
With probability of success at least $\frac{3}{4}$,  Algorithm~\ref{alg:normmala} gives an estimate $\wh{Z}$ of the normalizing constant $Z = \int e^{-f(x)} \ud x$ with relative error $\veps$ with query complexity 
\begin{equation*}
\Or\Bigl( MK \frac{d^2}{\veps^2} \log (dMK) \kappa \max \bigl\{ 1, \sqrt{\kappa / d} \bigr\} \Bigr) = \wt{\Or}\Bigl( \frac{d^2}{\veps^2} \kappa \max \bigl\{ 1, \sqrt{\kappa / d} \bigr\} \Bigr). 
\end{equation*}
\end{theorem}

\newcommand{\symatt}[3]{\matt{#1}{#2}{#2}{#3}}

\newcommand{\rmmeta}[0]{\max\bc{\fc{\ep^{\rc3}}{\ka^{\rc 6}}\pf{\mu}{d}^{\rc 6}\log^{-\rc 6}\prc{\ep},
\ep^{\fc 23} \pf{\mu}{d}^{\rc 3} \log^{-\rc 3}\prc{\ep}}}
\newcommand{\mletazero}[0]{\fc{c^{\rc{\be}}}{c_\si^{\fc2{\be}}}}
\newcommand{\mletak}[0]{\fc{\ep_b^{\rc \al}}{c_b^{\rc \al}}}
\newcommand{\mlNj}[0]{\fc{c_\si^2 \eta_0^{\fc{\be-1}2}\eta_j^{\fc{\be+1}2}}{\ep_\si^2}}
\newcommand{\mlqueries}[0]{T(\ep_b)\pa{\fc{c_\si^{\fc2{\be}}c^{1-\rc\be}}{\ep_\si^2} + \fc{c_b^{\rc \al}}{\ep_b^{\rc \al}}}}

\newcommand{\mletakvar}[0]{\fc{\ep_b^{\fc2\be}}{c_\si^{\fc2\be} L_g^{\fc2\be}}}
\newcommand{\mlNjvar}[0]{\fc{L_g^2c_\si^{1+\rc\be} c^{\rc2-\rc{2\be}}\eta_j^{\fc{\be+1}2}}{\ep_\si^2}}
\newcommand{\mlqueriesvar}[0]{T\pf{\ep_b}{L_g}c_\si^{\fc 2\be}\pa{\fc{L_g^2 c^{1-\rc\be}}{\ep_\si^2} + \fc{L_g^{\fc2\be}}{\ep_b^{\fc2\be}}}}

\section{Estimating the Normalizing Constant using Multilevel Langevin}
\label{s:ml}

In Section~\ref{ss:ml} we introduce multilevel Monte Carlo, a generic way to obtain a faster rate for estimating an expected value. 
Multilevel Monte Carlo reduces the variance in the estimate by simulating a SDE with multiple step sizes in a coupled fashion. We give guarantees for multilevel Monte Carlo for a general setting, assuming properties of the SDE and the coupling.
In Section~\ref{s:ml-uld} and~\ref{s:ml-uld-rmm} we apply the multilevel Monte Carlo to ULD and ULD with RMM, respectively. These two sections prove the two parts of Theorem~\ref{thm:expectation_main}. 
In Section~\ref{s:bias}, we introduce a truncation procedure to solve the technical issues mentioned in Section~\ref{subsec:technical}, namely that the function we are estimating is not Lipschitz. 
Finally in Section~\ref{s:ml-norm-uld} we apply multilevel ULD and ULD-RMM to normalizing constant estimation.

\subsection{The multilevel estimate}
\llabel{ss:ml}


We consider multilevel Monte Carlo for the following setting:
We wish to estimate $\E_{X\sim \rh} g(X)$, where $\rh$ cannot be sampled from exactly, but can be (approximately) sampled from by simulating a SDE for some time $T$. 
Suppose we have a discretization algorithm $\cal A$ that given time $T$ and step size $\eta$, simulates the SDE with step size $\eta$, making $\cO(T/\eta)$ queries (i.e., a constant number of queries per iteration), and returns a sample $X^\eta = x_T^\eta \sim \rh^\eta$. Smaller $\eta$ naturally gives more accurate samples, but it also requires more queries and takes longer time. 
Naively, we would just run $\cal A$ at a step size $\eta$ small enough so that $|\E_{X\sim \rh^{\eta}} g(X^\eta)-\E_{X\sim \rh} g(X)|\le \eph$, and take enough samples. If we need to take $\eta=\ep^{-\ga}$, then this gives a rate of $\cO\prc{\ep^{2+\ga}}$. 

Multilevel Monte Carlo method takes advantage of coupling of $\cal A$ at two step sizes to reduce the variance. Assume that we can run $\cal A$ coupled between two step sizes, to generate $(X^{\eta},X^{\eta/2})$ such that $\Var(g(X^\eta)-g(X^{\eta/2}))\ll \Var(g(X^\eta))$ decays sufficiently fast, 
multilevel Monte Carlo leads to a faster rate $\wt \cO\prc{\ep^2}$ for estimating $\EE_{X \sim \rho} g(X)$. 
The dependence on other parameters will also be improved. 

To achieve this, multilevel Monte Carlo uses the estimator  \begin{align}\llabel{e:ml-est}
\wh R:=
\rc{N_0} \sumo i{N_0} g(X_i^{\eta_0})
+\sumo jk \rc{N_j} \sumo i{N_j} [g(X_i^{\eta_0/2^{j}+}) - g(X_i^{\eta_0/2^{j-1}-})]
\end{align}
where $X_i^{\eta_0}$ are samples at the highest level (step size), and $(X_i^{\eta_0/2^j+}, X_i^{\eta_0/2^j})$ are coupled samples at level $j$. For larger $j$, the variance $\Var(g(X_i^{\eta_0/2^{j}+}) - g(X_i^{\eta_0/2^{j-1}-}))$ is smaller, so fewer samples are needed, offsetting the increased query complexity. 
We note that $\E\wh R = \E[g(X^{\eta_0/2^j})]$, so the bias is determined by the smallest step size. On the other hand, minimizing the variance requires optimizing the sample sizes $N_j$.

We work out non-asymptotic rates for multilevel Monte Carlo, given the guarantees on $\cal A$ (the rate of decay of the variance and bias of individual estimates in the step size $\eta$).
The result is similar to~\cite[Theorem 3.1]{giles2008multilevel}, which works out the asymptotic rates when the variance and bias 
follow a power law in $\eta$. 
However, we will need to work out the rates when the desired bias $\ep_b$ and variance $\ep_\si^2$ are different, because for our application of estimating the normalizing constant, we can tolerate a larger $\ep_\si$ than $\ep_b$ at each temperature. 

Note also the complication that in our setting, the bias depends not just on the step size, but also the time $T$. We simulate a SDE where $\rh$ is the stationary distribution, so running the algorithm for a finite time $T$ introduces some bias $\ep$, even as the step size $\eta\to 0$. Hence, we assume that the bias is bounded by $G(\eta)\vee \ep$, whenever $T\ge T(\ep)$, and need to set $T$ large enough. In our setting, the Markov processes will converge exponentially, so this only introduces a $\log\prc{\ep}$ factor.

\begin{algorithm}[h!]
\caption{Multilevel Monte Carlo}
\llabel{a:ml}
\begin{algorithmic}[1]
\REQUIRE Initial point $x_0$, time $T$, largest step size $\eta_0$, 
number of levels $k$, number of samples $N_0,\ldots, N_k$, function $f:\R^d\to \R$, function $g:\R^d\to \R$.
\REQUIRE Sampling algorithm $\cal A(x_0,f,\eta,T)$ which can give coupled samples $(x^\eta,x^{\eta/2})$ (or individual samples $x^\eta$).
\ENSURE Estimate of $\E_{x\sim \rho} g(x)$ where $\rho(\ud x)\propto e^{-f(x)}\dx$
\FOR{$1\le i\le N_0$}
  \STATE Run $\cal A$ with initial point $x_0$, function $f$, step size $\eta_0$, and time $T$ to obtain $X^{\eta_0}_i$.
\ENDFOR
\FOR{$1\le j\le k$}
	\FOR{$1\le i\le N_j$}
	  \STATE Run coupled $\cal A$ with initial point $x_0$, function $f$, step size $\eta=\eta_0/2^{j-1}$, and time $T$, to obtain $(X^{\eta-}_i,X^{\eta/2+}_i)$.
	\ENDFOR
\ENDFOR
\RETURN $\rc{N_0} \sumo i{N_0} g(X_i^{\eta_0})
+\sumo jk \rc{N_j} \sumo i{N_j} [g(X_i^{\eta_0/2^{j}-}) - g(X_i^{\eta_0/2^{j-1}+})]$.
\end{algorithmic}
\end{algorithm}


\begin{lem}\llabel{l:ml-est}\llabel{l:ml}
Let $\cal A$ be an algorithm that given a parameter $T$ (e.g. time) and $\eta>0$ (e.g., discretizations with step size $\eta$), returns $X^\eta$. Let $\rh^\eta$ be the distribution of $X^\eta$. Suppose also that $X\sim \rh$ (the distribution we are trying to approximate) and there are couplings between any two of the random variables. Suppose the following hold for any $\eta\le \eta_{\max}$:
\begin{enumerate}
\item
If $X^{\eta}$ and $X^{\eta'}$ are coupled, the variance satisfies $\Var[g(X^\eta)-g(X^{\eta'})]\le F(\eta)$ whenever $\eta'\le\fc{\eta}2$, where $F$ is a non-decreasing, non-negative function satisfying $\sumz j{\iy} \pf{F(\eta/2^j)}{\eta/2^j}^a\le C_F\pf{F(\eta)}{\eta}^a$ for some universal constant $C_F$ and any $a\in \{\rc 2,1\}$.
\item
The bias satisfies $|\E g(X^{\eta})-\E g(X)|\le G(\eta)\vee \ep$, for non-decreasing function $G$, whenever $T\ge T(\ep)$. 
\item
The variance satisfies 
$\Var[g(X)]\le c$.
\item
Algorithm $\cal A$ takes $\fc{T}{\eta}$ queries (e.g., to $\nb \log(\rh)$) to compute a sample $X^\eta$.
\end{enumerate}
Suppose $\eta_j = \fc{\eta_0}{2^j}$ and $\eta_0$, $\eta_k$, $N_j$, and $T$ are chosen so that the following hold:
\begin{itemize}
\item
$F(\eta_0)= c$ and $\eta_0\le \eta_{\max}$. 
\item
$G(\eta_k)\le \ep_b$.
\item
$N_j \ge \rc{\ep_\si^2} \sfc{F(\eta_0) \eta_j F(\eta_j)}{\eta_0}$.
\item
$T\ge T(\ep_b)$. 
\end{itemize}
Then the estimate~\eqref{e:ml-est}
satisfies $|\E \wh R - \E_\rh g|\le \ep_b$ and $\Var(\wh R)\le \ep_\si^2$. 
Taking $N_j$ to be the minimum possible, the number of queries needed is
\begin{align*}
Q&=T\pa{\fc{4C_F^2c}{\ep_\si^2 \eta_0} + \fc{2}{\eta_k}}
=\cO\pa{T\pa{\fc{c}{\ep_\si^2 \eta_0}+\rc{\eta_k}}}.
\end{align*}
\end{lem}
Note for example that the decay condition on $F$ is satisfied when $F(\eta) = C\eta^\be$ for some $\be>1$. This is the most favorable case in~\cite[Theorem 3.1]{giles2008multilevel}; reduced speedups are still available in the regime $\be\le 1$.

\begin{proof}
Let $T=T(\ep_b)$. 
The number of queries needed is $\sumo jk \fc{TN_j}{\eta_j}$. 

We claim that the total variance is $\Var(\wh R) \le 4\sumo jk \fc{F(\eta_j)}{N_j}$, and the bias is $|\E\wh R - \E_\rh g| \le G(\eta_k)\vee \ep_b$.

To see the expression for the variance, write
\begin{align*}
\wh R = 
\rc{N_0} \sumo i{N_0} g(X_i)+
\rc{N_0} \sumo i{N_0} \ba{g(X_i^{\eta_0}) -g(X_i)}
+\sumo jk \rc{N_j} \sumo i{N_j} [g(X_i^{\eta_0/2^{j}+}) - g(X_i^{\eta_0/2^{j-1}-})]
\end{align*}
so that the total variance is (the first two terms are not independent, but the others are)
\begin{align*}
\Var(\wh R)
\le 
\fc{2c}{N_0} + \fc{2F(\eta_0)}{N_0} + 
\sumo jk
\fc{F(\eta_j)}{N_j}
\le 4 \sumz jk \fc{F(\eta_j)}{N_j}
\end{align*}
since $\fc{c}{N_0} \le  \fc{F(\eta_0)}{N_0}$ by assumption on $\eta_0$.

For the bias, note that $|\E\wh R - \E_\rh g|= |\E g(X^{\eta_k})-\E_\rh g| \le G(\eta_k)\vee \ep_b$ by assumption.

To justify our choice of $N_j$, note that by Cauchy-Schwarz,
\begin{align*}
\ub{\pa{\sumz jk \fc{TN_j}{\eta_j}}}{\pat{number of time steps}}
\ub{\pa{\sumz jk \fc{F(\eta_j)}{N_j}}}{\pat{upper bound on variance}}
&\ge T \pa{\sumz jk  \sfc{F(\eta_j)}{\eta_j}}^2.
\end{align*}
If the bound on variance is kept constant, because the RHS does not depend on $N_j$, then the the number of steps is minimized when equality happens above. 
Equality happens when $N_j = K\sqrt{\eta_j F(\eta_j)}$ for some constant $K$.  When $N_j \ge K\sqrt{\eta_j F(\eta_j)}$
 the variance is bounded by
\begin{align*}
\Var(\wh R) = 
4\sumz jn \fc{F(\eta_j)}{N_j}
= 4\sumz jn \fc{F(\eta_j)}{K\sqrt{\eta_j F(\eta_j)}}
&\le 4\sumz jn \rc K\sfc{F(\eta_j)}{\eta_j} 
=\fc{4C_F}{K}\sfc{F(\eta_0)}{\eta_0}
\end{align*}
by assumption on the decay of $F$. By choosing $K=\fc{4C_F}{\ep_\si^2} \sfc{F(\eta_0)}{\eta_0}$, the variance is bounded by $\ep_\si^2$. Then the requirement on $N_j$ is $N_j \ge \fc{4C_F}{\ep_\si^2} \sfc{F(\eta_0)\eta_j F(\eta_j)}{\eta_0}$. 

It remains to compute the number of time steps. With the minimum choice of $N_j$, the number of time steps is
\begin{align*}
\sumz jk \fc{TN_j}{\eta_j}
&\le \sumz jk \fc{T}{\eta_j} \pa{
\fc{4C_F}{\ep_\si^2}\sfc{F(\eta_0)\eta_j F(\eta_j)}{\eta_0}
+1
}\\
&=
\sumz jk \fc{4TC_F}{\ep_\si^2}\sfc{F(\eta_0)F(\eta_j)}{\eta_0\eta_j }
+\sumz jk \fc{T}{\eta_j}\\
&\stackrel{\mathrm{(i)}}{\le}
\fc{4TC_F^2}{\ep_\si^2}\fc{F(\eta_0)}{\eta_0} + \fc{2T}{\eta_k}
\\
&\le T\pa{\fc{4C_F^2c}{\ep_\si^2 \eta_0} + \fc{2}{\eta_k}}.
\end{align*}
where (i) uses the assumption on decay of $F$ and the fact that $\fc{T}{\eta_k}$ is a decaying geometric series with largest term $\fc{T}{\eta_k}$.
\end{proof}

We put the lemma in a more convenient form for our applications.
\begin{lem}\llabel{l:ml2}
Suppose $g:\R^d\to \R$ is $L_g$-Lipschitz. 
Let $\cal A$ be an algorithm that given a parameter $T$ and $\eta>0$, returns $X^\eta$. Let $\rh^\eta$ be the distribution of $X^\eta$. Suppose also that $X^0\sim \rh^0$ (e.g., the continuous process with the same initial distribution) and $X\sim \rh$ (the distribution we are trying to approximate) and there are couplings between any two of the random variables. Suppose the following hold for any $\eta\le \eta_{\max}$:
\begin{enumerate}
\item
If $X^\eta$ and $X^0$ are coupled, then $\E[\ve{X^\eta-X^0}^2] \le F(\eta)$, where $F$ is a non-decreasing, non-negative function satisfying $\sumz j{\iy} \pf{F(\eta/2^j)}{\eta/2^j}^a\le C_F\pf{F(\eta)}{\eta}^a$ for some universal constant $C_F$ and any $a\in \{\rc 2,1\}$.
\item
If $T\ge T(\ep)$, then $W_2(\rh^\eta,\rh)^2 \le F(\eta)\vee \ep^2$.
\item
$\rh$ satisfies a Poincar\'e inequality with constant $c$. (In particular, this is satisfied for $c=\rc{\mu}$ if $\rh(\rdx) \propto e^{-f(x)}\dx$ and $f$ is $\mu$-strongly convex.)
\item
Algorithm $\cal A$ takes $\fc{T}{\eta}$ queries (e.g., to $\nb \log(\rh)$) to compute a sample $X^\eta$.
\end{enumerate}
Suppose $\eta_j = \fc{\eta_0}{2^j}$ and $\eta_0$, $\eta_k$, $N_j$, and $T$ are chosen so that the following hold:
\begin{itemize}
\item
$F(\eta_0)=\fc{c}{4}$, $\eta_0\le \eta_{\max}$.
\item
$F(\eta_k)\le \fc{\ep_b^2}{4L_g^2}$.
\item
$N_j \ge \fc{4L_g^2}{\ep_\si^2} \sfc{F(\eta_0) \eta_j F(\eta_j)}{\eta_0}$.
\item
$T\ge T\pf{\ep_b}{L_g}$
\end{itemize}
Then the estimate~\eqref{e:ml-est}
satisfies $|\E \wh R - \E_\rh g|\le \ep_b$ and $\Var(\wh R)\le \ep_\si^2$. 
Taking $N_j$ to be the minimum possible, the number of queries needed is
\begin{align*}
Q&
=\cO\pa{T\pa{\fc{cL_g^2}{\ep_\si^2 \eta_0}+\rc{\eta_k}}}.
\end{align*}
Moreover, we have $W_2(\rh^{\eta_k}, \rh)\le \fc{\ep_b}{L_g}$.
\end{lem}
\begin{proof}
We check that the conditions of Lemma~\ref{l:ml} are satisfied with $F(\eta)\mapsfrom 4L_g^2 F(\eta)$, $G(\eta)\mapsfrom 2L_g\sqrt{F(\eta)}$, $c\mapsfrom cL_g^2$, and $T(\ep)\mapsfrom T\pf{\ep}{L_g}$. Substituting then gives the parameters.
\begin{enumerate}
\item
Using the fact that $g$ is $L_g$ Lipschitz, Cauchy-Schwarz, and the Minkowski inequality,
\begin{align*}
\Var[g(X^\eta) - g(X^{\eta'})] &\le 
L_g^2 \E\ba{\ve{X^\eta-X^{\eta'}}^2}\\
&\le L_g^2 \pa{ \E\ba{\ve{X^\eta-X^{0}}^2}^{\rc 2}  + \E\ba{\ve{X^0-X^{\eta'}}^2}^{\rc 2}}^2
\le 4L_g^2 F(\eta).
\end{align*}
\item
Using the fact that $g$ is $L_g$ Lipschitz, for $T\ge T\pf{\ep}{L_g}$,
\begin{align*}
|\E g(X^\eta) - \E g(X)| &\le L_g W_2(X^\eta,X)\le L_g \pa{\sqrt{f(\eta)} \vee \fc{\ep}{L_g}}
=L_g\sqrt{F(\eta)}\vee \ep.
\end{align*}
\item
Since $\ve{\nb g(x)}\le L_g$, the Poincar\'e inequality implies that $\Var_\rh(g)\le 
\rc\mu\int_{\R^d} \ve{\nb g(x)}^2 \dx \le cL_g^2$. When $\rh(\rdx) \propto e^{-f(x)}$ and $f$ is $\mu$-strongly convex, it satisfies a Poincar\'e inequality by Bakry-\'Emery, Theorem~\ref{t:be}.
\item
This follows directly. 
\end{enumerate}
Finally, note that by choice of $\eta_k$, $W_2(\rh^{\eta_k}, \rh)\le F(\eta_k)\wedge \fc{\ep_b}{L_g} = \fc{\ep_b}{L_g}$.
\end{proof}

\subsection{Multilevel ULD}
\llabel{s:ml-uld}

Underdamped Langevin diffusion with parameters $\ga,u$ is given by the following SDE:
\begin{align*}
\ud v_t&= -\ga v_t \dt - u \nb f(x_t)\dt + \sqrt{2\ga u} \dB_t\\
\ud x_t&=v_t\dt
\end{align*}
where $x_t,v_t\in \R^d$ and $B_t$ is standard Brownian motion. Under mild conditions, the SDE is ergodic with  stationary distribution proportional to $e^{-\pa{f(x) + \rc{2u}\ve{v}^2}}$. Compared to overdamped Langevin dynamics on log-concave distributions, it is known to enjoy an improved rate of convergence in $W_2$ distance. Here, $v_t$ is thought of as velocity, and $-\ga v_t$ is a drag term. ULD is closely related to Hamiltonian Monte Carlo.

The discrete dynamics with step size $\eta$ can be described by
\begin{align*}
\ud v_t^\eta&= -\ga v_t^\eta \dt - u \nb f(x_{\fl{t/\eta}\eta}^\eta)\dt + \sqrt{2\ga u} \dB_t\\
\ud x_t^\eta&=v_t^\eta\dt.
\end{align*}
We will take $\ga=2$ and $u=\rc L$.
By integration, we can derive the explicit discrete-time update rule~\cite[Lemma 10]{cheng2017underdamped}:
\begin{align}
\llabel{e:vt+eta}
v_{t+\eta}^\eta &= 
e^{-2\eta} v_t^\eta + \fc{1}{L} \int_0^\eta e^{-2(\eta-s)} \nb f(x_t^\eta)\ud s +
\fc{2}{\sqrt L}\int_0^\eta e^{2(s-\eta)}\dB_{t+s}\\
&=e^{-2\eta} v_t^\eta + \fc{1}{2L} (1-e^{-2\eta}) \nb f(x_t^\eta) +
\fc{2}{\sqrt L}
\ub{\int_0^\eta e^{2(s-\eta)}\dB_{t+s}}{=:W_{1,t}^{\eta}}\\
\llabel{e:xt+eta}
x_{t+\eta}^\eta &= 
x_t^\eta+\rc{2}(1-e^{-2\eta}) v_t^\eta + \fc{1}{2L}\int_0^\eta \pa{1-e^{-2(\eta-s)}} \nb f(x_t^\eta)\ud s+ \fc1{\sqrt L}\int_0^\eta(1-e^{2(s-\eta)})\dB_{t+s}\\
&=x_t^\eta+\rc{2}(1-e^{-2\eta}) v_t^\eta + \fc{1}{2L}\pa{\eta - \rc{2}(1-e^{-2\eta})} \nb f(x_t^\eta)\ud s+ \fc1{\sqrt L}
\ub{\int_0^\eta(1-e^{2(s-\eta)})\dB_{t+s}}{=:W_{2,t}^\eta}
\end{align}
where all the instances of Brownian motion are the same. Let $G_{t}^\eta= \int_0^\eta e^{2s}\,dB_{t+s}$ and $H_t^\eta=\int_0^\eta\dB_{t+s}$. As calculated in~\cite[Lemma 5]{shen2019randomized}, 
\begin{align*}
\coltwo{G_t^\eta}{H_t^\eta} &\sim N\pa{\mathbf 0, \symatt{\rc 4(e^{4 \eta}-1)}{\rc 2(e^{2 \eta}-1)}{\eta}\ot I_d}\\
W_{1,t}^\eta &= e^{-2\eta} G_t^\eta\\
W_{2,t}^\eta &= H_t^\eta - e^{-2\eta} G_t^{\eta}.
\end{align*}
Define $S^\eta_{G_t^\eta, H_t^\eta}$ to be the map sending $(x_t^\eta,v_t^\eta)$ to $(x_{t+\eta}^\eta,v_{t+\eta}^\eta)$ as defined above. As shorthand, because the $\eta$ can be inferred, we write this as $S^{\eta}_{(G,H)_t}$. 

We define a coupling between the continuous and discrete dynamics, or between discrete dynamics with different step sizes, by having the processes share the same Brownian motion. We refer to this as synchronous coupling. When coupling the dynamics with step sizes $\eta$ and $\eta/2$, we have 
\begin{align*}
G_{t}^\eta &= \int_0^{\eta} e^{2s} \dB_{t+s}
= \int_0^{\eta/2} e^{2s} \dB_{t+s} + \int_{\eta/2}^{\eta} e^{2s} \dB_{t+s}\\
&= \int_0^{\eta/2} e^{2s} \dB_{t+s} + e^\eta \int_{0}^{\eta/2} e^{2s} \dB_{t+\fc{\eta}2+s}
= G_{t}^{\eta/2} + e^{\eta} G_{t+\eta/2}^{\eta/2}\\
H_{t}^\eta &= \int_0^{\eta} \dB_{t+s}
= \int_0^{\eta/2}  \dB_{t+s} + \int_{0}^{\eta/2} \dB_{t+\eta/2+s}
= H_{t}^{\eta/2} + H_{t+\eta/2}^{\eta/2}.
\end{align*}
This leads to the update in Algorithm~\ref{a:uld-couple}.

\begin{algorithm}[h!]
\caption{Coupled Underdamped Langevin Dynamics (ULD)}
\llabel{a:uld-couple}
\begin{algorithmic}[1]
\REQUIRE Initial point $x_0\in \R^d$, function $f:\R^d\to \R$ (with gradient access). 
\REQUIRE Time $T$ and step size $\eta$ OR bound on strong convexity $\mu$, condition number $\ka$, and desired accuracy $\ep$.
\ENSURE Coupled samples $(X^{\eta-}, X^{\eta/2+})$.
\IF{$\ep$ is given}
	\STATE Let $\eta = \fc{\ep}{208\ka}\sfc{\mu}d$.
	\STATE Let $T=\fc{\ka}2\log \pf{48(d/\mu)}{\ep}$.
\ENDIF
\STATE Let $t=0$.
\STATE Let $(x^\eta_0,v^\eta_0)=(x^{\fc{\eta}2}_0,v^{\fc{\eta}2}_0)=(x_0,0)$.
\WHILE{$t<T$}
			\STATE Draw $\coltwo{G_t^{\eta/2}}{H_t^{\eta/2}}, \coltwo{G_{t+\eta/2}^{\eta/2}}{H_{t+\eta/2}^{\eta/2}} \sim N\pa{\mathbf 0, \symatt{\rc 4(e^{2\eta}-1)}{\rc 2(e^{\eta}-1)}{\eta/2}\ot I_d}$.
			\STATE Let
			\begin{align*}
G_{t}^\eta &= G_{t}^{\eta/2} + e^{\eta} G_{t+\eta/2}^{\eta/2}\\
H_{t}^\eta &= H_{t}^{\eta/2} + H_{t+\eta/2}^{\eta/2}.
\end{align*}
			\STATE Let $(x^{\eta/2}_{t+\eta},v^{\eta/2}_{t+\eta})=S^{\eta/2}_{(G,H)_{t+\eta/2}}\circ S^{\eta/2}_{(G,H)_t}(x^{\eta/2}_{t},v^{\eta/2}_t)$.
			\STATE Let $(x^{\eta}_{t+\eta},v^{\eta}_{t+\eta})=S^{\eta}_{(G,H)_{t}}(x^\eta_{t},v^\eta_{t})$.
		\STATE Set $t\leftarrow t+\eta$.
\ENDWHILE
\STATE Output $(x^{\eta}_t, x^{\eta/2}_t)$.
\end{algorithmic}
\end{algorithm}

The main result on underdamped Langevin we will use is the following.
\begin{thm}[{Convergence of ULD,~\cite[Theorem 1]{cheng2017underdamped}}]\llabel{t:uld}
Suppose $f$ is twice continuously differentiable, $\mu$-strongly convex, and $L$-smooth, and let $\ka=\fc{L}{\mu}$. 
Let $\rh(dx)\propto e^{-f(x)}\dx$.

Let $\rh^\eta$ be the distribution of discretized underdamped Langevin with step size $\eta$ after time $T$, under the initial distribution $\de_{(x,v)=(x_0,0)}$. Let the initial distance to optimum $x^*=\amin f$ satisfy $\ve{x_0-x^*}\le D$. 
\begin{enumerate}
\item
Let $x^\eta$, $x^0$ be synchronously coupled trajectories from the discrete and continuous processes. Let $T\ge \fc{\ka}2\log \pf{24\sqrt{\fc d\mu+D^2}}{\ep}$, and $\rh^\eta$ be the distribution of $x_T^\eta$. Then for 
$\eta=O\prc{\ka}$ dividing into $T$,
\begin{align*}
\E [||x^\eta_T - x_T^0||^2] &\le {\cO}
\pa{
\fc d\mu \cdot \ka^2 \eta^2
}\\
W_2(\rh^\eta,\rh)^2 &\le 
\cO\pa{
\fc d\mu \cdot \ka^2 \eta^2
}\vee \ep^2
\end{align*}
\item
For step size $\eta \le \fc{\ep}{104\ka}\sfc{1}{d/\mu+D^2}$ and $T\ge \fc{\ka}2\log \pf{24\sqrt{\fc d\mu+D^2}}{\ep}$, we have $W_2(\rh^\eta,\rh)\le \ep$. The algorithm makes $\fc{T}{\eta}$ queries to $\nb f$.
\end{enumerate}
\end{thm}
\begin{proof}
The second part is~\cite[Theorem 1]{cheng2017underdamped}.\footnote{Note that they actually show the theorem with the $\sqrt{\fc d\mu + D^2}$ inside the log, although this is not reflected in their theorem statement.} Their proof 
essentially establishes the first part of the theorem: 
In their notation, $W_2(\rh^\eta,\rh)$ is  $W_2(p^{(n)}, p^*)$ where $n=\fc{T}{\eta}$. They show that $W_2(p^{(n)},p^*)\le T_1+T_2$, where $T_1\le \eph$ with the choice of $T$, and $T_2 \le 4\ka \eta \sfc{32\cdot 26 \pa{\fc d\mu + D^2}}{5}$.
This establishes the bound on $W_2(\rh^\eta,\rh)$.

For the bound on $\E [||x^\eta_T - x_T^0||^2]$, note that their bounds on Wasserstein distance 
come from synchronously coupling the continuous and discrete processes. In their notation, $q^{(n)}$ is the distribution of $(x^\eta,v^\eta)$ at the $n$th step, $p^{(n)}$ is the distribution of $x$ at the $n$th step, $\wt \Phi_\eta$ is one step of the discrete process, and $\Phi_\eta$ is the exact underdamped Langevin process for the same amount of time. The same induction in (9)--(10) of~\cite{cheng2017underdamped} shows that $W_2(\wt \Phi_\eta^{T/\eta} q^{(0)}, \Phi_\eta^{T/\eta} q^{(0)}) \le \rc{1-e^{-\eta/2\ka}}\eta^2 \sfc{8\cal E_K}5$ and $W_2(p^{(n)}, p^{\prime(n)})\le T_2:=\rc{1-e^{-\eta/2\ka}}\eta^2 \sfc{32\cal E_K}5$, where $p^{\prime (n)}$ is the distribution of the continuous process after $n$ steps. The bound on Wasserstein distance is attained by synchronous coupling of the two processes. Their bound $T_2 \le 4\ka \eta \sfc{32\cdot 26 \pa{\fc d\mu + D^2}}{5}$ then establishes the bound on $\E [||x^\eta_T - x_T^0||^2]$.
\end{proof}

The number of steps $\fc{T}{\eta}$ has a $\fc{\ka^2}{\ep}$ dependence on $\ka$ and $\ep$. We note that~\cite{DalalyanRiou-Durand} has a better dependence, $\ka^{\fc 32}\pa{\ka^{\rc 2} \wedge \fc{d^{\rc 2}}{\mu^{\rc 2} \ep^2}}$, and can be used to give better bounds in Theorem~\ref{t:muld-norm}. However, as ULD-RMM has faster running time (Theorem~\ref{t:rmm}), we will work with the simpler bound in~\cite{cheng2017underdamped}. The next Theorem gives the first part of Theorem~\ref{thm:expectation_main}.


\newcommand{\muldQ}[0]{\ka^2\sqrt d  \log \pa{\fc{L_g}{\ep_b}\cdot \sfc{d}{\mu}} 
 \pa{\fc{L_g^2}{\mu\ep_\si^2} + \fc{L_g}{\sqrt\mu \ep_b}}}
\begin{thm}[Rate of Multilevel ULD]\llabel{t:muld} Let $\rh(dx)\propto e^{-f(x)}\dx$, where $f:\R^d\to \R$ is $\mu$-strongly convex and $L$-smooth. Let $g:\R^d\to \R$ be $L_g$-Lipschitz. 
Suppose that $x_0$ satisfies $\ve{x_0-x^*}\le D=O\pa{\sfc{d}{\mu}}$.
Then 
Algorithm~\ref{a:ml} run using Algorithm~\ref{a:uld-couple} (ULD) started at $x_0$ with parameters 
\begin{align*}
\eta_0 &=\Te\prc{d^{\rc 2}\ka}&
T&\ge\fc{\ka}2 \log \pf{24L_g\sqrt{\fc d\mu +D^2}}{\ep_b}\\
\eta_k & = \Te\pa{\fc{\ep_b}{L_g \ka }\sfc{\mu}{d}}&
N_j &\ge \Om\pf{L_g^2 d^{\fc 34}\ka^{\fc 32}\eta_j^{\fc 32}}{\mu\ep_\si^2}
\end{align*}
outputs $\wh R$ such that 
$|\E \wh R - \E_{\rh} g|\le \ep_b$, and 
$\Var(\wh R)\le \ep_\si^2$.
This takes $\cO\pa{\muldQ}$ gradient evaluations.
Moreover, letting $\rh^{\eta}$ be the distribution of $x^{\eta}_T$, we have $W_2(\rh, \rh^{\eta_k})\le \fc{\ep_b}{L_g}$.

In particular, for $\ep\le \fc{L_g}{\sqrt \mu}$, taking $\ep_b = \ep_\si = \fc{\ep}{3}$, 
$\Pj\pa{|\wh R - \E_\rh g|>\ep}\le \rc 4$, and the algorithm uses
$\cO\pa{\fc{L_g^2\ka^2 \sqrt d}{\mu\ep^2}\log\pa{\fc{L_g}{\ep_b}\cdot \sfc{d}{\mu}} }$
gradient evaluations. 
\end{thm}
\begin{proof}
We check that the conditions of Lemma~\ref{l:ml2} hold with 
$F(\eta) = \fc{Cd\ka^2\eta^2}{\mu}$ (for some $C$), $c=\rc{\mu}$, and $T(\ep) = \fc{\ka}2 \log\pf{24\sqrt{\fc d\mu + D^2}}{\ep^2}$. Conditions 1, 2, and 4 follow from Theorem~\ref{t:uld}(1), and condition 3 follows since $f$ is $\mu$-strongly convex.

We choose $\eta_0$ so that $\fc{Cd\ka^2 \eta_0^2}{\mu} = F(\eta_0) = \rc{4\mu}$, leading to $\eta_0 = \Te\prc{d^{\rc 2} \ka}$. Note that we do have $\eta_0\le \eta_{\max} = \Te\prc{\ka}$. We choose $\eta_k$ so that $\fc{Cd\ka^2\eta_k^2}{\mu} \le \fc{\ep_b^2}{L_g^2}$, leading to $\eta_k = \cO\pa{\fc{\ep_b}{\ka L_g}\sfc{\mu}{d}}$. We choose $N_j$ so that 
\begin{align*}
N_j&\ge \fc{4L_g^2}{\ep_\si^2}\sfc{f(\eta_0)\eta_jf(\eta_j)}{\eta_0}
=\cO\pa{\fc{L_g^2}{\ep_\si^2}\sfc{\prc{\mu} \pf{d\ka^2\eta_j^3}{\mu}}{\prc{d^{\rc 2}\ka}}} =\cO \pf{L_g^2 d^{\fc 34}\ka^{\fc 32}\eta_j^{\fc 32}}{\mu\ep_\si^2}.
\end{align*}
We choose $T\ge T\pf{\ep_b}{L_g}$. 
Finally, the number of queries is 
\begin{align*}
Q&=\cO\pa{T\pa{\fc{cL_g^2}{\ep_\si^2 \eta_0}+\rc{\eta_k}}}
= \cO\pa{\ka \log\pf{L_g\sqrt{d/\mu}}{\ep_b} \pa{\fc{L_g^2/\mu}{\ep_\si^2 \prc{d^{\rc 2} \ka}}+\fc{\ka L_g}{\ep_b}\sfc{d}{\mu}} }\\
&=\cO\pa{\muldQ}.
\end{align*}

The last part follows since
\begin{align*}
\Pj\pa{|\wh R - \E_\rh g|>\ep}
&\le \Pj\pa{|\wh R - \E \wh R|>\frac{2\ep}{3}}\le \rc 4.\qedhere
\end{align*}
\end{proof}

\subsection{Multilevel ULD-RMM}
\llabel{s:ml-uld-rmm}
In the integral formulation of the dynamics~\eqref{e:vt+eta} and~\eqref{e:xt+eta}, the difference between the continuous and discrete dynamics is that in the continuous dynamics, we have the current gradient $\nb f(x_{t+s})$ instead of the gradient at the last time step $\nb f(x_t^\eta)$. The idea of the randomized midpoint method (RMM)~\cite{shen2019randomized} is to estimate the integrals by their value at $s=\al\eta$ for a uniformly random $\alpha\in [0,1]$, instead of at $s=0$. This reduces the bias caused by the one-step numerical quadrature with the price of increasing the standard deviation, which accumulates much slower than the bias in the numerical integration. This is in fact similar to our choice of $\veps_b \ll \veps_{\sigma}$ later for using the multilevel Monte Carlo method combined with annealing. The estimate of $x_{t+\al \eta}$, which we denote by $y_t^\eta$, is obtained using the discretization with step size $\al\eta$.
The update is given by 
\begin{align*}
y_t^\eta&=x_t^\eta + \rc 2 (1-e^{-2\al \eta})v_t^\eta - \rc 2 u \pa{\al \eta - \rc 2 \pa{1-e^{-2\al \eta}}}\nb f(x_t^\eta) + \rc{\sqrt L} W_{1,t}^\eta\\
x_{t+\eta}^\eta&= x_t^\eta + \rc 2 (1-e^{-2\eta})v_t^\eta - \fc\eta{2L} (1-e^{-2(1-\al)\eta}) \nb f(y_t^\eta) + \rc{\sqrt L} W_{2,t}^\eta\\ 
v_{t+\eta} &= v_t^\eta e^{-2\eta} - u\eta e^{-2(1-\al)\eta} \nb f(y_{t}^\eta) + \fc2{\sqrt L} W_{3,t}^\eta 
\end{align*}
where
\begin{align*}
W_{1,t}^\eta &= H_1 - e^{-2\al \eta}G_1&
W_{2,t}^\eta &=(H_1+H_2) - e^{-2\eta} (G_1+G_2)&
W_{3,t}^\eta &=e^{-2\eta}(G_1+G_2)
\end{align*}
and 
\begin{align*}
G_{1,t}^\eta &=\int_0^{\al \eta} e^{2s}\dB_{t+s}&
H_{1,t}^\eta &=\int_0^{\al \eta} \dB_{t+s}\\
G_{2,t}^\eta &=\int_{\al \eta}^\eta e^{2s}\dB_{t+s}&
H_{2,t}^\eta &=\int_{\al \eta}^\eta \dB_{t+s}.
\end{align*}
Writing $G_{i}=G_{i,t}^\eta$ and $H_i = H_{i,t}^\eta$, define $R^\eta_{\al, G_1,H_1,G_2,H_2}$ to be the map sending $(x_t^\eta,v_t^\eta)$ to $(x_{t+\eta}^\eta,v_{t+\eta}^\eta)$ as defined above. 

To define the coupled dynamics, note that once we have selected $\al_1$ and $\al_2$ for step size $\eta/2$ for time steps $t$ and $t+\eta/2$ respectively, one way to define a uniformly random $\al\in [0,1]$ is to take $\fc{\al_1}2$ or $\fc{1+\al_2}2$ each with probability $\rc 2$. This coupling has the advantage that we have $t+\al \eta = t+\al_1 \fc{\eta}2$ or $t+\fc{\eta}2 + \al_2 \fc{\eta}2$, so we can calculate the $W_{i,t}^\eta$ in terms of quantities already computed. (This coupling is out of convenience only; it is the fact that we use the same Brownian motion that reduces the variance, not the fact that $\al$ is coupled to $\al_1$ and $\al_2$.) A straightforward calculation gives the updates for coupled ULD-RMM, Algorithm~\ref{a:uld-rmm-couple}. For ease of notation we drop the subscripts and superscripts for $G$ and $H$.

\begin{algorithm}[h!]
\caption{Coupled Underdamped Langevin Dynamics with Randomized Midpoint Method (ULD-RMM)}
\llabel{a:uld-rmm-couple}
\begin{algorithmic}[1]
\REQUIRE Initial point $x_0\in \R^d$, function $f:\R^d\to \R$ (with gradient access). 
\REQUIRE Time $T$ and step size $\eta$ OR bound on strong convexity $\mu$, condition number $\ka$, and desired accuracy $\ep$.
\ENSURE Coupled samples $(X^{\eta-}, X^{\eta/2+})$.
\IF{$\ep$ is given}
	\STATE Let $\eta = c\rmmeta$, where $c$ is a small enough universal constant.
	\STATE Let $T=2\ka\log\pf{20}{\ep^2}$.
\ENDIF
\STATE Let $t=0$.
\STATE Let $(x^\eta_0,v^\eta_0)=(x^{\fc{\eta}2}_0,v^{\fc{\eta}2}_0)=(x_0,0)$.
\WHILE{$t<T$}
	\STATE Let $\al_1,\al_2$ be random numbers in $[0,1]$: $\al_1,\al_2\sim U([0,1])$.
	\STATE 
	Draw $\coltwo{G_1^{(i)}}{H_1^{(i)}} \sim N\pa{\mathbf 0,\symatt{\rc 4(e^{2\al \eta}-1)}{\rc 2(e^{\al \eta}-1)}{\al \eta/2}\ot I_d}$ for $i=1,2$.
	\STATE Draw 
	$\coltwo{G_2^{(i)}}{H_2^{(i)}} \sim N\pa{\mathbf 0, \symatt{\rc 4 (e^{2\eta}-e^{2\al \eta})}{\rc 2 (e^{\eta}-e^{\al \eta})}{(1-\al)\eta/2}\ot I_d}$ for $i=1,2$.
	\IF{random coin flip $=$ heads}
		\STATE Set $\al=\fc{\al_1}2$
		\begin{align*}
			G_1 &= G_1^{(1)}&
			H_1 &= H_1^{(1)}\\
			G_2 &= G_2^{(1)} + e^{\eta} (G_1^{(2)}+G_2^{(2)})&
			H_2 &= H_2^{(1)} + H_1^{(2)} + H_2^{(2)}
		\end{align*}
	\ELSE
		\STATE Set $\al = \fc{1+\al_2}2$ and
		\begin{align*}
			G_1 &= G_1^{(1)}+G_2^{(1)}+e^{\eta}G_1^{(2)}&
			H_1 &= H_1^{(1)} + H_2^{(1)}+H_1^{(2)}\\
			G_2 &= e^{\eta}G_2^{(2)}&
			H_2 &= H_2^{(2)}.
		\end{align*}
	\ENDIF
	\STATE Let $(x^{\eta/2}_{t+\eta},v^{\eta/2}_{t+\eta})=R^{\eta/2}_{\al_2,G_1^{(2)},H_1^{(2)},G_2^{(2)},H_2^{(2)}}
	\circ R^{\eta/2}_{\al_1,G_1^{(1)},H_1^{(1)},G_2^{(1)},H_2^{(1)}}
	(x^{\eta/2}_t,v^{\eta/2}_t)$.
	\STATE Let $(x^{\eta}_{t+\eta},v^{\eta}_{t+\eta})=R^{\eta}_{\al, G_1,H_1,G_2,H_2}(x^\eta_{t},v^\eta_{t})$.
	\STATE Set $t\leftarrow t+\eta$.
\ENDWHILE
\STATE Output $(x^{\eta}_t, x^{\eta/2}_t)$.
\end{algorithmic}
\end{algorithm}

This gives the following improved rates.

\begin{thm}[{Convergence of ULD-RMM, \cite[Theorem 3]{shen2019randomized}}]\llabel{t:rmm}
Suppose $f$ is twice continuously differentiable, $\mu$-strongly convex, and $L$-smooth, and let $\ka=\fc{L}{\mu}$. 
Let $\rh(\rdx)\propto e^{-f(x)}\dx$.

Let 
$\rh^\eta$ be the distribution of the Randomized Midpoint Method for ULD with step size $\eta$ after time $T$, under the initial distribution $\de_{(x,v)=(x^*,0)}$. 
\begin{enumerate}
\item
Let $x^\eta$, $x^0$ be synchronously coupled points from the discrete and continuous processes. 
Let $T\ge 2\ka \log\pf{20(d/\mu)}{\ep^2}$, and  $\rh^\eta$ be the distribution of $x^\eta_T$. 
For $\eta$ smaller than some constant,
\begin{align*}
\E [||x^\eta_T - x_T||^2] &\le {\cO}
\pa{
\pa{\fc{d\ka \eta^6}{\mu} + \fc{d\eta^3}{\mu}} \log\pf{\sqrt{d/\mu}}\ep
}\\
W_2(\rh^\eta,\rh)^2&\le 
{\cO} \pa{
\pa{\fc{d\ka \eta^6}{\mu} + \fc{d\eta^3}{\mu}} \log\pf{\sqrt{d/\mu}}\ep
}\vee \ep^2
\end{align*}
\item
Let $c>0$ be a small enough constant.
For step size $\eta\le c
\min\bc{
\fc{\ep^{\rc3}}{\ka^{\rc 6}\log^{\rc 6}\pf{\sqrt{d/\mu}}{\ep}}\pf{\mu}{d}^{\rc 6},
\fc{\ep^{\fc 23}}{\log^{\rc 3}\pf{\sqrt{d/\mu}}{\ep}} \pf{\mu}{d}^{\rc 3} 
}$,
and time $T\ge 2\ka \log\pf{20(d/\mu)}{\ep^2}$, 
$W_2(\rh^\eta,\rh)\le \ep$. 
The algorithm makes $\fc{2T}{\eta}$ queries to $\nb f$.
\end{enumerate}
\end{thm}
\begin{proof}
The second part is exactly~\cite[Theorem 3]{shen2019randomized}. 

In their notation, $(x_n,v_n)$ is the $n$th iterate of their algorithm, and $(y_n,w_n)$ is the $n$th step of the exact ULD, started from a random point from the stationary distribution. 
Examining their proof, they show that 
\begin{align*}
&\quad \E[||x_N-y_N||^2 + ||(x_N+v_N)-(y_N+w_N)||^2]\\ 
&\quad \le 
e^{-\fc{N\eta}{2\ka}} \ub{\E[||x_0-y_0||^2 + ||(x_0+v_0)-(y_0+w_0)||^2] }{\le \fc{5d}{\mu}}+ 
\Or\pa{\pa{\fc{\ka d \eta^6}{\mu} + \fc{d\eta^3}{\mu}}\log\pf{d/\mu}{\ep^2}} \\
&\quad \quad + \Or(\ka\eta^7 + \eta^3)\E[||x_N-y_N||^2 + ||(x_N+v_N)-(y_N+w_N)||^2]
\end{align*}
For $\eta=O(\ka^{-\rc 7})$, we have that the last term is $\le \rc 2\E[||x_N-y_N||^2 + ||(x_N+v_N)-(y_N+w_N)||^2]$, so 
\begin{align*}
\E[||x_N-y_N||^2 + ||(x_N+v_N)-(y_N+w_N)||^2] &\le 
e^{-\fc{N\eta}{2\ka}}  \fc{5d}{\mu} +
\Or\pa{\pa{\fc{\ka d \eta^6}{\mu} + \fc{d\eta^3}{\mu}}\log\pf{d/\mu}{\ep^2}}.
\end{align*}
By choice of $T$ (or $N$), this term is $\le \eph$. This establishes the bound on $W_2(\rh^\eta,\rh)$. 

Finally, note that we can replace $(y_n,w_n)$ by the exact ULD started with the same initial condition. Then the same derivation holds, except that the first term is 0. This shows the bound on $\E[||x^\eta_T-x_T||^2]$.
\end{proof}

Combining Theorem~\ref{t:rmm} with Lemma~\ref{l:ml2}, we can prove the second part of Theorem~\ref{thm:expectation_main}.


\begin{thm}[Rate of Multilevel ULD-RMM]\llabel{t:muld-rmm} Let $\rh(dx)\propto e^{-f(x)}\dx$, where $f:\R^d\to \R$ is $\mu$-strongly convex and $L$-smooth. Let $g:\R^d\to \R$ be $L_g$-Lipschitz. 
Then 
Algorithm~\ref{a:ml} run using Algorithm~\ref{a:uld-rmm-couple} (ULD) started at $x^*$ with parameters 
\begin{align*}
\eta_0 &= \Te\pa{\rc{d^{\rc 6}\ka^{\rc 6}\log\pa{\fc{L_g^2}{\ep_b^2}\cdot \fc{d}{\mu}}^{\rc 6}}\wedge \rc{d^{\rc 3}\log\pa{\fc{L_g^2}{\ep_b^2}\cdot \fc{d}{\mu}}^{\rc 3}}}&
T&=\cO\pa{\ka \log\pa{\fc{L_g^2}{\ep_b^2}\cdot \fc{d}{\mu}}}\\
\eta_k &= \Te\pa{\fc{\ep_b^{\rc 3} \mu^{\rc 6}}{d^{\rc 6} \ka^{\rc 6} L_g^{\rc 3}\log\pa{\fc{L_g^2}{\ep_b^2}\cdot \fc{d}{\mu}}^{\rc 6}}
\wedge \fc{\ep_b^{\fc 23} \mu^{\rc 3}}{d^{\rc 3} L_g^{\fc 23}\log\pa{\fc{L_g^2}{\ep_b^2}\cdot \fc{d}{\mu}}^{\rc 3}}}&
N_j &\ge \Om \pa{\fc{L_g^2}{\ep_\si^2}\sfc{f(\eta_0)\eta_jf(\eta_j)}{\eta_0}}
\end{align*}
outputs $\wh R$ such that 
$|\E \wh R - \E_{\rh} g|\le \ep_b$, and 
$\Var(\wh R)\le \ep_\si^2$.
This takes 
\begin{multline*}
\cO\Biggl(
\pa{\ka^{\fc 76}d^{\rc 6}\log\pa{\fc{L_g^2}{\ep_b^2}\cdot \fc{d}{\mu}}^{\fc 76} + \ka d^{\rc 3}\log\pa{\fc{L_g^2}{\ep_b^2}\cdot \fc{d}{\mu}}^{\fc 43}}
\fc{L_g^2}{\mu\ep_\si^2}  \\
+
\fc{\ka^{\fc 76}d^{\rc 6}L_g^{\fc 13}}{\ep_b^{\fc 13}}\log\pa{\fc{L_g^2}{\ep_b^2}\cdot \fc{d}{\mu}}^{\fc 76} + \fc{\ka d^{\rc 3}L_g^{\fc 23}}{\ep_b^{\fc 23}}\log\pa{\fc{L_g^2}{\ep_b^2}\cdot \fc{d}{\mu}}^{\fc 43}
\Biggr)
\end{multline*}
gradient evaluations.
Moreover, letting $\rh^{\eta}$ be the distribution of $x^{\eta}_T$, we have $W_2(\rh, \rh^{\eta_k})\le \fc{\ep_b}{L_g}$.

In particular, for $\ep\le \fc{L_g}{\sqrt \mu}$, 
taking $\ep_b = \ep_\si = \fc{\ep}{2}$, 
$\Pj\pa{|\wh R - \E_\rh g|>\ep}\le \rc 4$, and the algorithm uses
$\cO
\pf{L_g^2\ka^{\fc 76}d^{\rc 6}\log\pa{\fc{L_g^2}{\ep_b^2}\cdot \fc{d}{\mu}}^{\fc 76} + \ka d^{\rc 3}\log\pa{\fc{L_g^2}{\ep_b^2}\cdot \fc{d}{\mu}}^{\fc 43}}{\mu\ep^2}
$
gradient evaluations. 
\end{thm}
\begin{proof}
We check that the conditions of Lemma~\ref{l:ml2} hold with 
$F(\eta) = C\pa{\fc{d\ka \eta^6}{\mu} + \fc{d\eta^3}{\mu}} \log\pa{\fc{L_g^2}{\ep_b^2}\cdot \fc{d}{\mu}}$ (for some $C$), $c=\rc{\mu}$, and $T(\ep) = 2\ka \log\pf{20(d/\mu)}{\ep^2}$. Conditions 1, 2, and 4 follow from Theorem~\ref{t:rmm}(1), and condition 3 follows since $f$ is $\mu$-strongly convex.

We choose $\eta_0$ so that 
\begin{align*}
C\pa{\fc{d\ka \eta_0^6}{\mu} + \fc{d\eta_0^3}{\mu}} \log\pa{\fc{L_g^2}{\ep_b^2}\cdot \fc{d}{\mu}} &= F(\eta_0) = \rc{4\mu}
\end{align*}
and $\eta_k$ so that 
\begin{align*}
\fc{\ep_b^2}{4L_g^2} \ge C\pa{\fc{d\ka \eta_k^6}{\mu} + \fc{d\eta_k^3}{\mu}} \log\pa{\fc{L_g^2}{\ep_b^2}\cdot \fc{d}{\mu}}&= F(\eta_k) \\
\Leftarrow
\fc{d\ka \eta_k^6}{\mu} \log\pa{\fc{L_g^2}{\ep_b^2}\cdot \fc{d}{\mu}} &\le \cO\pf{\ep_b^2}{L_g^2}\\
\text{and }\fc{d\eta_k^3}{\mu} \log\pa{\fc{L_g^2}{\ep_b^2}\cdot \fc{d}{\mu}} &\le \cO\pf{\ep_b^2}{L_g^2},
\end{align*}
leading to the given bounds on $\eta_0$ and $\eta_k$. We choose $T$ so that $T\ge T\pf{\ep_b}{L_g}$. We do have that $\eta_0\le \eta_{\max}=\Te(1)$.
Substituting the bounds on $\eta_0$ and $\eta_k$ into $Q=\cO\pa{T\pa{\fc{L_g^2/\mu}{\ep_\si^2 \eta_0}+\rc{\eta_k}}}$ gives the bound on the number of queries.
\end{proof}

\subsection{Truncation error and bias}
\llabel{s:bias}
There is a technical point that the ratio $g$ is not Lipschitz, as it grows exponentially for large $\ve{x}$; however, because large $x$'s are very unlikely under $\rh$, the expected value of $g$ changes very little if we replace it by a ``clamped" version of $g$ (Lemma~\ref{l:g-tail}).
\begin{lem}[Truncation error]\llabel{l:g-tail}
Suppose that $f:\R^d\to \R$ is a $\mu$-strongly convex function, $\rh$ is a probability measure on $\R^d$ with $\rh(\dx) = \rc{Z}e^{-\pa{\rc 2 \fc{\ve{x}^2}{\si^2}+f(x)}}\dx$, and $g(x) = \exp\pa{\fc{\ve{x}^2}{2\si^2(1+\al^{-1})}}$ for $\al \in (0,\iy]$. 
Let $\rh'$ be the probability distribution with $\dd{\rh'}{\rh}\propto g(x)$, and $\ol r = \E_{x\sim \rh'}\ve{x}$. 
For any $r\ge \ol r$,
\begin{align*}
\fc{\int_{\ve{x}\ge r} g(x) \rh(dx)}{\E_{x\sim \rh} g(x)}
&\le \exp\pa{-\rc 2 \pa{\rc{\si^2(1+\al)} + \mu}(r-\ol r)^2}.
\end{align*}

\end{lem}
Note that we allow $\al=\iy$, in which case $\al^{-1}=0$, $g(x) = \exp\pf{\ve{x}^2}{2\si^2}$, and the bound is $\exp\pa{-\fc{\mu(r-\ol r)^2}{2}}$. 
\begin{proof}
Note that $\rc{\si^2} - \rc{\si^2(1+\al^{-1})} = \fc{1}{\si^2(\al+1)}$, so $\rh'$ is $\pa{\rc{\si^2(\al+1)}+\mu}$-strongly convex. Then for any $r$,
\begin{align*}
\fc{\int_{\ve{x}\ge r} g(x) \rh(dx)}{\int_{\R^d} g(x) \rh(dx)}
&= \Pj_{x\sim \rh'} (\ve{x}\ge r). 
\end{align*}•
By Theorem~\ref{t:be} and~\ref{t:lsi-conc} on the 1-Lipschitz function $\ve{x}$, we have the concentration bound
\begin{align*}
\Pj_{x\sim \rh'}(\ve{x} \ge r) &\le 
\exp\pa{-\rc 2 \pa{\rc{\si^2(1+\al)} + \mu}(r-\ol r)^2}.
\end{align*}\qedhere
\end{proof}
\begin{lem}[Bias calculation]\llabel{l:bias-var}\llabel{l:bias}
Suppose that $f:\R^d\to \R^d$ is a convex function, $\rh$ is a probability measure on $\R^d$ with $\rh(\rdx) = \rc{Z}e^{-\pa{\rc 2 \fc{\ve{x}^2}{\si^2}+f(x)}}\dx$, and $g(x) = g_r(\ve{x})$, where $g_r(r):=e^{\fc{r^2}{2\si^2(1+\al^{-1})}}$. Let $\ol r$ be as in Lemma~\ref{l:g-tail}. 
Suppose one of the following hold.
\begin{enumerate}
\item
Suppose $r^+=\ol r + c\si$ for $c\ge 0$. Let $L_h = \fc{\ol r + c\si}{\si^2(1+\al^{-1})} \exp\pa{4\al^2 d+\fc{\ol rc}{\si(1+\al^{-1})} + \fc{c^2}{2(1+\al^{-1})}}$ and $\ep_1 = \exp\pa{-\fc{c^2}{2(1+\al)}}$.
\item
Suppose $f$ is $\mu$-strongly convex, $\si^2 \ge \fc{2}{\mu}$, $\al=\iy$, and $r^+=\ol r+ \fc{c}{\sqrt{\mu}}$ for $c\ge 0$. 
Let $L_h=\fc{\ol r + \fc{c}{\sqrt\mu}}{\si^2}\exp\pa{\fc{4d}{\mu \si^4} + \fc{\ol rc}{\si^2\sqrt{\mu}}+\fc{c^2}{2\si^2\mu}}$ and $\ep_1=\exp\pa{-\fc{c^2}2}$. 
\end{enumerate}
Define 
$h(y) = g(y) \wedge g_r(r^+)$.
Then $h$ is $L_h\cdot \E_\rh g$-Lipschitz and 
\begin{align*}
\fc{\ab{\E_{x\sim \wt \rh}h(x)
- \E_{x\sim \rh}g(x)}}{\E_{x\sim \rh} g(x)} &\le 
\ep_1 + L_h W_2(\rh,\wt \rh).
\end{align*}
\end{lem}
\begin{proof}
By the triangle inequality,
\begin{align*}
\fc{\ab{\E_{x\sim \wt\rh}h(x)
- \E_{x\sim \rh}g(x)}}{\E_{x\sim \rh} g(x)}
&\le 
\fc{\ab{\E_{x\sim \rh}[h(x)-g(x)]}}{\E_{x\sim \rh} g(x)}
+\fc{\ab{\E_{x\sim \wt \rh}h(x) - \E_{x\sim \rh} h(x)}}{\E_{x\sim \rh} g(x)}.
\end{align*}

In either case, the first expression is bounded by Lemma~\ref{l:g-tail}:
\begin{align*}
\fc{\ab{\E_{x\sim \rh}[h(x)-g(x)]}}{\E_{x\sim \rh} g(x)} &\le \ep_1
\end{align*}
To bound the second expression, we note that $h$ is Lipschitz with constant $\max_{\ve{x}\le r^+}\ve{\nb g(x)}$. Thus by Kantorovich-Rubinstein duality, 
\begin{align*}
\fc{\ab{\E_{x\sim \wt \rh}h(x) - \E_{x\sim \rh} h(x)}}{\E_{x\sim \rh}g(x)}
 &\le \fc{W_1(\rh, \wt \rh) \max_{\ve{x}\le r^+}\ve{\nb g(x)}}{\E_{x\sim \rh}g(x)} 
\le W_2(\rh, \wt \rh)\fc{\max_{\ve{x}\le r^+}\ve{\nb g(x)}}{\E_{x\sim \rh}g(x)}
\end{align*}
It remains to show the bound $\fc{\max_{\ve{x}\le r^+}\ve{\nb g(x)}}{\E_{x\sim \rh}g(x)}\le L_h$. We consider the two cases separately.

\paragraph{Case 1.} 
First, we compare the numerator to $g_r(\ol r)$. Let $\rh'$ be the probability density with $\dd{\rh'}{\rh} = g(x)$. Then $\E_{x\sim \rh'} g(x) = \fc{\E_{\rh}(g^2)}{\E_\rh g}$, so
\begin{align*}
\E_{\rh} g &= 
\E_{\rh} g \fc{E_\rh g}{\E_{\rh}(g^2)} \E_{\rh'}g=
\fc{\pa{\E_\rh g}^2}{\E_{\rh} (g^2)} \E_{\rh'} g\ge \exp(-4\al^2d)\E_{\rh'}g
\end{align*}
using~\eqref{e:vari} and Lemma~\ref{lem:varbound}. 
Now, by definition of $\ol r$ in Lemma~\ref{l:g-tail} because $g_r$ is convex, $\E_{x\sim \rh'}g(x) = \E_{x\sim \rh'} g_r(\ve{x})
\ge g_r(\E_{x\sim \rh'}\ve{x}) = g_r(\ol r)$. Hence
\begin{align*}
\fc{\max_{\ve{x}\le r^+} \ve{\nb g(x)}}{\E_{x\sim \rh}g(x)}
& =
\fc{\ddd x g_r(x)|_{x=r^+}}{\E_{x\sim \rh}g(x)}
\le 
\fc{\fc{r^+}{\si^2(1+\al^{-1})}g_r(r^+)}{\exp(-4\al^2d)g_r(\ol r)}\\
&=\fc{r^+}{\si^2 (1+\al^{-1})} \exp\pa{4\al^2 d+\fc{r^{+2}-\ol r^2}{2\si^2 (1+\al^{-1})}}\\
&=\fc{\ol r + c\si}{\si^2(1+\al^{-1})} \exp\pa{4\al^2 d+\fc{\ol rc}{\si(1+\al^{-1})} + \fc{c^2}{2(1+\al^{-1})}}=L_h.
\end{align*}

\paragraph{Case 2.} Similar to the first case,
\begin{align*}
\E_{\rh} g &= \fc{\pa{\E_\rh g}^2}{\E_{\rh} g^2} \E_{\rh'} g\ge \exp\pa{-\fc{4d}{\mu \si_M^2}}\E_{\rh'}g
\ge \exp\pa{-\fc{4d}{\mu \si_M^2}} g_r(\ol r)
\end{align*}
using Lemma~\ref{lem:varboundM}, noting that the condition on $\si^2$ is satisfied. 
We now have
\begin{align*}
\fc{\max_{\ve{x}\le r^+} \ve{\nb g(x)}}{\E_{x\sim \rh}g(x)}
& =
\fc{\ddd x g_r(x)|_{x=r^+}}{\E_{x\sim \rh}g(x)}
\le 
\fc{\fc{r^+}{\si^2}g_r(r^+)}{\exp\pa{-\fc{4d}{\mu \si^4}}g_r(\ol r)}\\
&=\fc{r^+}{\si^2} \exp\pa{\fc{4d}{\mu \si^4}+\fc{r^{+2}-\ol r^2}{2\si^2}}\\
&=\fc{\ol r + c/\sqrt{\mu}}{\si^2} \exp\pa{\fc{4d}{\mu \si^4}+ \fc{\ol rc}{\si^2\sqrt{\mu}}+\fc{c^2}{2\si^2\mu}}=L_h. \qedhere
\end{align*}
\end{proof}

\newcommand{\simax}[0]{4\pf{\sqrt d\vee \sqrt{\log\prc{\ep}}}{\mu}\pa{1\vee \rc{\sqrt\mu}}}
\begin{cor}\llabel{c:bias}
Keep the setup of Lemma~\ref{l:bias}. Then 
\begin{enumerate}
\item 
If
$\al \le \fc{\log2}{2\sqrt d\log \prc{\ep}}$ and 
$r^+\in \ol r+ \si \sqrt{(1+\al)\log \prc{\ep}} [\sqrt 2,2],$ then 
in Lemma~\ref{l:bias}(1), $L_{h}\le \fc{112 e}{\si}$.
\item 
If
$\si^2\ge \simax$, $r^+\in \ol r +\rc{\sqrt \mu} \sqrt{\log\prc{\ep}}[\sqrt 2,2]$, 
and $\ep\le \rc4$, then in Lemma~\ref{l:bias}(2), $L_{h}\le 2e^2\sqrt \mu$.
\end{enumerate}
In either case, $\fc{|\E_{\tilde\rh}h - \E_{\rh}g|}{|\E_{\rh}g|}\le \ep_1 + L_{h}W_2(\rh, \wt\rh)$.
\end{cor}
\begin{proof}
To show (1), 
write $r^+=\ol r+ c\si$. Then $c\ge \sqrt{2(1+\al) \log \prc{\ep_1}}$, so we have $e^{-\fc{c^2}{2(1+\al)}}\le \ep_1$.
By Lemma~\ref{l:mode-mean},
\begin{align*}
\ol r &\le \si\sqrt{1+\al} (\sqrt d + 2\sqrt{2\log 2})\le 5\si\sqrt d.
\end{align*}
Noting that $c\le 2\sqrt{(1+\al) \log \prc{\ep}}$, 
\begin{align*}
\fc{\ol r+c\si}{\si^2(1+\al^{-1})}
&\le \fc{5\sqrt d}{\si(1+\al^{-1})} + \fc{2\al \sqrt{\log \prc{\ep}}}{\si}\\
\fc{\ol r c}{\si(1+\al^{-1})} &\le 10\al \sqrt{d\log\prc{\ep}}\\
\fc{c^2}{2(1+\al^{-1})} &\le 2\al\log\prc{\ep}.
\end{align*}•
Substituting into the definition of $L_{h}$ in Lemma~\ref{l:bias}(1), 
\begin{align*}
L_{h} &=
\fc{\ol r + c\si}{\si^2(1+\al^{-1})} \exp\pa{4\al^2 d+\fc{\ol rc}{\si(1+\al^{-1})} + \fc{c^2}{2(1+\al^{-1})}}\\
&\stackrel{\mathrm{(i)}}{\le} \pa{\fc{5\sqrt d}{\si(1+\al^{-1})} + \fc{2\al\sqrt{\log \prc{\ep_1}}}{\si}}
\exp\pa{4\al^2 d+10\al \sqrt{d\log\prc{\ep}}+2\al \log \prc{\ep}}\\
&\stackrel{\mathrm{(ii)}}{\le}
\rc{\si}\pa{\fc 52+2\al \sqrt{\log \prc{\ep}}} \exp\pa{1+10\al \sqrt{d\log \prc{\ep}}}\prc{\ep}^{2\al}\\
&\stackrel{\mathrm{(iii)}}{\le} 
\fc{7}{2\si} \exp(1+5\log 2) 2=\fc{112e}{\si},
\end{align*}
where (i) follows from substitution, (ii) follows from $\al\le \rc{2\sqrt d}$, and (iii) follows from $\al \le \fc{\log 2}{2\sqrt d\log \prc{\ep}}$.

To show (2), write $r^+=\ol r+\fc{c}{\sqrt \mu}$. Then $c\ge \sqrt{2\log \prc{\ep}}$, so we have $\exp\pa{-\fc{\mu(r^+-\ol r)^2}2}=
\exp\pa{-\fc{c^2}{2}}\le \ep$.
By Lemma~\ref{l:mode-mean},
\begin{align*}
\ol r &\le \rc{\sqrt{\mu}} (\sqrt d + 2\sqrt{2\log 2})\le 5\sfc{d}{\mu}
\end{align*}
Noting that $c\le 2\sqrt{\log\prc{\ep}}$, 
\begin{align*}
\fc{\ol r+\fc{c}{\sqrt{\mu}}}{\si^2}
&\le \fc{5\sqrt d + 2\sqrt{\log \prc{\ep}}}{\si^2\sqrt \mu}
\stackrel{\mathrm{(i)}}{\le} \fc 74
\\
\fc{4d}{\mu\si^4} &\stackrel{\mathrm{(ii)}}{\le}
\rc 4
\\
\fc{\ol r c}{\si^2\sqrt{\mu}} &\le
\fc{5\sqrt{d\log\prc{\ep}}}{\si^2\mu}
\stackrel{\mathrm{(iii)}}{\le} \fc 54\\
\fc{c^2}{2\si^2\mu} &\le  \fc{2\log\prc{\ep}}{\si^2\mu}
\stackrel{\mathrm{(iv)}}{\le} \fc 24.
\end{align*}
where in (i) we use $\si^2 \ge 4\fc{\sqrt d}{\sqrt \mu}$ and $\si^2 \ge 4 \fc{\sqrt{\log\prc{\ep}}}{\sqrt \mu}$, in (ii) 
we use $\si^2 \ge 4 \fc{\sqrt d}{\sqrt\mu}$, in 
 (iii) we use $\si^2\ge 4\cdot \fc{\sqrt d}{\sqrt \mu}\cdot \fc{\sqrt{\log\prc{\ep}}}{\sqrt \mu}$, and in (iv) we use $\si^2\ge 4\cdot \fc{\sqrt{\log\prc{\ep}}}{\sqrt \mu}\cdot \fc{\sqrt{\log\prc{\ep}}}{\sqrt \mu}$.
Substituting into the definition of $L_{h}$ in Lemma~\ref{l:bias}(2), \begin{align*}
L_{h} &=
\fc{\ol r + \fc{c}{\sqrt{\mu}}}{\si^2} \exp\pa{\fc{4d}{\mu\si^4}+\fc{\ol rc}{\si^2\sqrt{\mu}} + \fc{c^2}{2\si^2\sqrt{\mu}}}\\
&\le  \fc 74 \exp\pa{\rc4 + \fc 54 + \fc 24}\le 2e^2.
\end{align*}

In either case, by Lemma~\ref{l:bias}, 
$\fc{|\E_{\wt\rh}h - \E_{\rh}g|}{|\E_{\rh}g|}\le \ep_1 + \ep_2$.
\end{proof}

\subsection{Estimating the normalizing constant} 

\llabel{ss:ml-norm}
\llabel{s:ml-norm-uld}

Before stating the main algorithm, let us first understand how errors in individual annealing steps can be composed to give the final error for estimating the normalizing constant.

\begin{lem}\llabel{l:skeleton}\llabel{l:combine}
Suppose the following hold.
\begin{enumerate}
\item (Estimate of partition function at highest temperature)
$\fc{\wh Z_1}{Z_1}\in [e^{-\ep_1},e^{\ep_1}]$.
\item (Bias of ratio) For $1\le i\le M$, letting $\wt R_i = \E \wh R_i$, 
$|\wt R_i - R_i|\le \fc{\ep_2 R_i}{2M}$.
\item (Variance of ratio) For $1\le i\le M$, $\wh R_i$ is independent with $\Var(\wh R_i) \le \fc{\ep_3^2\wt R_i^2}{40 M}$.
\end{enumerate}
Then  $\Pj\pa{\fc{\wh Z_1 \prodo iM \wh R_i}{Z_1\prodo iM R_i }\nin [e^{-(\ep_1+\ep_2+\ep_3)},e^{\ep_1+\ep_2+\ep_3}] }\le \fc18$.
\end{lem}
\begin{proof}
From (2) we get that $\fc{\wt R_i}{R_i} \in [1-\fc{\ep_2}{2M}, 1+\fc{\ep_2}{2M}]\subeq [e^{-\fc{\ep_2}M},e^{\fc{\ep_2}M}]$.

From (3) and  Lemma~\ref{l:prod}, $\Pj\pa{\prodo iM \fc{\wh R_i}{\wt R_i}\nin [e^{-\ep_3},e^{\ep_3}]}\le \fc{5\ep_3^2 M}{40 \ep_3^2 M} = \fc 1{8}$.

Factoring $\fc{\wh Z_1 \prodo iM \wh R_i}{Z_1\prodo iM R_i }
 = \fc{\wh Z_1}{Z_1} \cdot \prodo iM \fc{\wt R_i}{R_i}
 \cdot \prodo iM \fc{\wh R_i}{\wt R_i}$, the result now follows.
\end{proof}

We are now ready to introduce the main algorithm for estimating the normalizing constant. Algorithm~\ref{a:ml-norm} first estimates the thresholds $r_i^+$ to cut off $g_i$ in Lemma~\ref{l:bias} so that the resulting estimate has bias $\cO(\ep)$. Then it calls the Multilevel Monte Carlo algorithm at each temperature with the truncated functions $h_i$. We can choose which Monte Carlo algorithm to use; we will consider both the ULD and ULD-RMM algorithms. Note that an alternative to estimating $\ol r_i^+ = \E_{X\sim \rh_{i+1}} \ve{X}$ separately is to use the samples obtained from the multilevel procedure; we only estimate it separately to make the proof simpler.

\begin{algorithm}[h!]
\caption{Multilevel Monte Carlo for normalizing constant estimation}
\llabel{a:ml-norm}
\begin{algorithmic}[1]
\REQUIRE Initial point $x_0$, function $f:\R^d\to \R$, initial temperature $\si_1$, final temperature $\si_{\max}$, multiplier $\al$, desired accuracy $\ep$.
\REQUIRE Algorithm $\cal A(x_0,f)$ which: (1) given $(\eta,T)$, returns coupled samples $(X^\eta,X^{\eta/2})$, (2) given $\ep$, returns samples $\wt X\sim \wt \rh$ such that $W_2(\rh,\wt\rh)\le \ep$, where $\rh(\rdx) \propto e^{-f(x)}\dx$.
\REQUIRE Algorithm $\cal B\pa{L,\mu,L_h,\ep_b,\ep_\si}$ to set parameters $(T,\eta_0,k,N_0,\ldots, N_k)$ for the multilevel Monte Carlo.
\ENSURE Estimate of $Z=\int_{\R^d} e^{-f(x)}\dx$.
\STATE Let $\wh Z = \wh Z_1 = \pa{2\pi \si_1^2}^d$.
\STATE Let the number of levels be $M=\ce{\log_{1+\al}\pf{\si_{\max}^2}{\si_1^2}}+1$.
\STATE For each $1\le i\le M$ let Let $\si_{i} = \si_1(1+\al)^{(i-1)/2}$.
\FOR{$1\le i\le M-1$}
  \STATE Let $f_{i+1}(x) = \fc{\ve{x}^2}{2\si_{i+1}^2} + f(x)$.
  \STATE Run algorithm $\cal A\pa{x_0,f_{i+1},\ep=\fc{\si_i}8}$ to obtain $S=2^{10}M$ samples $x_i$, and let $\wh r_i = \rc{S}\sumo j{S} \ve{x_j}$.
  \STATE Let $r_i^+ = \wh r_i + \si_i\sqrt{2(1+\al)\log\pf8{\ep} }+ \rc 4$.
\ENDFOR
\STATE Run algorithm $\cal A\pa{x_0,f,\ep=\fc{1}{8\sqrt\mu}}$ to obtain $S=2^{10}M$ samples $x_j$, and let $\wh r_M = \rc{S}\sumo j{S} \ve{x_j}$.
\STATE Let $r_M^+ = \wh r_M + \rc{\sqrt{\mu}}\sqrt{2\log\pf8{\ep}} + \rc 4$.
\FOR{$1\le i\le M$}
    \STATE If $i=M$ set $\al\leftarrow \infty$.
	\STATE Let $g_i(x) = \exp\pf{\ve{x}^2}{2\si_i^2(1+\al^{-1})}$.
	\STATE Let $h_i(x) = g_i(x) \wedge  \exp\pf{r_i^{+2}}{2\si_i^2(1+\al^{-1})}$.
	\STATE Run Algorithm~\ref{a:ml} (Multilevel Monte Carlo) on functions $f_i$, $h_i$ with sampling algorithm $\cal A$ and with parameters set by $\cal B\pa{L+\rc{\si_i^2},\mu+\rc{\si_i^2},L_{h_i},\fc{\ep}{16M}, \fc{\ep}{128\sqrt{M}}}$, to obtain estimate $\wh R_i$ of $\E_{\rh_i} h_i(x)$.
	\STATE Set $\wh Z \leftarrow \wh Z \wh R_i$.
\ENDFOR
\RETURN $\wh Z$
\end{algorithmic}
\end{algorithm}


To prove the correctness of Algorithm~\ref{a:ml-norm}, we rely on guarantees proved in Theorem~\ref{t:muld} and Theorem~\ref{t:muld-rmm}, as well as the truncation in Section~\ref{s:bias}. The final ingredient is to show that Algorithm~\ref{a:ml-norm} estimates $r_i^+$ and $r_M^+$ correctly for the truncation in Section~\ref{s:bias} to work.

\begin{lem}\label{l:r}
Suppose $\al\le \rc4$ and $\si_M^2 \ge \rc{\mu}$. 
In Algorithm~\ref{a:ml-norm}, with probability $\ge \fc 78$ both the following hold:
\begin{enumerate}
\item
For $1\le i<M$, $r_i^+\in \ol r_i + \si_i \sqrt{(1+\al)\log\prc{\ep}}[\sqrt 2,2]$.
\item
$r_M^+\in \ol r_M + \rc{\sqrt\mu} \sqrt{\log\prc{\ep}}[\sqrt 2,2]$.
\end{enumerate}
\begin{proof}
Let $\mu_i$ be the strong convexity constant of $f_i$.
Let $\wt \rh_{i+1}$ be the distribution of the output of $\cal A\pa{x_0,f_{i+1}, \ep = \fc{\si_i}8}$.
By guarantee of algorithm $\cal A$ and the fact that $\ved$ is 1-Lipschitz, $|\E_{x\sim \wt\rh_{i+1}}\ve{x}-\E_{x\sim \rh_{i+1}}\ve{x}|\le \fc{\si_i}8$.

Now  
\begin{align*}
\Var_{x\sim \wt\rh_{i+1}}(\ve{x})
&\le 
\E_{x\sim \wt \rh_{i+1}}\ba{\pa{\ve{x} - \E_{y\sim \rh_{i+1}}\ve{y}}^2}\\
&\stackrel{\mathrm{(i)}}\le  \inf_{(x,y)\in \cal C(\wt\rh_{i+1},\rh_{i+1})} \ba{
\E_{(x,y)} \ba{(\ve{x}-\ve{y})^2}^{\rc 2} + \E_{y\sim \rh_{i+1}} \ba{\pa{\ve{y} - \E_{y\sim \rh_{i+1}} \ve{y}}^2}^{\rc 2} 
}^2\\
&=\pa{W_2(\wt\rh_{i+1},\rh_{i+1})+\Var_{y\sim \rh_{i+1}}(\ve{y})^{\rc 2}}^2\\
&\stackrel{\mathrm{(ii)}}\le
\pa{\fc{\si_i}{8} + \fc 54 \si_i}^2\le \pa{\fc{11}8\si_i}^2 \le 2\si_i^2 
\end{align*}
where in (i) we use Minkowski's inequality and in (ii) we use the fact that 
$\Var_{x\sim \rh_{i+1}}(\ve{x})\le \rc{\mu_{i+1}}$ by Theorem~\ref{t:be}, and
for $\al\le \rc 4$, 
$\Var_{x\sim \rh_{i+1}}(\ve{x})\le 
\rc{\mu_{i+1}}\le \si_{i+1}^2 \le \si_i^2(1+\al)^2\le \pf 54^2 \si_i^2$. 
Then since $S=2^{10}M$,
\begin{align*}
\Var_{x_j\sim \wt\rh_{i+1}}\pa{\rc{S}\sumo j{S}\ve{x_j}}
&\le \fc{2\si_i^2}{2^{10}M}
= \fc{\si_i^2}{2^9M}.
\end{align*}
Thus by the triangle inequality and the bound on the bias,
\begin{align*}
\Pj\pa{\ab{\wh r_i - \E_{x\sim \rh_{i+1}}\ve{x}}\ge \fc{\si_i}4}
&\le 
\Pj\pa{\ab{\wh r_i - \E\wh r_i}\ge \fc{\si_i}8}
\le \fc{\si_i^2/(2^9M)}{\si_i^2/2^6}
\le \rc{8M}.
\end{align*}
The analogous statement for $i=M$ follows similarly with $f_{M+1}= f$ and $\rh_{M+1}=\rh$ by noting $\Var_{x\sim \rh}(\ve{x}) \le \rc{\mu} \le \si_M^2$, using the assumption on $\si_M^2$.
By the union bound, 
letting $s_i = \begin{cases}
\si_i, &1\le i< M\\
\rc{\sqrt{\mu}}, & i=M,
\end{cases}$
we have
$\Pj\pa{\forall i\in [1,M], 
\ab{\wh r_i - \E_{x\sim \rh_{i+1}}\ve{x}}\le \fc{s_i}4}\ge \fc 78$.
Under this event, for $1\le i\le M-1$,
\begin{align*}
r_i^+&\in \ol r_i +\si_i\pa{\sqrt{2(1+\al)\log \prc{\ep_1}}+\ba{0,\rc 2}}
\subeq \ol r_i + \si_i\sqrt{(1+\al)\log \prc{\ep}} [\sqrt 2,2],
\end{align*}
and for $i=M$,
\begin{align*}
r_M^+&\in \ol r_M + \rc{\sqrt{\mu}}\pa{\sqrt{2\log \prc{\ep_1}}+\ba{0,\rc 2}}
\subeq \ol r_M + \rc{\sqrt{\mu}}\sqrt{\log \prc{\ep}} [\sqrt 2,2]\qedhere
\end{align*}
\end{proof}
\end{lem}

Finally we are ready to state and prove the main theorems.

\begin{thm}[Multilevel ULD for estimating the normalizing constant]\llabel{t:muld-norm}
Let $f(x)$ be $\mu$-strongly convex and $L$-smooth. Let $Z=\int_{\R^d} e^{-f(x)}\dx$. 
Let $\al = \fc{\log 2}{2\sqrt d \log\pf{8}{\ep}}\wedge \rc 4$, 
$\si_1 = \fc{\ep}{8dL}$, and $\si_{\max} = 4\pf{\sqrt d\vee \sqrt{\log\pf8{\ep}}}{\mu}\pa{1\vee \rc{\sqrt\mu}}$.
Algorithm~\ref{a:ml-norm} with Algorithm~\ref{a:uld-couple} as the sampling algorithm  $\cal A$, with parameters set by Theorem~\ref{t:muld} computes $\wh Z$ such that with probability $\fc 34$, $\fc{\wh Z}{Z}\in [1-\ep,1+\ep]$. The number of queries to $\nabla f(x)$ is $\wt \cO\pf{d^{\fc 32}\ka^2}{\ep^2}$.
\end{thm}
\begin{proof}
Let $\mu_{i}$, $L_i$, $\ka_i$ be the strong convexity constant, smoothness constant, and condition number of $f_i$. 
Note that $\ka_i =\fc{L_i}{\mu_i}\le \fc{L+\rc{\si_i^2}}{\mu + \rc{\si_i^2}}\le \ka$, so we can always bound the dependence on $\ka_i$ by $\ka$; we will use this fact implicitly. 
Let $\wt\rh_i$ be the distribution of $x^{\eta_k}_T$, where $\eta_k,T$ are the smallest step size and time for the $i$th temperature. 
Let $\wh R_i$ be the estimate at the $i$th temperature, $\wt R_i := \E \wh R_i = \E_{\wt \rh_i}h_i$, and $R_i = \E_{\rh_i}g_i$. For ease of computation, let $\ep_1$ and $\ep_2\le \rc4$ be such that $\si_1 = \fc{\ep_1}{dL}$ (our assumption has $\ep_1=\fc{\ep}{8}$) and $\si_{\max} =  4\pf{\sqrt d\vee \sqrt{\log\pf1{\ep_2}}}{\mu}\pa{1\vee \rc{\sqrt\mu}}$ and $\al = \fc{\log 2}{2\sqrt d \log\pf{1}{\ep_2}}\wedge \rc 4$ (our assumption has $\ep_2=\fc{\ep}8$).

By assumption on $\al$ and $\si_{\max}$, by Lemma~\ref{l:r}, with probability $\ge\fc 78$, Corollary~\ref{c:bias}(1) is satisfied for $(r^+,\ol r)=(r_i^+,\ol r_i)$ for $1\le i\le M-1$ and (2) is satisfied for $(r^+,\ol r)=(r_M^+,\ol r_M)$. Then $L_{h_i}=\cO\prc{\si_i}$ and  $L_{h_M}=\cO\pa{\sqrt{\mu}}$. In either case, $L_{h_i}=\cO(\sqrt{\mu_i})$ and $h_i/R_i$ is $L_{h_i}$-Lipschitz. For the rest of the proof, we will condition on the event that the hypothesis of Corollary~\ref{c:bias} are satisfied.

By Corollary~\ref{c:bias}, $\fc{|\wt R_i-R_i|}{R_i}\le \ep_2 + L_h W_2(\rh, \wt \rh_i)$. 
In order to make $\Var(\wh R_i) \le \fc{\ep_2^2R_i^2}{256M}=: \ep_\si^2R_i^2$ and $|\wt R_i - R_i|\le \fc{\ep_2 R_i}{2M} =:\ep_bR_i$, by Theorem~\ref{t:muld}, the number of queries required is
\begin{align*}
Q=
\cO\pa{
\ka_i^2\sqrt d  \log \pa{\fc{L_{h_i}}{\ep_b}\cdot \sfc{d}{\mu_i}} 
 \pa{\fc{L_{h_i}^2}{\mu_i\ep_\si^2} + \fc{L_{h_i}}{\sqrt{\mu_i} \ep_b}}
}
&= 
\cO\pa{\ka^2 \sqrt d \log\pf{\sqrt dM}{\ep_2}\pa{\fc{M}{\ep_2^2} + \fc{M}{\ep_2}}}
\end{align*}
where we substitute $\ep_b$ and $\ep_\si$ and use $L_{h_i}=\cO(\sqrt{\mu_i})$. 

Also by Theorem~\ref{t:muld}, $W_2(\rh_i, \wt \rh_i)\le \fc{\ep_b}{L_{h_i}}$,  so $\fc{|\wt R_i - R_i|}{R_i}\le \ep_2 + L_{h_i} W_2(\rh_i, \wt \rh_i)\le \ep_2+\fc{\ep_2}{2M}\le \rc 2$, where in the last step we use $\ep_2\le \rc 4$.
Hence $\wt R_i \ge \rc 2 R_i$ and 
$\Var(\wh R_i) \le \fc{\ep^2R_i^2}{256M}\le \fc{\ep^2 \wt R_i^2}{64M}$.

By choice of $\si_1$, by Lemma~\ref{l:start}, $1\le \fc{\hat Z_1}{Z_1}\le \rc{1-\ep_1}\le e^{2\ep_1}$. We also have $|\wt R_i - R_i|\le \fc{\ep_2 R_i}{2M}$ and $\Var(\wh R_i) \le \fc{\ep_2^2 \wt R_i^2}{64M}$.
By Lemma~\ref{l:combine}, $\Pj\pa{\fc{\wh Z}{Z}\nin [e^{-(2\ep_1+2\ep_2)},e^{(2\ep_1+2\ep_2)}}\le \rc 8$. Taking $\ep_1=\ep_2=\fc{\ep}8$ as in our assumptions, and recalling that we conditioned on an event of probability $\ge \fc 78$, we have that $\fc{\wh Z}{Z}\in [1-\ep,1+\ep]$ with probability $\ge \fc 34$.

The total number of levels is $M=\log_{1+\al}\pf{\si_M^2}{\si_1^2}=\wt \cO(\sqrt d)$. The total query complexity is $QM =\wt \cO \pa{\ka^2 \sqrt d M^2} =  \wt \cO\pf{\ka^2 d^{\fc 32}}{\ep^2}$.
\end{proof}

\llabel{s:ml-norm-uld-rmm}

\begin{thm}[Multilevel ULD-RMM for estimating the normalizing constant]\llabel{t:muld-rmm-norm}
Let $f(x)$ be $\mu$-strongly convex and $L$-smooth. Let $Z=\int_{\R^d} e^{-f(x)}\dx$. 
Define $\al$, $\si_i$, and $\si_{\max}$ as in Theorem~\ref{t:muld-norm}, and let $x_0=x^*=0$.
Algorithm~\ref{a:ml-norm} with Algorithm~\ref{a:uld-rmm-couple} as the sampling algorithm $\cal A$, with parameters set by Theorem~\ref{t:muld-rmm} computes $\wh Z$ such that with probability $\fc 34$, $\fc{\wh Z}{Z}\in [1-\ep,1+\ep]$. The number of queries to $\nabla f(x)$ is $\wt \cO\pf{d^{\fc 43}\ka + d^{\fc 76}\ka^{\fc 76}}{\ep^2}$. 
\end{thm}
Note that we assume $x_0=x^*$ as Theorem~\ref{t:rmm} makes that assumption; however, we note that we can use gradient descent to approximately find $x^*$, and that the analysis of~\cite{shen2019randomized} can tolerate a warm start.
\begin{proof}
The proof is the same as Theorem~\ref{t:muld-norm}. The only difference is that the number of queries at a level is given by Theorem~\ref{t:muld-rmm} instead:
\begin{align*}
Q&=\cO\Biggl(\Biggl(\ka^{\fc 76}d^{\rc 6}\log\pa{\fc{L_g^2}{\ep_b^2}\cdot \fc{d}{\mu}}^{\fc 76} + \ka d^{\rc 3}\log\pa{\fc{L_g^2}{\ep_b^2}\cdot \fc{d}{\mu}}^{\fc 43}\Biggr)
\fc{L_g^2}{\mu\ep_\si^2} \\
& \qquad \qquad +
\fc{\ka^{\fc 76}d^{\rc 6}L_g^{\fc 13}}{\ep_b^{\fc 13}}\log\pa{\fc{L_g^2}{\ep_b^2}\cdot \fc{d}{\mu}}^{\fc 76} + \fc{\ka d^{\rc 3}L_g^{\fc 23}}{\ep_b^{\fc 23}}\log\pa{\fc{L_g^2}{\ep_b^2}\cdot \fc{d}{\mu}}^{\fc 43}\Biggr)\\
&=
\cO\pa{\ka^{\fc 76}d^{\rc 6}
\pa{\fc{M}{\ep^2} + \fc{M^{\fc 13}}{\ep^{\fc 13}}}
+ \ka d^{\rc 3}
\pa{\fc{M}{\ep^2} + \fc{M^{\fc 23}}{\ep^{\fc 23}}}
}
\end{align*}
The total query complexity is $QM = \wt \cO((d^{\fc 13}\ka + d^{\fc 16}\ka^{\fc 76}) M^2) = \wt\cO\pf{d^{\fc 43}\ka + d^{\fc 76}\ka^{\fc 76}}{\ep^2}$.
\end{proof}

\section{Proof of Lowerbound}
\label{sec:lowerbound_appendix}

In this section we prove the lowerbound. More precisely we prove Theorem~\ref{thm:lowerbound} below:

\begin{thm*}[Theorem \ref{thm:lowerbound}] For any fixed constant $\gamma > 0$, 
for large enough $d$, 
given query access to gradient or function value of a function $f:\R^d \to \R$ that is 1.5-smooth and $0.5$-strongly convex, any algorithm that makes $o\left(d^{1-\gamma}\ep^{-(2-\gamma)}\right)$ queries cannot estimate the normalizing constant $Z = \int_{\R^d} e^{-f(x)} dx$ within a multiplicative factor of $1\pm\ep$ with probability more than $3/4$.
\end{thm*}


As we explained earlier, we will first prove a lowerbound when the dimension is a small constant $k$. 

\begin{thm*}[Theorem~\ref{thm:lowerbound_constant}] For any fixed integer $k > 0$, given query access to gradient or function value of a function $f:\R^k \to \R$ that is 1.5-smooth and $0.5$-strongly convex, any algorithm that makes $o(\ep^{-\frac{2}{1+4/k}})$ queries cannot estimate the normalizing constant $Z = \int_{\R^k} e^{-f(x)} dx$ within a multiplicative factor of $1\pm\ep$ with probability more than $3/4$.
\end{thm*}

Theorem~\ref{thm:lowerbound_constant} relies on an information theoretic approach, whose core is based on the well-known result on biased coin:

\begin{claim} \label{clm:biasedcoin}
Given independent samples of a random variable $X$, where $X$ is drawn from Bernoulli distribution with either $p=1/2+\delta$ or $p=1/2-\delta$, any algorithm that looks at $o(1/\delta^2)$ samples of $X$ cannot decide which distribution $X$ is drawn from without probability better than $1/2+c$ for any constant $c>0$.
\end{claim}

This is very standard and we give a proof here just for completeness.

\begin{proof}
Let $Y$ and $Z$ be two Bernoulli random variables with $p_Y = 1/2+\delta$ and $p_Z = 1/2-\delta$ of being 1 respectively. Then the KL-divergence between these two distributions is $KL(Y\|Z) \le \cO(\delta^2)$. Let $Y^n$ and $Z^n$ be $n$ independent samples of $Y$ and $Z$; by a property of KL divergence we know $KL(Y^n\|Z^n) = nKL(Y\|Z) \le \cO(n\delta^2)$. When $n = o(1/\delta^2)$, $KL(Y^n\|Z^n) = o(1)$. Finally by Pinsker's inequality we know the TV-distance between $Y^n$ and $Z^n$ is at most $\sqrt{KL(Y^n\|Z^n)/2} = o(1)$. Therefore it is impossible to distinguish between $Y^n$ and $Z^n$ with any probability $1/2+c$ for constant $c>0$.
\end{proof}

The proof of Theorem~\ref{thm:lowerbound_constant} proceeds by constructing a hard distribution with many independent cells. Intuitively, we start from a basic function $f_0(x) = \frac{\|x\|^2}{2}$ and will modify it in the cube $[-1/\sqrt{k}, 1/\sqrt{k}]^k$. The cube is going to be partitioned into $n$ cells by partitioning each dimension as $n^{1/k}$ intervals of length $2l$ each, where $l := 1/(\sqrt{k}n^{1/k})$. As explained in the main text we will use $I_i$ to denote the $i$-th interval, and a $k$-tuple $(i_1,i_2,...,i_k)\in\{1,2,...,n^{1/k}\}^k$ to denote a cell $I_{i_1}\times I_{i_2}\times \cdots \times I_{i_k}$ in $\R^k$.

To ensure that we can modify each cell independently, we will first construct a function on a cube whose function value, gradient and Hessian vanishes on the boundary.

\paragraph{Construction of function $q$}

First, the lowerbound construction needs a function $q$ which we use to modify the initial function $f_0$. We construct such a $q$ function in the following lemma: 

\begin{lemma}\label{lem:qproperty}
There exists a function $q:[-1,1]^k\to \R$ that satisfies
\begin{enumerate}
    \item For any $x\in[-1,1]^k$ with at least one coordinate $x_i = \pm 1$, $q(x) = 0$, $\nabla q(x) = 0$ and $\nabla^2 q(x) = 0$.
    \item For any $x\in[-1,1]^k$, $0 \le q(x) \le 1$, $\norm{\nabla^2 q(x)}\le 36k$. 
    \item For any $x\in[-1/2,1/2]^k$, $q(x) \ge 3^{-k}$.
\end{enumerate}
\end{lemma}

\begin{proof}
We construct $q$ as a product of individual coordinates. Let $p:[-1,1]\to \R$ be the function $p(x) = (1+x)^3(1-x)^3$. It is easy to verify that $p(-1) = p'(-1) = p''(-1) =  p(1) = p'(1) = p''(1) = 0$, $0 \le p(x) \le 1$ and $p(x)\ge 1/3$ when $x\in[-1/2,1/2]$.

Now we define $q(x) = p(x_1)p(x_2)\cdots p(x_k)$. If any coordinate $x_i$ $(i=1,2,\ldots,k)$ is $1$ or $-1$, we have $q(x) = 0$ because $p(x_i) = 0$. The gradients $\frac{\partial q}{\partial x_i} = p'(x_i)p(x_1)p(x_2)\cdots p(x_{i-1})p(x_{i+1})\cdots p(x_k) = 0$; for any $j\ne i$, $\frac{\partial q}{\partial x_j}$ has a factor of $p(x_i)$ so it is also 0. Similarly, all the second order partial derivatives will have a factor of $p(x_i), p'(x_i)$ or $p''(x_i)$, so the Hessian is also 0. Therefore we have verified Property 1.

For Property 2, we observe that for $i\ne j$, $\frac{\partial^2 q}{\partial x_i\partial x_j} (x) = p'(x_i)p'(x_j)\prod_{t\ne i,j} p(x_t)$. It is easy to verify that $|p'(x_i)| \le 6$ for any value of $x_i\in[-1,1]$, therefore $|\frac{\partial^2 q}{\partial x_i\partial x_j} (x)| \le 36$. Similarly, we also know for any $i$, $|\frac{\partial^2 q}{(\partial x_i)^2}(x)|\le 36$. Therefore, the Hessian matrix $\nabla^2 q(x)$ is a $k\times k$ matrix with entries no larger than $36$, so we have $\norm{\nabla^2 q(x)} \le \norm{\nabla^2 q(x)}_F \le 36 k.$

Property 3 follows immediately from $p(x)\ge 1/3$ when $x\in[-1/2,1/2]$.
\end{proof}

Using such a function, in each cell we can just add a multiple of (scaled and shifted version of) this function. We can choose the multipliers independently without worrying about the smoothness of the original function because of properties of $q$. This allows us to construct functions as in Lemma~\ref{lem:function_construct}.

\begin{lem*}[Lemma \ref{lem:function_construct}]
For any $n$ where $n^{1/k}$ is an integer, and $l = 1/(\sqrt{k}n^{1/k})$. For each cell $\tau = (i_1,...,i_k)$, let $v_\tau$ be its center. Construct the function $f(x)$ as
\[
f(x) = \left\{\begin{array}{cl}f_0(x), & \mbox{cell $\tau$ is of type 1} \\ f_0(x)+c_\tau q\left(\frac{1}{l}(x-v_\tau)\right), &\mbox{cell $\tau$ is of type 2.}\end{array}\right.
\]
Here $q$ is the function constructed in Lemma~\ref{lem:qproperty}.
There exists a way to choose $c_\tau$'s such that no matter what types each cell has, the family of functions satisfies the following properties:
\begin{enumerate}
    \item $f(x)$ is $1.5$-smooth and $0.5$-strongly convex.
    \item The normalizing constant $Z_f = \int_{\R^k} e^{-f(x)}dx = (2\pi)^{k/2} - C\frac{n_2}{n}$, where $n_2$ is the number of type-2 cells, and $C$ is at least $\Omega\left(l^2\right)$.
\end{enumerate}
\end{lem*}

\paragraph{Proof of Lemma~\ref{lem:function_construct}} Using the construction of $q$, one can select a type for each of the cell and construct a corresponding function as in Lemma~\ref{lem:function_construct}. We give the proof of the lemma here:

\begin{proof}
First, by Lemma~\ref{lem:qproperty}, the $q$ function has 0 value, gradient and Hessian at the boundary. Therefore the function value, gradient and Hessian of $f(x)$ agrees with $f_0(x)$ on the boundary. As a result, the function we construct is still twice differentiable on every point. 

For any cell $\tau$, by Lemma~\ref{lem:qproperty} the function $q\left(\frac{1}{l}(x-v_\tau)\right)$ for $x\in \tau$ has Hessian bounded by $\frac{36k}{l^2}$. We will make sure that every $c_\tau$ is bounded by $\frac{l^2}{72k}$, so the function $c_\tau q\left(\frac{1}{l}(x-v_\tau)\right)$ has a Hessian with spectral norm at most $1/2$. Since $\nabla^2 f(x) = \nabla^2 f_0(x) + c_\tau \nabla^2 q\left(\frac{1}{l}(x-v_\tau)\right)$, by standard matrix perturbation bounds, the Hessian of $f$ always satisfies $0.5I\preceq \nabla^2 f \preceq 1.5I$, which implies $f(x)$ is $1.5$-smooth and $0.5$-strongly convex.

For the second property, note that $f(x) \ge f_0(x)$ as both $c_\tau$ and $q$ are positive. Therefore $\int e^{-f(x)}dx$ is always smaller than $\int e^{-f_0(x)}dx$. For each cell $\tau$, let \[
C_\tau = \int_{x\in \tau} \left[\exp(-f_0(x)) - \exp\left(-f_0(x) + \frac{l^2}{72k}q\left(\frac{1}{l}(x-v_\tau)\right)\right)\right]dx.
\] Therefore $C_\tau$ is the amount of decrease in normalizing constant if we choose $c_\tau = \frac{l^2}{72k}$ (the maximum allowed value). We first show a lowerbound on $C_\tau$:

\begin{align*}
    C_\tau & = \int_{x\in \tau} \left[\exp(-f_0(x)) - \exp\left(-f_0(x) + \frac{l^2}{72k}q\left(\frac{1}{l}(x-v_\tau)\right)\right)\right]dx\\
    & = \int_{x\in \tau} e^{-f_0(x)} \left[1 - \exp\left(-\frac{l^2}{72k}q\left(\frac{1}{l}(x-v_\tau)\right)\right)\right]dx \\
    & \ge \int_{\|x-v_\tau\|_\infty \le l/2} e^{-f_0(x)} \left[1 - \exp\left(-\frac{l^2}{72k}q\left(\frac{1}{l}(x-v_\tau)\right)\right)\right]dx \\
    & \ge \int_{\|x-v_\tau\|_\infty \le l/2} e^{-f_0(x)} \pa{1 - \exp\pa{-\frac{l^2}{72k3^k}}}dx \\
    & \ge \int_{\|x-v_\tau\|_\infty \le l/2} e^{-1} \pa{1 - \exp\pa{-\frac{l^2}{72k3^k}}}dx \\
    & = \Omega(l^{2+k}) =p \Omega\left(\frac{l^2}{k^{k/2}n}\right). 
\end{align*}

Let $\tau^*$ be the cell with the smallest $C_{\tau^*}$, set $c_{\tau^*} = \frac{l^2}{72k}$. Set all the $c_\tau$'s carefully in $[0,\frac{l^2}{72k}]$ so that the decrease in every cell is equal to $C_{\tau^*}$ (this is always possible because the amount of decrease is continuous and monotonically increasing with respect to $c_\tau$), and we have the second property. 
\end{proof}

\paragraph{Proof of Theorem~\ref{thm:lowerbound_constant}}

Now we are ready to prove the lowerbound Theorem~\ref{thm:lowerbound_constant} for a constant number of dimensions.

\begin{proof}
Fix an desired accuracy $\delta$ small enough, choose $n\ge 100/\delta^2$ and make sure $n^{1/k}$ is an integer (when $\delta < 1$ we still have $n = \cO(1/\delta^2)$).

Consider two distributions of functions $\cF_1$ and $\cF_2$. In $\cF_1$, each cell is of type 1 with probability $1/2+\delta$ independently, in $\cF_2$, each cell is of type 1 with probability $1/2-\delta$ independently. After the types of cells are decided, function $f$ is constructed according to Lemma~\ref{lem:function_construct}.

Clearly, querying any point of $f(x)$ (whether the query is on function value or gradient) can give information about at most one cell. Therefore by Claim~\ref{clm:biasedcoin}, any algorithm that makes $o(1/\delta^2)$ queries will not be able to distinguish whether the function comes from $\cF_1$ or $\cF_2$ with probability better than $0.6$.

On the other hand, by standard concentration bounds and the fact that $n\ge 100/\delta^2$, we know with at least $0.99$ probability functions in $\cF_1$ has at most $n(1-\delta)/2$ type 2 cells, and functions in $\cF_2$ has at least $n(1+\delta)/2$ type 2 cells. By Lemma~\ref{lem:function_construct}, we know with probability at least $0.99$, the normalizing constant $Z \ge (2\pi)^{k/2} - C(1-\delta)/2 =: \theta_1$ for $f\sim \cF_1$, and $Z\le (2\pi)^{k/2} - C(1+\delta)/2 =: \theta_2$ for $f\sim \cF_2$. Therefore, if an algorithm can estimate the normalizing constant with accuracy better than $\sqrt{\theta_1/\theta_2} - 1$ with probability $3/4$, it is going to be able to distinguish $\cF_1$ and $\cF_2$ with probability better than $0.6$, which is impossible.

Now, by Lemma~\ref{lem:function_construct}, we know $C = \Omega(l^2/k^{k/2})$, therefore $\theta_1/\theta_2 = 1 + \Omega(C\delta/(2\pi)^{k/2}) = 1 + \Omega(l^2\delta/(2\pi k)^{k/2})$. Using the fact that  $l = 1/(\sqrt{k}n^{1/k})$ and $n = \Theta(1/\delta^2)$, we know $\sqrt{\theta_1/\theta_2} - 1 = \Omega\left(\frac{\delta^{1+4/k}}{k(2\pi k)^{k/2}}\right)$. The Theorem follows by choosing $\delta$ such that $\ep = \Theta \left(\frac{\delta^{1+4/k}}{k(2\pi k)^{k/2}}\right)$. When $k$ is a constant this gives the desired trade-off.
\end{proof}

\paragraph{Proof of Theorem~\ref{thm:lowerbound}}
Finally we extend Theorem~\ref{thm:lowerbound_constant} to Theorem~\ref{thm:lowerbound}.

\begin{proof}[Proof of Theorem~\ref{thm:lowerbound}]
The proof is very similar to Theorem~\ref{thm:lowerbound_constant}. Fix a constant $k$ depending only on $\gamma$ that we will determine later. We will break the $d$ coordinates of input $x$ into $d'= \lfloor d/k\rfloor$ groups of size $k$ each (ignoring the remainder). Let $x_{S_i}$ be the input $x$ restricted to the $i$-th group of coordinates.  The function we construct will be a sum of functions $f(x) = \sum_{i=1}^{d'} f_i(x_{S_i})$.

Fix an desired accuracy $\delta$ small enough, choose $n\ge 100/\delta^2$ and make sure $n^{1/k}$ is an integer (when $\delta$ is small enough we still have $n = \cO(1/\delta^2)$). 

Consider two distributions of functions $\cF_1$ and $\cF_2$ same as in the proof of Theorem~\ref{thm:lowerbound_constant}. When $f \sim \cF_1$, construct $f_1,f_2,...,f_{d'}$ independently using Lemma~\ref{lem:function_construct}, where each cell is of type 1 with probability $1/2+\delta$; when $f\sim \cF_2$, construct $f_1,f_2,...,f_{d'}$ independently using Lemma~\ref{lem:function_construct}, where each cell is of type 1 with probability $1/2-\delta$.

It is easy to see that the normalizing constant for $f$ is the product of normalizing constant of $f_1,f_2,...,f_{d'}$. By construction in Lemma~\ref{lem:function_construct} and calculations in Theorem~\ref{thm:lowerbound_constant}, there exists a constant $Z$ such that the normalizing constant for $f_i$ is $Z(1+\Omega\left(\frac{\delta^{1+4/k}}{k(2\pi k)^{k/2}}\right))$ with probability at least 0.99 when $f\sim \cF_1$, and $Z(1-\Omega\left(\frac{\delta^{1+4/k}}{k(2\pi k)^{k/2}}\right))$ with probability at least 0.99 when $f\sim \cF_2$. When $\delta^{1+4/k}d \le 1/5$, by Lemma~\ref{lem:prodvar} we know with probability at least 0.99, the normalizing constant for $f\sim \cF_1$ is at least $Z^{d'}(1+\Omega\left(\frac{\delta^{1+4/k}d}{k(2\pi k)^{k/2}}\right))=:\theta_1$, and the normalizing constant for $f\sim \cF_2$ is at most $Z^{d'}(1-\left(\frac{\delta^{1+4/k}d}{k(2\pi k)^{k/2}}\right))=:\theta_2$. 
When the number of queries is $o(1/d\delta^2)$, no algorithm can distinguish between these two distributions, which means no algorithm can estimate the normalizing constant with accuracy better than $\sqrt{\theta_1/\theta_2}-1 = \Theta\left(\frac{\delta^{1+4/k}d}{k(2\pi k)^{k/2}}\right)$. 

If we set $\ep = \Theta\left(\frac{\delta^{1+4/k}d}{k(2\pi k)^{k/2}}\right)$, then (when $k$ is a constant that only depends on $\gamma$) any algorithm that uses $o\left(d^{\frac{1-4/k}{1+4/k}}\epsilon^{-\frac{2}{1+4/k}}\right)$ queries cannot estimate the normalizing constant with multiplicative error $1\pm \ep$ with probability better than $3/4$. Finally, we choose $k = \lceil 8/\gamma\rceil$, so $\frac{2}{1+4/k} \ge 2-\gamma$ and $\frac{1-4/k}{1+4/k} \ge 1-\gamma$, which gives the guarantee in the theorem.
\end{proof}
\section{Quadrature Method for Estimating the Normalizing Constant}
\label{s:quad}
Alternative to the Monte Carlo strategy as discussed, for lower
dimensions, a deterministic quadrature scheme for
$Z = \int e^{-f(x)} \ud x$ might be computationally less expensive. 

First, we recall that for $X \in \RR^d$ a random variable distributed according to a logconcave
distribution with $\EE(\norm{X}^2) \leq R^2$. Restricted the support of
$X$ to a ball with radius $2R \log (1/ \veps)$ captures at least
$1 - \veps / 2$ fraction of the mass. Thus it suffices to integrate $e^{-f(x)}$ inside a square $Q_{R_0}$ centered at the origin of radius 
$R_0 = 2 \sqrt{d / \mu} \log(1/\veps)$. 


Inside the square $Q_{R_0}$, we  use a trapezoidal quadrature rule
with grid spacing $h$ to integrate $e^{-f(x)}$. Denote the estimate from quadrature as $S_h$, the error  is bounded from above by
\begin{equation}\label{eq:quaderror}
  \Bigl\lvert \int_{Q_{R_0}} e^{-f(x)} \ud x - S_h \Bigr\rvert \leq C\, \mathrm{vol}(Q_{R_0}) h^2 d^2 \max_{x \in Q_{R_0}} \norm{\nabla^2 (\exp(-f(x)))}. 
\end{equation}
The Hessian of $e^{-f(x)}$ can be bounded from above by 
\begin{equation*}
  \begin{aligned}
    \max_{x \in Q_{R_0}} \norm{\nabla^2(\exp(-f(x)))} & = \max_{x \in Q_{R_0}}
    \bigl( \norm{ \nabla^2 f(x)}  + \norm{\nabla f(x)}^2 \bigr) \exp(-f(x))  \\
    & \leq \max_{x \in Q_{R_0}}
    \bigl( L  + L^2 \norm{x}^2 \bigr) \exp(-\frac{\mu}{2} \norm{x}^2) \\
    & \leq L + \frac{2}{e} \frac{L^2}{\mu} \\
    & \leq C L (1 + \kappa).
  \end{aligned}
\end{equation*}
Thus, to make the right hand side of \eqref{eq:quaderror} error $\veps$, we need
\begin{equation*}
  h \leq C d^{-1} L^{-1/2} ( 1 + \kappa)^{-1/2}\, \mathrm{vol}(Q_{R_0})^{-1/2} \veps^{-1/2}. 
\end{equation*}
The number of quadrature points is given by 
\begin{equation*}
  \begin{aligned}
    N & = \Or\bigl(\mathrm{vol}(Q_{R_0})^{1+d/2} L^{d/2} ( 1 + \kappa)^{d/2} \veps^{d/2} \bigr) \\
    & = \wt{\Or}\Bigl( \Bigl(\frac{d}{\mu}\Bigr)^{d/2 + d^2/4} d^d L^{d/2} (1 + \kappa)^{d/2} \veps^{d/2} \Bigr).
\end{aligned}
\end{equation*}
While this complexity has a better dependence in $\veps$ for low
dimension ($ d \leq 3$), the dependence in dimension is much worse
than that of the Monte Carlo method.

\section{Tools and Auxiliary Lemmas}

We note some concentration results and functional inequalities for log-concave distributions.

\begin{lem}[Concentration around mode for log-concave distributions]\label{l:lcc}
\label{l:conc-mode}
Suppose $f:\R^d\to \R$ is a convex $\rc{\si^2}$-strongly convex function with minimum at 0,  
and let $\rh$ be a probability measure on $\R^d$ with $\rh(dx) \propto e^{-f(x)}\dx$.  
Then for any $r$, $\Pj_{x\sim \rh}(\ve{x}\ge r) \le \Pj_{x\sim N(0,\si^2)}(\ve{x}\ge r)$.
\end{lem}
\begin{proof}
Without loss of generality, $f(0)=0$.
Using spherical coordinates, we have 
\begin{align}
\Pj_{x\sim \rh} (\ve{x}\ge r) & = \fc{\int_{\bS^{d-1}} \int_r^\iy s^{d-1} e^{-f(sv)}\ud s\ud \bS^{d-1}(v)}{\int_{\bS^{d-1}} \int_0^\iy s^{d-1} e^{-f(sv)}\ud s\ud \bS^{d-1}(v)}
\end{align}
Let
\begin{align}
A(v) &= \int_r^\infty  s^{d-1}e^{-f(sv)}\ud s\ud \bS^{d-1}(v)& 
C &=\int_r^\infty s^{d-1}e^{-\fc{s^2}{2\si^2}}\ud s\ud \bS^{d-1}(v)\\
B(v) &= \int_0^r s^{d-1}f(sv)\ud s\ud \bS^{d-1}(v) &
D &=\int_0^r s^{d-1}e^{-\fc{s^2}{2\si^2}}\ud s\ud \bS^{d-1}(v).
\end{align}
We will show that $\fc{A(v)}{B(v)}\le \fc{C}{D}$. Then $\fc{A(v)}{A(v)+B(v)}\le \fc{C}{C+D}$, so 
\begin{align}
 &\fc{\int_{\bS^{d-1}} \int_r^\iy s^{d-1} e^{-f(sv)}\ud s\ud \bS^{d-1}(v)}{\int_{\bS^{d-1}} \int_0^\iy s^{d-1} e^{-f(sv)}\ud s\ud \bS^{d-1}(v)}
=\fc{\int_{\bS^{d-1}}A(v)\ud \bS^{d-1}(v)}{\int_{\bS^{d-1}}A(v)+B(v)\ud \bS^{d-1}(v)}\\
&
 \le \sup_{v\in \bS^{d-1}} \fc{A(v)}{A(v)+B(v)}\le \fc{C}{C+D} = \Pj_{x\sim N(0,\si^2I_d)}(\ve{x}\ge r).
\end{align}

It suffices to show $\fc{A(v)}{B(v)}\le \fc CD$. For this, we first prove the following claim: If $a,c$ are positive functions on $\Om_1$, and $b,d$ are nonnegative functions on $\Om_2$, then
$\fc{\int_{\Om_1} a(x)\dx}{\int_{\Om_2} b(x)\dx}\le \fc{\int_{\Om_1}c(x)\dx}{\int_{\Om_2}d(x)\dx}$.

To see the claim, note that 
$$
\fc{\int_{\Om_1}c(x)\dx}{\int_{\Om_2}d(x)\dx}
=\fc{\int_{\Om_1}a(x)\cdot \fc{c(x)}{a(x)}\dx}{\int_{\Om_2}b(x)\cdot \fc{d(x)}{b(x)}\dx}
 \ge \fc{\int_{\Om_1}a(x)\dx \cdot \inf_{\Om_1}\fc ca}{\int_{\Om_2}b(x)\dx\cdot  \inf_{\Om_1}\fc db}
 \ge \fc{\int_{\Om_1}a(x)\dx}{\int_{\Om_2}b(x)\dx}.
 $$
 
Now we show that the claim implies $\fc{A(v)}{B(v)}\le \fc CD$. We have
$$
\inf_{s\in [r,\iy)} e^{-\fc{s^2}{2\si^2}+f(s)} 
\le \sup_{s\in [0,r]} e^{-\fc{s^2}{2\si^2}+f(s)}
$$
because $f(s)-\fc{s^2}{2\si^2}$ is an increasing function. Thus the claim implies that $\fc{A(v)}{B(v)}\le \fc CD$.
\end{proof}

\begin{lem}\label{l:d-mode-mean}\label{l:mode-mean}
Let $f:\R^d\to \R$ be a $m$-strongly convex function and let $\rh(dx)\propto e^{-f(x)}\dx$. Let $x^*=\amin_x f(x)$ be the mode. 
Then 
\begin{align*}
\E_{x\sim \rh}\ve{x-x^*} &\le \fc{1}{\sqrt m} (\sqrt d + 2\sqrt{2\log 2}).
\end{align*}
\end{lem}

\begin{proof}
By Lemma~\ref{l:conc-mode} and the $\chi^2$ tail bound from~\cite{laurent2000adaptive},
\begin{align*}
&\Pj\pa{\ve{x-x^*}^2\ge\rc m\pa{ d+2(\sqrt{d\log 2}+\log 2)}}\\
&\le 
\Pj_{x\sim N\pa{0,\fc{\si^2}mI_d}}
\pa{\ve{x-x^*}^2\ge \rc m\pa{ d+2(\sqrt{d\log 2}+\log 2)
}}\le \rc 2
\end{align*}
so $\Pj\pa{\ve{x-x^*} \ge \rc{\sqrt{m}}\pa{\sqrt{d}+\sqrt{2\log 2}}}\le\rc 2$.

By Theorem~\ref{t:be} and Theorem~\ref{t:lsi-conc}, 
\begin{align}
\Pj\pa{\ve{x-x^*}\le \E \ve{x-x^*}-\fc{c}{\sqrt m}}
&\le \exp\pa{-\fc{c^2}2}.
\end{align}
Taking $c=\sqrt{2\log 2}$, we get this is $\le \rc 2$.

Hence the sets 
$\set{x}{\ve{x-x^*}\le 
\rc{\sqrt m}\pa{\sqrt d+\sqrt{2\log 2}}}$ 
and
$\set{x}{\ve{x-x^*} \ge \E \ve{x-x^*} - \fc{\sqrt{2\log 2}}{\sqrt m}}$ 
must interesect, so
\begin{align*}
\E\ve{x-x^*} - \fc{\sqrt{2\log 2}}{\sqrt m} &\le \rc{\sqrt m}\pa{\sqrt d + \sqrt{2\log 2}},
\end{align*}
as needed.
\end{proof}


\begin{thm}[{Bakry-\'Emery~\cite{bakry1985diffusions,bakry2013analysis}}]\label{t:be}
Suppose $f$ is $\mu$-strongly convex. Then $\rh(dx) \propto e^{-f(x)}\dx$ satisfies a Poincar\'e inequality with constant $\rc{\mu}$ ($\rc{\mu}\int_{\R^d} \ve{\nb g(x)}^2 \dx \ge  \Var_\rh(g)$ for all $g$ where the integral is defined) and a log-Sobolev inequality with constant $\rc{\mu}$.
\end{thm}

\begin{thm}[{Log-Sobolev inequality implies Gaussian measure concentration,~\cite[(5.4.2)]{bakry2013analysis}}]\label{t:lsi-conc}
Suppose $\rh(dx)$ is a distribution on $\R^d$ that satisfies a log-Sobolev inequality with constant $C$. Let $g:\R^d\to \R$ be $L$-Lipschitz. Then 
\begin{align*}
\Pj\pa{|g-\E_{\rh} g|\ge r} &\le 2\exp\pa{-\fc{r^2}{2CL^2}}.
\end{align*}
\end{thm}

\end{document}